\documentclass{emulateapj}

\def\arcdeg{\hbox{$^\circ$}}

\def\arcsec{\hbox{$^{\prime\prime}$}}
\def\deg2{\hbox{$\rm deg^{2}$}}

\def\lsim{\mathrel{\rlap{\lower4pt\hbox{\hskip1pt$\sim$}}\raise1pt\hbox{$<$}}}                
\def\gsim{\mathrel{\rlap{\lower4pt\hbox{\hskip1pt$\sim$}}\raise1pt\hbox{$>$}}}                

\lefthead{Drake et~al.}
\righthead{Optical Transient CSS100217}

\begin{document}
\title{The Discovery and Nature of Optical Transient CSS100217:102913+404220$^{\star \dagger}$}

\author{A.J.~Drake\altaffilmark{1}, S.G.~Djorgovski\altaffilmark{1}, 
A.~Mahabal\altaffilmark{1}, J.~Anderson\altaffilmark{2}, 
R.~Roy\altaffilmark{3}, V.~Mohan\altaffilmark{4}, S.~Ravindranath\altaffilmark{4}\\
D.~Frail\altaffilmark{5}, S.~Gezari\altaffilmark{6}, James D.~Neill\altaffilmark{1},
L.C.~Ho\altaffilmark{7}, J.L.~Prieto\altaffilmark{7}, D.~Thompson\altaffilmark{8}, J.~Thorstensen\altaffilmark{9},\\ 
M.~Wagner\altaffilmark{8}, R.~Kowalski\altaffilmark{10}, J.~Chiang\altaffilmark{11}, J.E.~Grove\altaffilmark{12}, 
F.K.~Schinzel\altaffilmark{13}, D.L.~Wood\altaffilmark{12},  L.~Carrasco\altaffilmark{14},\\
E.~Recillas\altaffilmark{14}, L.~Kewley\altaffilmark{15}, K.N.~Archana\altaffilmark{16,17}, Aritra Basu\altaffilmark{17}, 
Yogesh Wadadekar\altaffilmark{17}, Brijesh Kumar\altaffilmark{3},\\ A.D.~Myers\altaffilmark{18},
E.S.~Phinney\altaffilmark{1}, R.~Williams\altaffilmark{1}, M.J.~Graham\altaffilmark{1}, 
M.~Catelan\altaffilmark{19}, E.~Beshore\altaffilmark{10}, S.~Larson\altaffilmark{10}\\ 
and E.~Christensen\altaffilmark{20}
}

\altaffiltext{1}{California Institute of Technology, 1200 E. California Blvd, CA 91225, USA}
\altaffiltext{2}{STSCI, 3700 San Martin Drive Baltimore, MD 21218}
\altaffiltext{3}{Aryabhatta Research Institute of Observational Sciences, Manora Peak, Nainital - 263129, Uttarakhand, India}
\altaffiltext{4}{IUCAA, Postbag 4, Ganeshkhind, Pune 411007, India}
\altaffiltext{5}{NRAO,  Campus Building 65, 949 North Cherry Avenue, Tucson, AZ 85721-0655}
\altaffiltext{6}{Johns Hopkins University, Department of. Physics \& Astronomy, 366 Bloomberg Center, 3400 N. Charles Street Baltimore, MD 21218}
\altaffiltext{7}{Carnegie Observatories, 813 Santa Barbara St., Pasadena, CA 91101}
\altaffiltext{8}{LBT, University of Arizona, 933 N. Cherry Ave, Room 552, Tucson, AZ 85721}
\altaffiltext{9}{Dartmouth College, 6127 Wilder Laboratory,  Hanover, NH 03755-3528}
\altaffiltext{10}{The University of Arizona, Department of Planetary Sciences,  Lunar and Planetary Laboratory, 1629 E. University Blvd, Tucson AZ 85721, USA}
\altaffiltext{11}{W.W. Hansen Experimental Physics Laboratory, Kavli Institute for Particle Astrophysics and Cosmology,
SLAC National Accelerator Laboratory,  Stanford University, Standford, CA 94305}
\altaffiltext{12}{Space Science Division, Naval Research Laboratory, Washington, D.C. 20375.}
\altaffiltext{13}{Max-Planck-Institut f\"ur Radioastronoie, Auf dem H\"ugel 69, 53121 Bonn, Germany}
\altaffiltext{14}{INAOE, Tonantzintla, Puebla, Mexico.}
\altaffiltext{15}{IFA, 640 North Aohoku Place, \#209, Hilo, Hawaii 96720-2700}
\altaffiltext{16}{School of Computer Sciences, Mahatma Gandhi University, Kottayam 686 560, India}
\altaffiltext{17}{National Centre for Radio Astrophysics, TIFR, Post Bag 3, Ganeshkhind, Pune 411007, India.}
\altaffiltext{18}{Department of Astronomy, University of Illinois at Urbana-Champaign, Urbana IL 61801}
\altaffiltext{19}{Pontificia Universidad Cat\'olica de Chile, Departamento de Astronom\'ia y Astrof\'isica, 
Av. Vicu\~na Mackena 4860, 782-0436 Macul, Santiago, Chile}
\altaffiltext{20}{Gemini Observatory, Casilla 603, La Serena, CL, Chile}

\altaffiltext{$\star$}{Some of the data presented herein were obtained at the W.M. Keck Observatory, which is operated as a scientific
partnership among the California Institute of Technology, the University of California and the National Aeronautics
and Space Administration. The Observatory was made possible by the generous financial support of the W.M. Keck
Foundation.}
\altaffiltext{$\dagger$}{
Based on observations made with the NASA/ESA Hubble Space Telescope, obtained at the Space Telescope Science Institute,
which is operated by the Association of Universities for Research in Astronomy, Inc., under NASA contract NAS 5-26555.
These observations are associated with program 12117.
}

\begin{abstract} 
  
  We report on the discovery and observations of the extremely luminous optical transient CSS100217:102913+404220
  (CSS100217 hereafter).  Spectroscopic observations showed this transient was coincident with a galaxy at redshift
  $z=0.147$, and reached an apparent magnitude of $V\sim16.3$.  After correcting for foreground Galactic extinction we
  determine the absolute magnitude to be $\rm M_V =-22.7$ approximately 45 days after maximum light.  Based on our
  unfiltered optical photometry the peak optical emission was $L = 1.3 \times 10^{45}$erg s$^{-1}$,
  and over a period of 287 rest-frame days had an integrated bolometric luminosity of $1.2 \times 10^{52}$ erg.
  
  Analysis of the pre-outburst SDSS spectrum of the source shows features consistent with a Narrow-line Seyfert1
  (NLS1) galaxy. High-resolution HST and Keck followup observations show the event occurred within 150pc of nucleus of the
  galaxy, suggesting a possible link to the active nuclear region. However, the rapid outburst along with photometric 
  and spectroscopic evolution are much more consistent with a luminous supernova.
  Line diagnostics suggest that the host galaxy is undergoing significant star formation.
  
  We use extensive follow-up of the event along with archival CSS and SDSS data to investigate the three most likely
  sources of such an event; 1) an extremely luminous supernova; 2) the tidal disruption of a star by the massive nuclear
  black hole; 3) variability of the central AGN.  We find that CSS100217 was likely an extremely luminous type IIn
  supernova that occurred within range of the narrow-line region of an AGN. We discuss how similar events may have
  been missed in past supernova surveys because of confusion with AGN activity.

\end{abstract}

\section{Introduction}

Exploration of the time domain is now one of the most rapidly growing and exciting areas of astrophysics.  This vibrant
observational frontier has fueled the advent of the new generation of digital synoptic sky surveys, which cover the sky
many times, as well as the necessity of using robotic telescopes to respond rapidly to transient events (Paczynski
2000). However, the discovery of transient astronomical events is by no means new to astronomy with phenomena such as
supernovae being discovered and documented for centuries (Zhao et al.~2006). Transient events themselves have been
observed on timescales from seconds; eg. GRBs (Klebesadel et al.~1973), to years; eg. supernovae (Rest et al.~2009) and
AGN (Ulrich et al.~1997).  Predictions of new types of observable astrophysical phenomena continue to be theorized and
made detectable by advancing technology. For example, the possibility that Massive Compact Halo Objects (MACHOs) could 
be observable due to gravitational microlensing was theorized by Paczynski (1986) and soon proven by Alcock et
al.~(2003) and Aubourg et al.~(1993). Such discoveries set the stage for larger time domain surveys with broader goals.
 
A systematic exploration of the observable parameter space in the time domain is very likely to lead to many new
discoveries (e.g., Djorgovski et al. 2001ab, and references therein).  For example, many types of transient events have
been theoretically predicted, yet remain to be convincingly observed.  Among these are GRB orphan afterglows (Nakar,
Piran, \& Granot 2002), and the electromagnetic counterparts to gravitational-wave inspiral events, where close binaries
coalesce to release a surge of gravitational radiation (Abadie et al.~2010). Nevertheless, the recent detections of rare
types of transients, including candidate pair-instability supernovae (Gal-Yam et al.~2010), tidal disruption events
(Gezari et al. 2009a, van Velzen et al. 2010), and supernova shock breakouts (Soderberg et al.~2008; Schawinski et
al.~2008) show the promise of current and future large transient surveys.

Recent optical surveys searching for transient phenomena include ROTSE (Akertlof et al.~2003), SDSS (Sasar et al.~2007),
Palomar Quest (PQ; Djorgovski et al.~2008), the Catalina Real-time Transient Survey (CRTS; Drake et al.~2009), Palomar
Transient Factory (PTF; Rau et al.~2009), and the Panoramic Survey Telescope \& Rapid Response System (PanSTARRS; Hodapp
et al.~2004). In the near future additional surveys such as SkyMapper (Keller et al.~2007) and the Large Synoptic Survey
Telescope (LSST; Ivezic et al.~2008) will begin operation.  At radio wavelengths work is being undertaken by the low
frequency array (LOFAR; Rottgering 2003), the Allen Telescope Array (ATA, Croft et al.~2009) and will soon begin with
the Australian Square Kilometre Array Pathfinder (ASKAP; Johnston et al.~2007).  At high energies ongoing satellite
searches for transients include the Fermi Large Area Telescope (LAT; Atwood et al.~2009), Swift (Barthelmy et al.~2005),
and Galaxy Evolution Explorer (GALEX; Martin et al.~2005), Rossi X-ray Timing Explorer (RXTE; Jahoda et al. 1996), and
the Monitor of All-sky X-ray Image (MAXI, Ueno et al.~2008).

As the temporal and spatial coverage of transient surveys increases the prospects for discovery of rare events on short
timescales. To this end, the new generation of transient surveys are increasingly working on real-time analysis
and detection. In late 2007, the Catalina Real-time Transient Survey (Drake et al.~2009) simultaneously began real-time
analysis and notification of events in images taken by the Catalina Sky Survey NEO search (CSS; Larson et al. 2003). The
CRTS transient survey currently analyzes data from three telescopes operated by CSS. These telescopes cover $\sim 1800$
square degrees on the sky per night to a depth ranging from $V=19$ to 21.5. New objects are automatically flagged in
real-time and filtered to isolate genuine optical transients from artifacts and other noise sources. In order to
maximize discovery potential all CRTS transients are immediately distributed publicly as
VOEvents\footnote{http://crts.caltech.edu/ and SkyAlert, http://www.skyalert.org/}.  For each of the few dozen transient
events detected per night a portfolio of historical observational information is extracted from past surveys from radio
to gamma ray wavelengths. This information allows most events to be classified into a few broad types of phenomena
including supernovae, blazars and CV outbursts. Objects which are of uncertain nature are examined in greater detail
using Virtual Observatory services such as DataScope\footnote{http://heasarc.gsfc.nasa.gov/cgi-bin/vo/datascope/init.pl}.  Events that
remain a poor match for known types phenomena are of particular interest and are the most closely scrutinized.
 
\section{The Discovery of CSS100217}

On February 17th 2010 
we discovered transient event CSS100217 during the course of the Catalina Real-time Transient
Survey.  This event was flagged as unusual at discovery as the object had a past spectrum from the Sloan Digital Sky
Survey (SDSS, Abazajian et al. 2009) that resembled a Seyfert, yet the outburst was uncharacteristically rapid and large
for an AGN.  Additionally, there was no detection in archival FIRST and NVSS radio data covering the object's location.
The lack of any radio source suggests that the variability was not being powered by a jet, as seen with optically
variable blazars. Given the unexpected nature of the event, we immediately scheduled photometric and spectroscopic
follow-up of the event. 

\section{Multi-wavelength Observations}

Following the discovery of the clearly energetic event we undertook follow-up observations in 
X-rays, UV, optical, near-IR and radio wavelengths, as well as a targeted archival gamma ray search. 
In Table 1, we present the sequence and nature of the follow-up observations we obtained. In the 
following section we will discuss the details of each set of observations as well as historical data
for the source galaxy. We will then combine the data to interpret the nature of the event
in relation to known types of transients and make concluding remarks about the source.

\subsection{UV, Optical and near-IR Photometry}

Following CSS100217's discovery, unfiltered observations continued as part of the CSS survey.  All CSS photometry is
routinely transformed to V-magnitudes by using between 10 and 100 G-type dwarf calibration stars measured in each eight
square degree image.  These calibration stars are pre-selected using 2MASS near-IR
data\footnote{http://www.ipac.caltech.edu/2mass/releases/allsky/doc/explsup.html}.  The magnitudes for each calibration
star are transformed to V following Bessell \& Brett (1988) and the zero point for each field is derived. The scatter in
the V magnitude for the calibration stars is typically $< 0.05$ magnitudes (Larson et al.~2003).
To improve the quality of the photometry for CSS100217 we created a high signal-to-noise ratio template
image and carried out image subtraction (Tomaney \& Crotts 1996) on all the CSS images. This process
reduces the photometric dependence on external calibrators, but can introduce uncertainty due to the 
subtraction process. Based on our analysis the zero-point uncertainty is approximately 0.1 magnitudes.

\begin{figure}{
\epsscale{0.9}
\plottwo{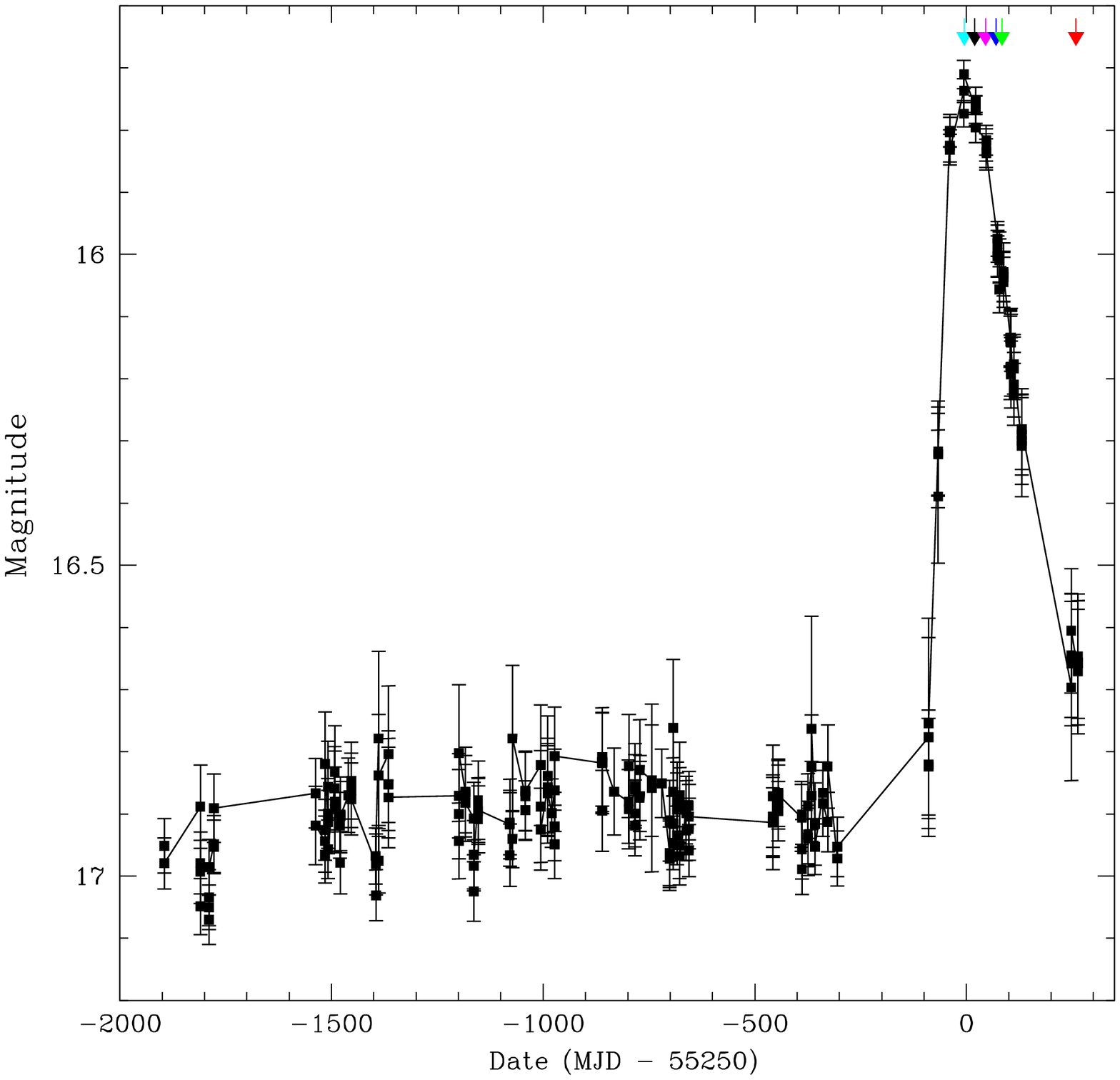}{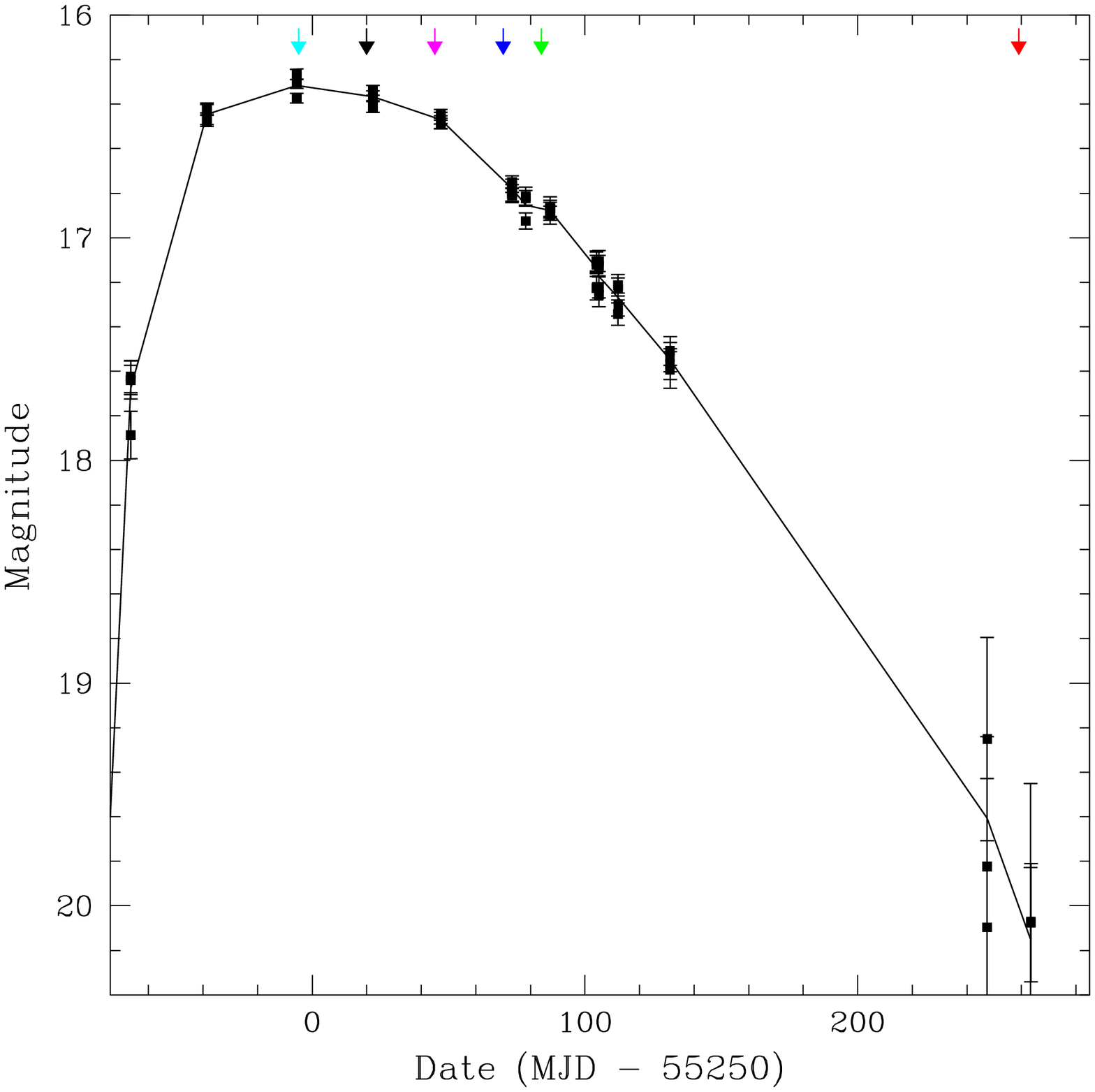}
\caption{\label{LC} 
  The $V_{CSS}$ lightcurves of CSS100217 taken with the 0.7m Catalina Schmidt telescope with respect to Modified Julian
  Date and maximum light.  Left: the full CSS light curve covering host and event. Right: the event light curve after
  subtracting the galaxy flux.  The dates at which the IGO, P200, APO, MDM, and Keck follow-up spectra were observed are
  marked with arrows.
}
}
\end{figure}

In Figure \ref{LC}, we present the CSS lightcurve of the transient plus host galaxy flux determined from the template image.
In order to determine the brightness of the transient we carried out image subtraction (Tomaney \& Crotts 1996) on the
CSS data using a combined high signal-to-noise template image produced from observations taken before the outburst. The
flux of the transient source is thus in units of the template image flux.  We determine the brightness of the transient
by calibrating the flux using the magnitudes of stars measured in the template image. CSS photometry is available in
Table 2.
SDSS photometric observations taken in late 2002 and early 2003 exhibit $r$ and $i$ magnitudes of 17.6 and 17.3,
respectively and is given in Table 3.
Additional measurements in the USNO-B1.0 catalog (Monet et al.~2003) with epoch 1977.1, list the object with magnitudes 
B2=17.6, R2 = 17.4, I=17.2. This suggests that the object was relative stable on a very long timescale.

Optical follow-up was taken soon after discovery with the Palomar 1.5m telescope (P60) in gunn $g$,$r$,$i$,$z$ filters.
Data was also taken in Johnson U,B,V and Cousins R,I using the 1m Sampurnanand telescope
in Nainital, India.  Data were reduced by performing PSF photometry using DAOPHOT (Stetson 1987). The photometry was
calibrated using Landolt (1992) standard stars fields PG1047+003 and PG1323-085. Ten bright isolated stars within the
field of CSS100217 were used as local standards and to derive the zero points for the images of the transient.
To supplement the optical data, and constrain the possibility that this was a tidal disruption event (TDE), we requested
Swift ToO time to observe the transient at X-ray and UV wavelengths. These observations clearly showed that the object
was bright at UV wavelengths and a source was detected in X-rays.

\begin{figure}{
\epsscale{0.8}
\plotone{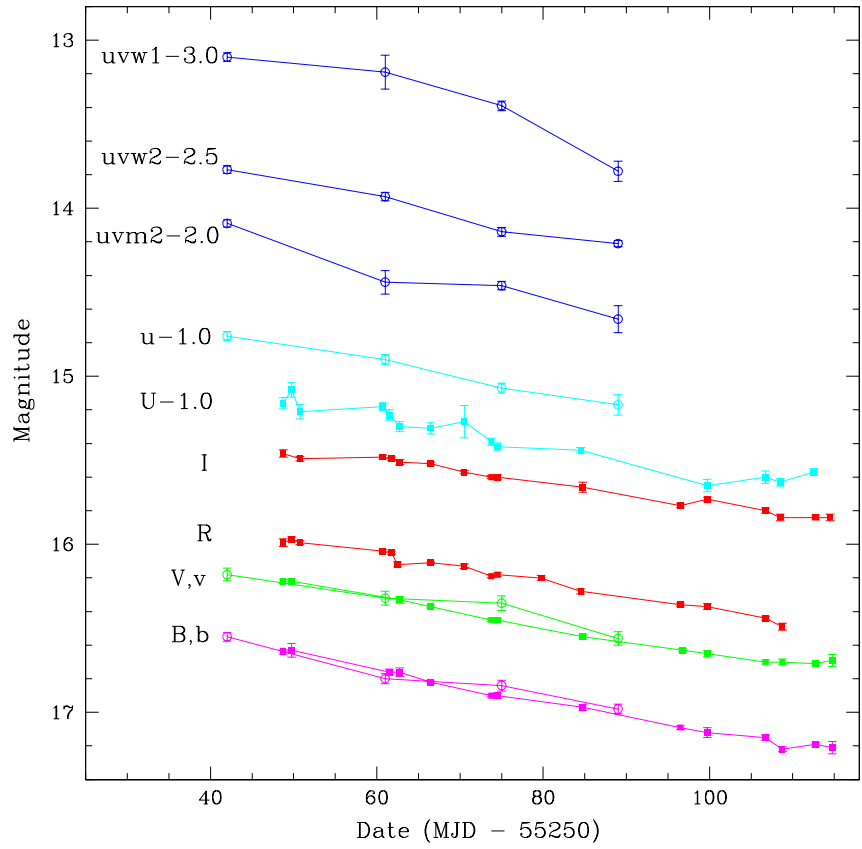}
\caption{\label{Gal} 
The multicolour evolution of CSS100217 near peak from  the Swift space telescope 
and ARIES 1m. Swift UVOT observations are in uvw1, uvw2, uvm2, u, v and b bands.
ARIES observations are Johnson U, B, V and Cousins R and I bands.
}
}
\end{figure}

In Figure \ref{Gal}, we present the photometry of the CSS100217 during the early decline phase in six Swift filters $uvw1$,
$uvw2$, $uvm2$, $u$, $b$, $v$ and ground-based $U$, $B$, $V$, $R$, $I$. The $B$ and $b$, and $V$ and $v$ magnitudes are
in excellent agreement. However, the magnitudes vary between $U$ and $u$ as the Bessell filter has central wavelength
3663 {\AA} and effective width 650 {\AA} while the Swift $U$ filter is bluer and broader (central wavelength 3465 {\AA},
width 785 {\AA}). Variations in the SED between these two filters likely causes most of the observed difference.
Li et al.~(2006) gives the transformation from Swift filters to standard Johnson filters.  For b and v the difference
from standard filters is of order 0.02 magnitudes. However, for objects such as CSS100217 with $u-v < 0$ the difference
noted by Li et al.~(2006) is significant and the transformation to standard magnitudes is very poorly constrained.
Therefore, we have not attempted to transform the Swift magnitudes to the standard system.

From the CSS photometry we find that the peak luminosity occurred on February 23rd, 2010 (MJD 55250).  Hereafter we
denote this date as $Tp$.  We fit the decline rate in each photometric filter over the range of dates from $Tp + 42$ to
$Tp + 112$. 
The observed decline of CSS100217 plus the host galaxy is close to linear and varies between bands from $0.012$ mag
day$^{-1}$ in uvm2 to $0.0064$ mag day$^{-1}$ in $I$. The rapid decline at bluer wavelengths is consistent with an
outburst that is cooling with time.  The rate of decline is consistent with the range observed for type IIn supernovae
(Trundle et al.~2009).  However, we note that the true decline rate of the transient is much greater as the photometry
includes both the galaxy and transient light. In order to determine the true decline rate for CSS100217 we transformed the 
SDSS photometry of the host galaxy to Bessell magnitudes.
As the spectral energy distribution of the host galaxy varies from that expected from stars, we preformed the 
transformation in two ways that follow Jester et al.~(2005). Firstly, we used the transformations derived for QSOs
with redshift $z < 2.1$ and secondly for stars ($R-I_c < 1.15$ and $U-B < 0$). 
Using the relation for stars we obtain host galaxy flux $U=17.55$, $B=18.17$, $V=17.77$, $R_c=17.25$ and 
$I_c = 16.71$. With the QSO transformation we obtain $U=17.51$, $B=18.07$, $V=17.77$, $R_c=17.38$ 
and $I_c = 16.88$. We note that the difference in magnitudes using the two different methods is 
maximum for $I_c$ (0.17 mag) and for $V$ produces the same result. Hereafter we adopt the transformation f
or stars but note that the use of the other set of equations will produce a small constant shift.

After subtracting the host flux contribution from each measurement we combined uncertainties in photometry in
quadrature, with the conversion errors noted by Jester et al.~(2005), then fit each with a simple linear decline over
the time between $Tp+42$ and $Tp+112$. The following declines rates were found: $\Delta U = 0.0121 (0.0017)$, $\Delta B
=0.0134 (0.0018)$, $\Delta V = 0.0112 (0.0026)$, $\Delta R= 0.0132 (0.0021)$, $\Delta I = 0.00985 (0.0019)$.  The slope
derived from the CSS lightcurve during this period was $\Delta V_{CSS}=0.0117 (0.0022)$, and thus is very close to that
observed in the filtered $V$ photometry. For the Palomar 1.5m photometry in the gunn filter system matches that of SDSS,
therefore, we subtract the SDSS host magnitudes. The host-subtracted filtered optical photometry from Palomar 
1.5m and AIRES 1m is presented in Table 3, as is the Swift photometry for this event.

Based on our filtered photometry, the CSS transformed V magnitudes ($V_{CSS}$) lie very close to $R$ magnitudes.  The
decline rate is also very similar. The $V_{CSS}$ values are slightly brighter than filtered $V$ magnitudes ($\Delta V <
0.2$ mag). The departure of CSS photometry from Bessell $V$ is due to the non-stellar SED of the transient source
combined with the galaxy.  In particular, the strong $\rm H_{\alpha}$ emission lies within the CSS transmission
sensitivity as well as the overlap between $R$ and $I$ filter response (well beyond $V$ filter response).

The foreground Galactic extinction for this event is $A_V= 0.046$, based on Schlegel et al.~(1998) reddening maps.  We
do not see evidence for host galaxy extinction in the spectra of event, suggesting this is a small effect. Thus we only
correct for foreground extinction.  Applying the V-band reddening correction and a K-correction of 0.15 magnitudes, 
to account for the effective rest-frame bandwidth, we find the peak luminosity was $\rm M_{V\,CSS} = -23.0$.  We 
adopt $H_0 = 72$ km s$^{-1}$ Mpc$^{-1}$, $\Omega_{\Lambda}= 0.73$ and $\Omega_{m}= 0.27$ with the host galaxy's
redshift of $z=0.147$.  To derive k-correction in filtered photometry near peak we first subtracted the SDSS spectrum 
from the IGO follow-up spectrum observed near peak. Using the Palomar spectrum that was taken near maximum light 
($Tp+20$ days), we derive the K-corrections in the $V$ and $R$ filters of -0.01 and 0.04, respectively.
As the filter transmission range of our other filtered observations goes beyond that of this spectrum (3950 - 9050\AA),
k-corrections are not derived for other filters. 

The extinction and K-corrected peak magnitudes measured at $Tp + 40$ are $\rm M_V =-22.7\pm 0.3$ and $\rm
M_R =-22.8 \pm 0.3$.  Here we have combined estimates for uncertainties in the true colour variation near peak (0.1),
the photometric zero-points (0.15), the K-corrections (0.1), and the colour transformations from SDSS filters (0.17).
For the Palomar 1.5m gunn photometry taken at $Tp + 7$, we use the IGO spectrum taken at $Tp-5$ and find a K-correction
of $Kr = -0.16$ and thus $M_r=-22.8 \pm (0.3)$. Here there is no need to transform the filter system and the
observations are not long after maximum. However, the photometric uncertainties are larger. For the other gunn
measurements taken at this time we find $M_g=-22.8$, $M_r=-23.0$, $M_i=-23.2$ and $M_z=-23.1$, without K-correction.
Similarly, for AIRES photometry, taken at $Tp + 40$, we find $\rm M_U =-22.9$, $\rm M_B =-22.3$ and $\rm M_I =-23.3$,
without K-correction.

The peak brightness in V-band is similar to that observed for the type IIn supernovae SN 2008fz (Drake et al.~2010a,
$\rm M_{V}= -22.3$) and type IIL supernova SN 2008es ($\rm M_{V}= -22.2$, Gezari et al.~2009b; $\rm M_{V}= -22.3$,
Miller et al.~2009).  Another luminous type-IIn, SDWFS-MT-1 (aka SN 2007va) was detected in Spitzer data and was
observed with $\rm M_{[4.5]} \sim -24.2$ in $4.5 \mu m$ Spitzer/IRAC band (Kozlowski et al.~2010).

\begin{figure}{
\epsscale{0.9}
\plotone{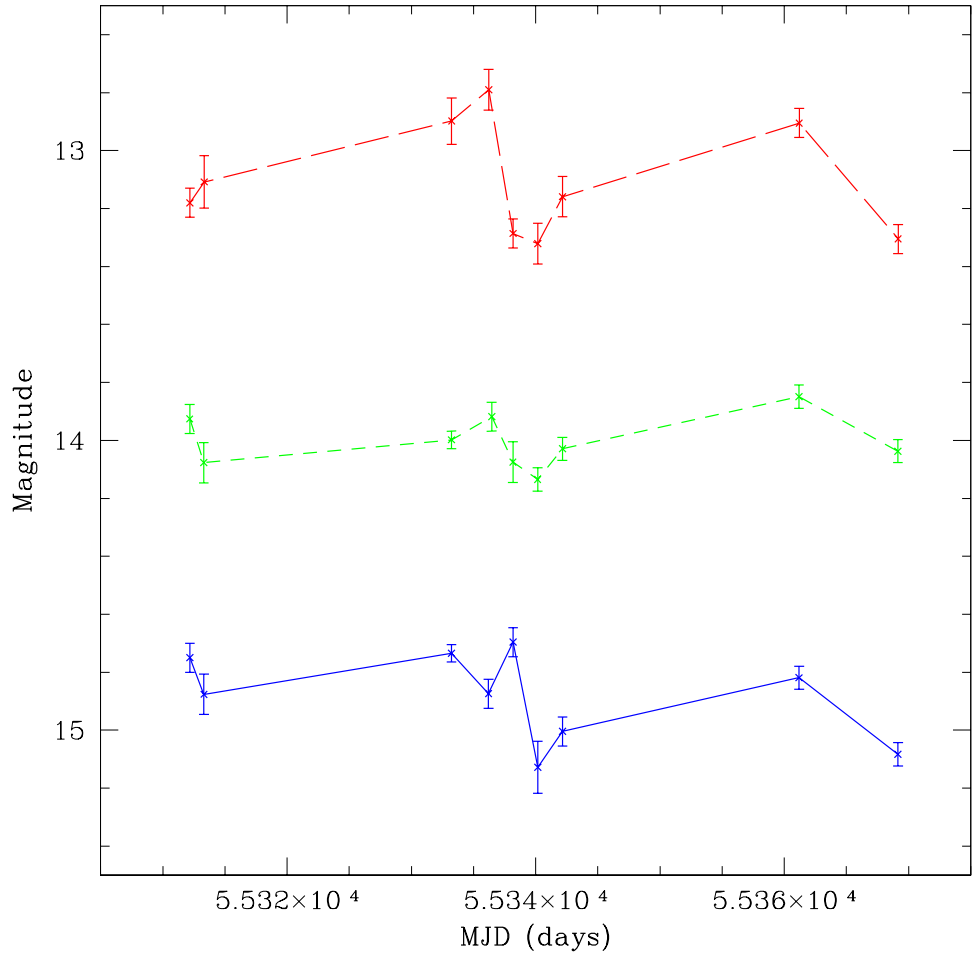}
\caption{\label{PhotIR}
The near-IR lightcurves of CSS100217 in J, H and Ks bands.
The short-dashed line presents the observed $\rm K_{S}$-band values, the long-dashed line 
connects H-band measurements, and the solid-line J-band magnitudes.
}
}
\end{figure}

Near-IR $J$, $H$ and $Ks$ observations were carried out with CANICA, a NIR camera equipped with a 1024 x 1024 pixel
Hawaii array, at the 2.1m telescope of the Guillermo Haro Observatory located in Cananea, Sonora, Mexico. Data was
reduced using standard procedures.  As some observations were carried out on partially cloudy nights, differential
photometry for the object and field stars was carried out. For the latter, we adopted the photometric values listed in
the 2MASS All-Sky Catalog of Point Sources (Skrutskie et al.~2006). In Figure \ref{PhotIR}, we present the near-IR
measurements and in Table 3 we include the host-subtracted near-IR photometry.  The peak observed apparent brightness of
CSS100217 in near-IR, after subtracting the host brightness using 2MASS magnitudes, is $J \sim 15.0$, $H \sim 14.1$ \&
$Ks \sim 13.0$.  The corresponding absolute magnitudes are thus $\rm M_J= -24.2$, $\rm M_H= -25.1$ and $\rm M_{Ks} =
-26.2$ (again without K-correction).  However, we note that there may be significant uncertainty in the near-IR
brightness of the transient.  Once the event has fully faded it will be possible to determine the host contribution more
accurately and thus the event's peak brightness in each filter. Unlike the optical data there is little evidence for a
decline in luminosity.
This effect may be attributed to cooling of expanding material. The object is brighter than 2MASS in the near-IR
follow-up observations by $\Delta J \sim 1.6$, $\Delta H \sim 1.45$, and $\Delta K \sim 1.35$ magnitudes.  Strong
near-IR excess has been observed in many type IIn supernovae and has been interpreted as thermal emission from dust in a
pre-existing circumstellar nebula (Gerardy et al.  2002).  If CSS100217 is a type IIn supernova, near-IR emission is
expected to increase at first and then gradually decline over the coming years.

The host galaxy, SDSS J102912.58+404219.7, was serendipitously observed by the GALEX All-Sky Imaging Survey on 2004
January 24 with FUV=19.52 $\pm$ 0.17 and NUV = 18.97 $\pm$ 0.08 mag, and by the Medium Imaging Survey on 2010 January 29
with NUV=17.078 $\pm$ 0.009 mag (Gezari et al.~2010).  A request for follow-up GALEX observations was made after these
archival observations revealed that the event had brightened in the NUV filter by 1.9 mag, $\sim$ 1.5 months before the
optical maximum.  We obtained NUV imaging on 2010 April 17 and 29, which measured NUV = 17.756 $\pm$ 0.012 and 17.881
$\pm$ 0.015 mag, respectively, indicating that the event had faded by 0.8 mag in the NUV 3 months later.  The GALEX
magnitudes are in the AB system, and have been corrected for the energy lost in a 6$\arcsec$ radius aperture, and for
Galactic extinction.  We also obtained NUV grism observations on 2010 April 17 and 2010 April 29 and detected emission
near a rest wavelength of $\lambda$1910 with a FWHM of 3900 $\pm$ 500 km s$^{-1}$ after correcting for the instrumental
resolution.  The emission is very likely due to the CIII] $\lambda$1909 emission line which is commonly seen in NLS1
galaxies, and is also associated with circumstellar emission ejecta in type IIn supernovae (Fransson et al.~2005; Cooke
et al.~2010). We did not detect any variability between the two epochs of spectra taken 12 days apart.

\subsection{Optical Spectra}

\subsubsection{The Nature of the Host Galaxy}

\begin{figure}{
\epsscale{0.9}
\plotone{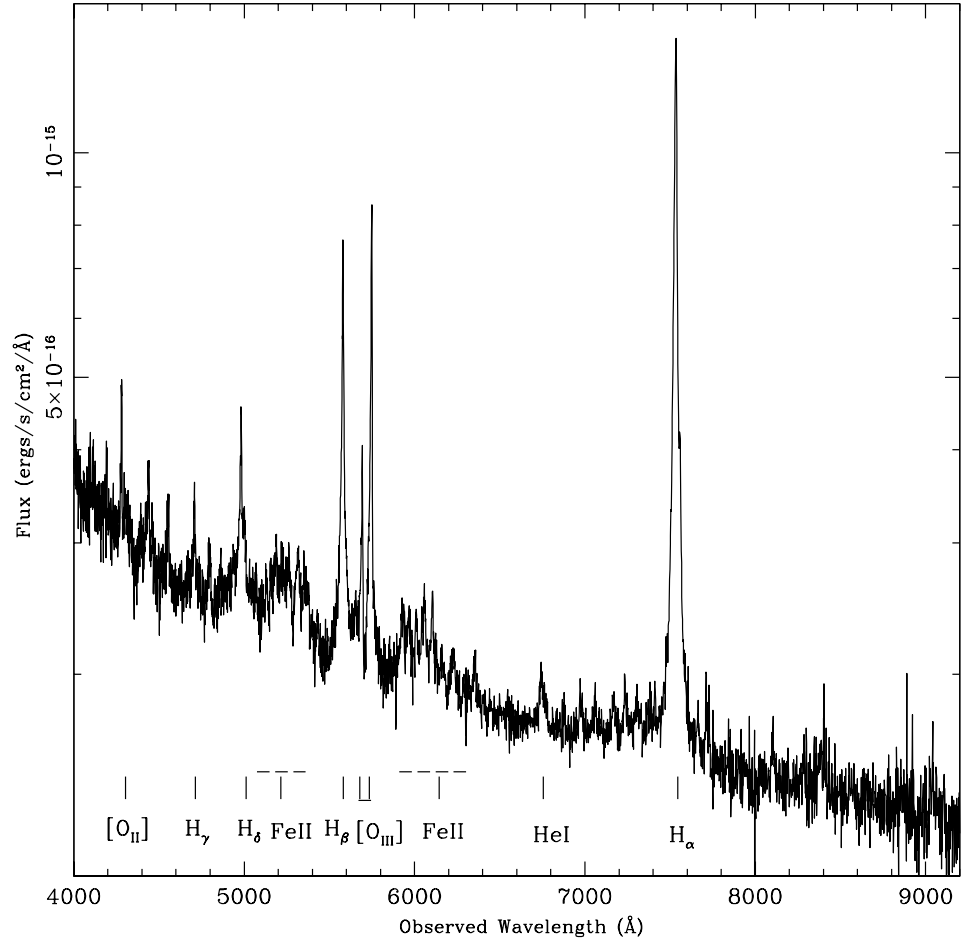}
\caption{\label{SDSSSpec}
The archival SDSS DR7 spectrum of the host galaxy to CSS100217 (SDSS J102912.58+404219.7).
}
}
\end{figure}

When CSS100217 was discovered it was immediately recognized as an unusual object. Analysis of the SDSS DR7 (Abazajian et
al.~2009) spectrum of the host galaxy (SDSS J102912.58+404219.7), suggested that the object was a Seyfert. However, the
rate and degree of variability had never been encountered during the prior years of the CRTS transient survey, nor in
previous transient searches carried out by the Palomar-Quest Survey (Djorgovski et al.~2008).  In Figure \ref{SDSSSpec},
we present the spectrum obtained by the Sloan Digital Sky Survey in December 2002.  The spectrum is reminiscent of an
AGN with clear strong Balmer, [OII] and [OIII] emission lines. AGN viewed away from the line-of-sight of the jet exhibit
smooth lightcurves that increase in variability on long timescales (Webb \& Malkan 2000). In contrast optical
variability of many magnitudes can be seen in blazars (Bauer et al.~2009) which have broad lines and are associated with
strong radio sources and have smooth featureless continua during their outbursts.  In this case the emission lines
observed from the galaxy were relatively narrow and no radio detection has been made in past radio surveys by FIRST
(1.4GHz, Becker, White, \& Helfand 1995), WENSS (326MHz, Rengelink et al. 1997), and NVSS (1.4GHz, Condon et al. 1998),
at the level of 1 mJy/beam.

\begin{figure}{
\epsscale{0.9}
\plottwo{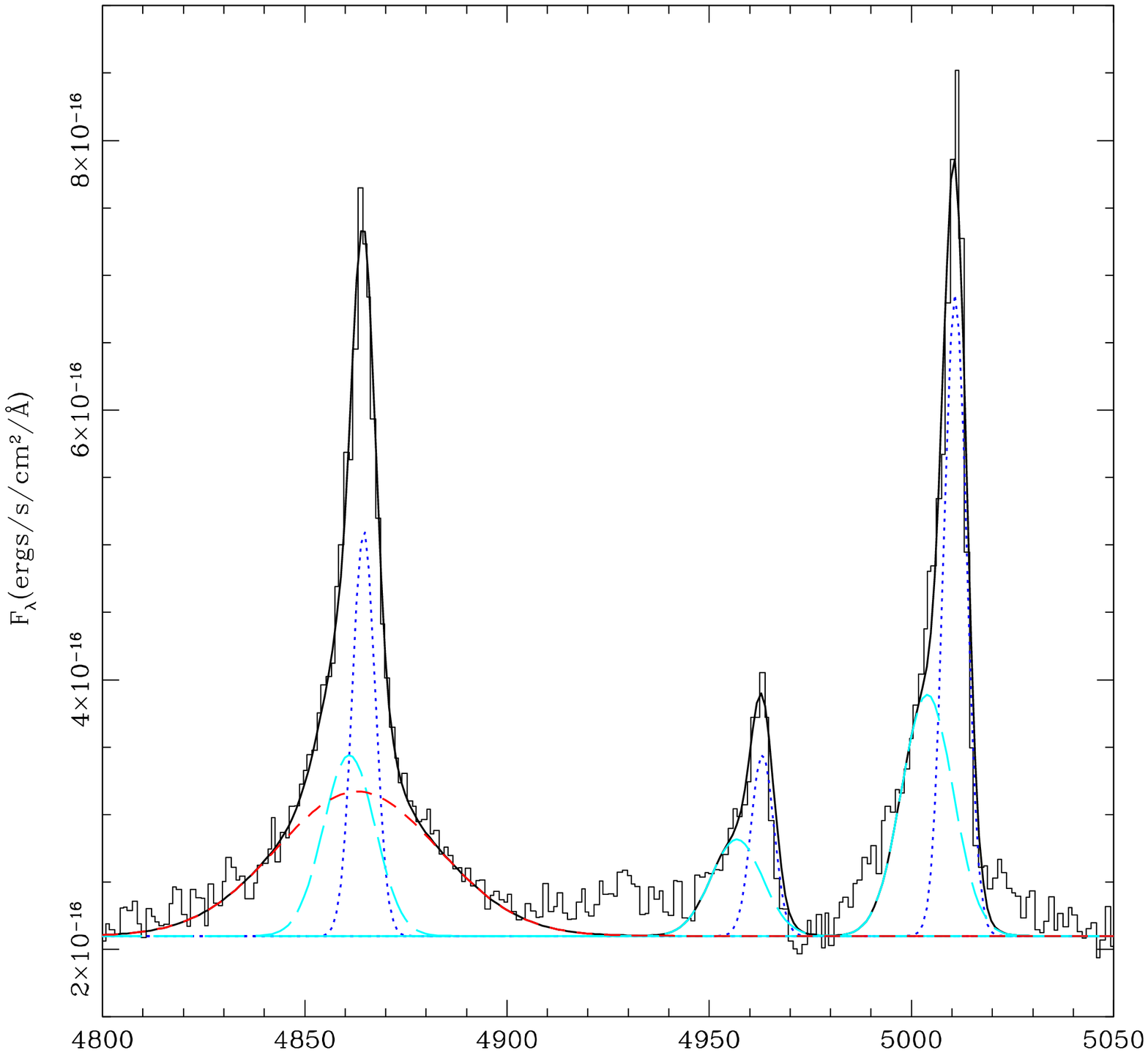}{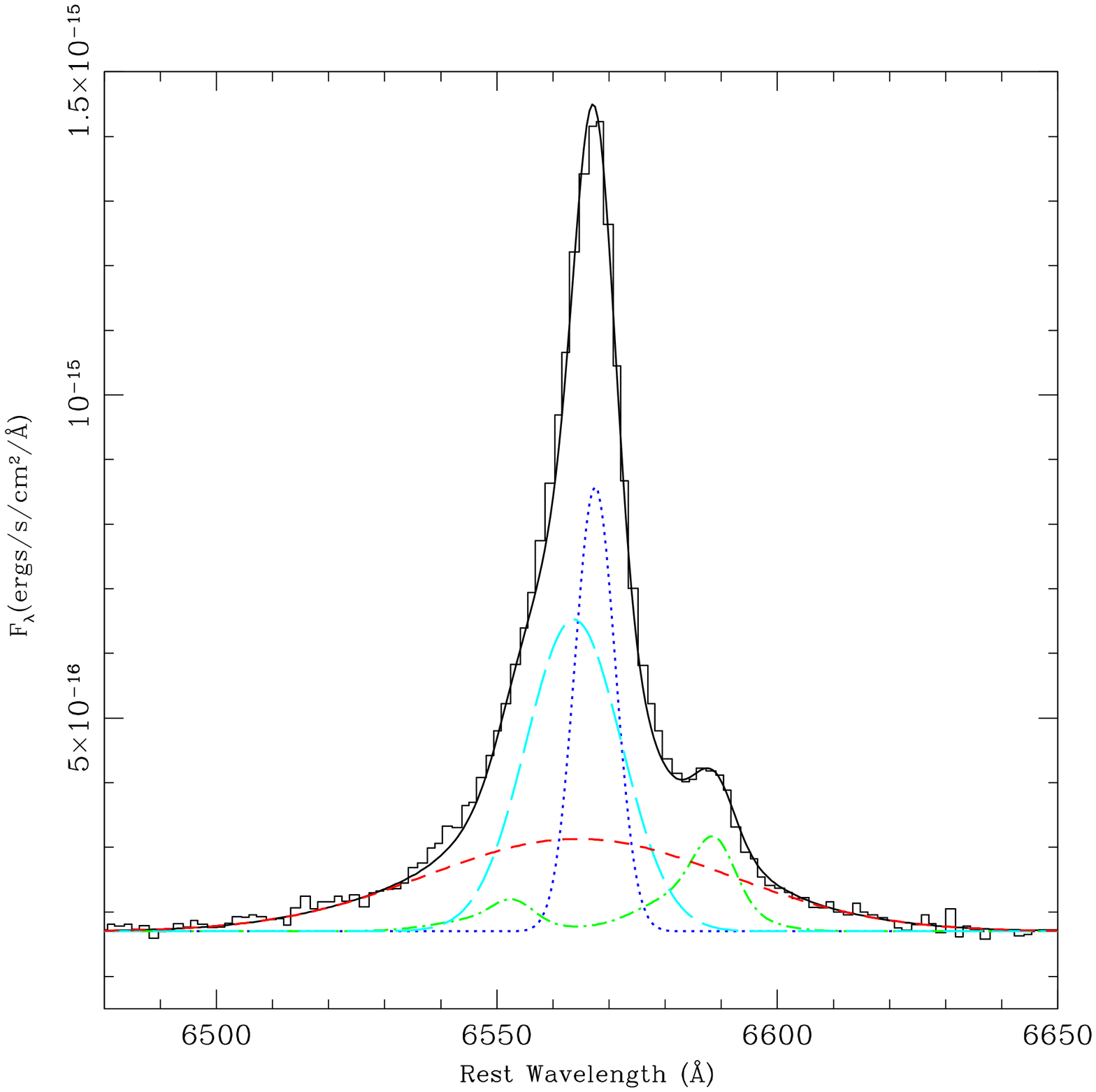}
\caption{\label{Emiss} 
Host galaxy emission line fits.
Dotted-line: narrow components of $H_{\beta}$, $H_{\alpha}$ and $\rm [OIII]$. 
Long-dashed line: intermediate velocity components. Short-dashed line:broad 
$H_{\alpha}$ and $H_{\beta}$ components. Solid-line: overall fit.
}
}
\end{figure}

The emission lines features observed in the SDSS spectrum are very clearly asymmetric. In addition, the Balmer lines
appear to have a broad component.  We decomposed the SDSS spectrum to determine the nature of the SDSS source.  For
$H_{\beta}$ we see three clear significant components with narrow, medium, and broad velocity widths. There were clearly
systematic offset between the broad, medium and narrow line components.  For [OIII] we found just narrow and medium
width components that were consistent with the $H_\beta$ emission features.  For the $H_{\alpha}$ region the line
fitting process is more complex because of the presence of blended NII emission lines.  Firstly we fit the line complex
with individual [NII] 6548 $\rm \AA$ and 6583 $\rm \AA$ lines of fixed $1/3$ flux ratio plus $H_\alpha$ emission lines
considering just a narrow and broad component. The fit result was quite poor as the $H_{\alpha}$ has broad wings and an 
asymmetric peak. The [NII] lines were also poorly modeled.  We then decided to add an medium width component to the
[NII] and $H_\alpha$ emission to match that expected given the $H_{\beta}$ and [OIII] lines. Next simultaneously fit 
the $H_\beta$ and $H_\alpha$ with three velocity components, and the [OIII] and [NII] with two components. In Figure
\ref{Emiss}, we present the fits to these lines. In Table 4, we present the flux, velocity and central wavelength values 
obtained for the emission lines including both models for the $H_{\alpha}$ region.

\begin{figure}{
\epsscale{0.9}
\plottwo{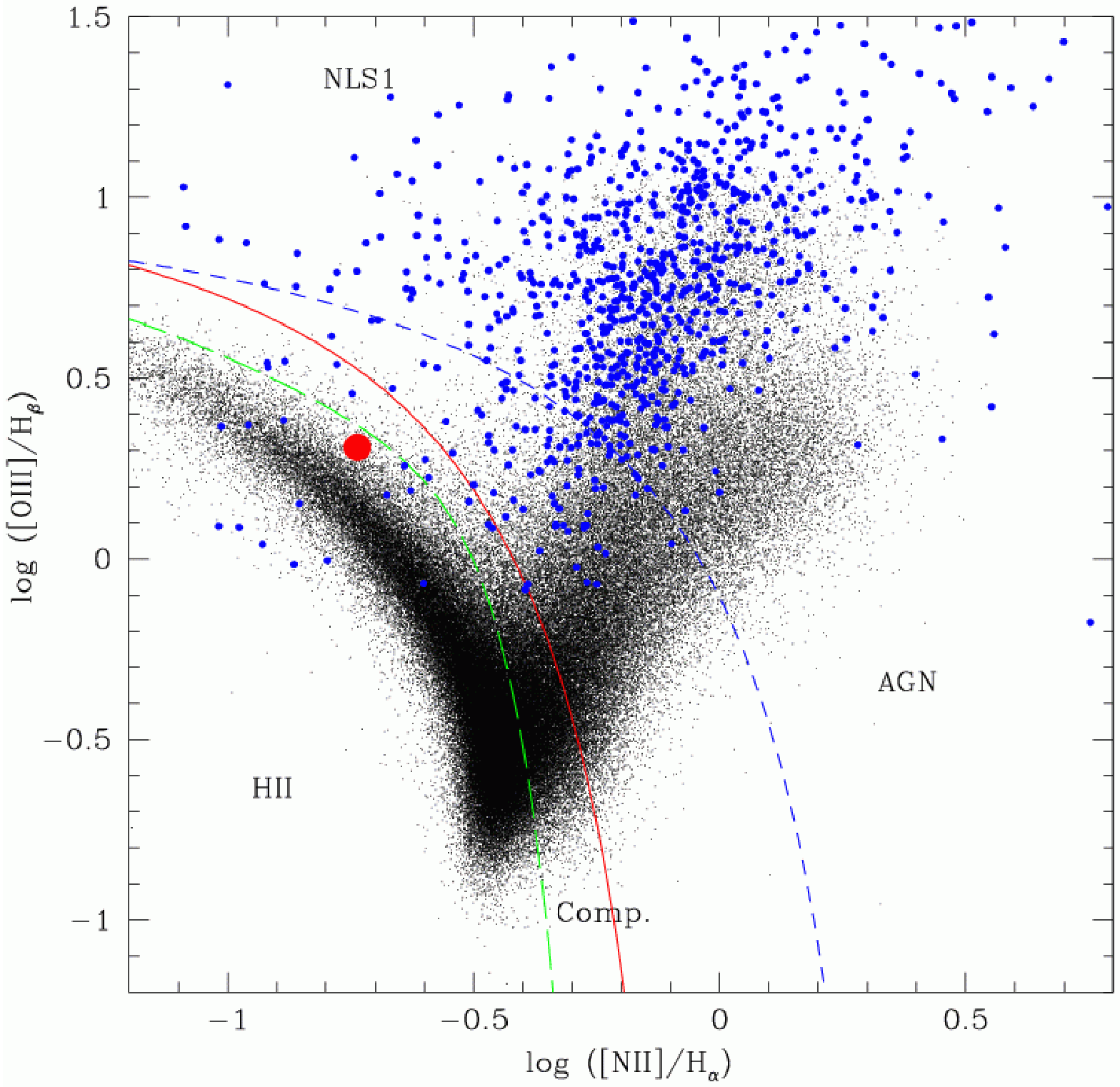}{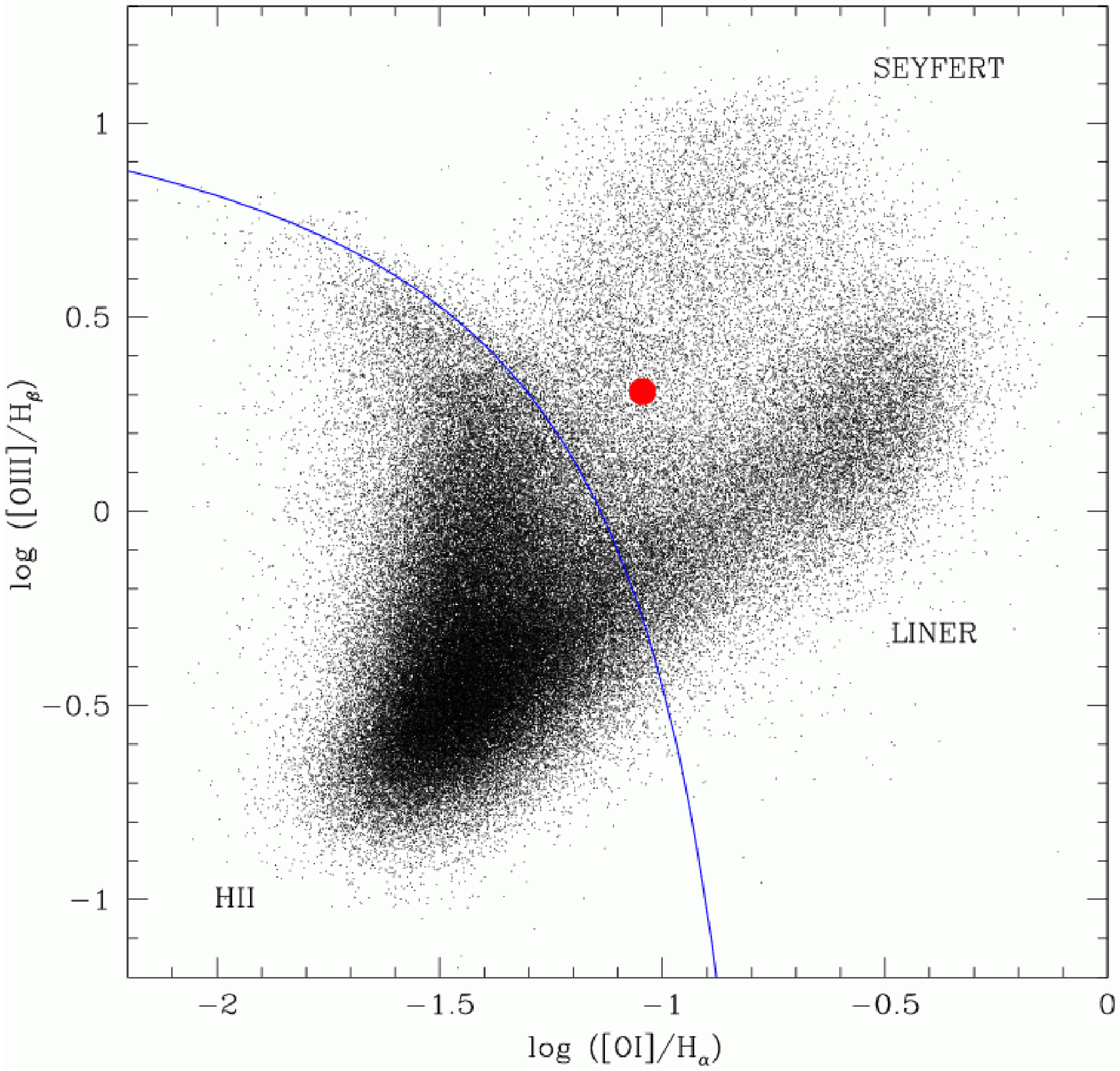}
\caption{\label{AGN}
Comparison of the host galaxy abundances relative to SDSS emission-line galaxies.
The large solid dot shows the location for CSS100217 based on values from Table 4.
Left: The large dots give the locations of known NLS1 galaxies from Zhou et al.~(2006).
The dashed line shows the Kewley et al.~(2001) theoretical devision between
starburst and AGN galaxies. The solid-line shows the devision found by Kauffmann et al.~(2003)
and the long-dashed lines shows the division determined by Stasinska et al.~(2006).
Right: The dashed line shows the Kewley et al.~(2001) theoretical devision between
starburst and AGN galaxies.
}
}
\end{figure}

In Figure \ref{AGN}, we present the standard Baldwin, Phillips, \& Terlevich (1981) (hereafter BPT) emission-line
diagnostic diagrams used to separate AGN from starburst galaxies.  We include points using the fit values given in Table
4, along with the emission line galaxies presented in the MPA/JHU version of the SDSS DR4 catalog (Kauffmann et
al.~2003; Alderman-McCarthy et al.~2006).  Here we have selected galaxies largely following Kewley et al.~(2006).  That
is, only galaxies having emission lines with signal-to-noise ratio $> 3$ in each line species and $S/N > 20$ in the
spectrum were selected.  Galaxies were selected in the redshift range $0.04 < z < 0.17$.  The sample consists of $\sim
135,000$ objects.  For comparison, we have also plotted the NLS1 from Zhou et al.~(2006) that contain flux measurements
for $H_\alpha$, $H_\beta$ and $\rm [NII]$.

We have not included the standard [OIII]/$H_{\beta}$ vs.~SII/$H_{\alpha}$ diagnostic diagram as we found that the
location of redshifted [SII] corresponds to the location of a strong sky emission line in the raw SDSS spectrum. We
found that the ratio of SII/$H_{\alpha}$ is much lower than observed in other SDSS emission line galaxies.  This is
likely due to poor subtraction of the strong sky line.  In Figure \ref{AGN}, we show the theoretical demarcation lines
separating emission-line galaxies as determined by Kewley et al.~(2001) as well as the empirical demarcation lines of
Kauffmann et al.~(2003) and Stasinska et al.~(2006). The galaxy appears in the starburst region of the
[OIII]/$H_{\beta}$ vs.~[NII]/$H_{\alpha}$ but in either starburst or AGN region for [OIII]/$H_{\beta}$
vs.~[OI]/$H_{\alpha}$. This suggests that the object may be a combination object having both starburst and AGN features.

\begin{figure}{
\epsscale{0.9}
\plotone{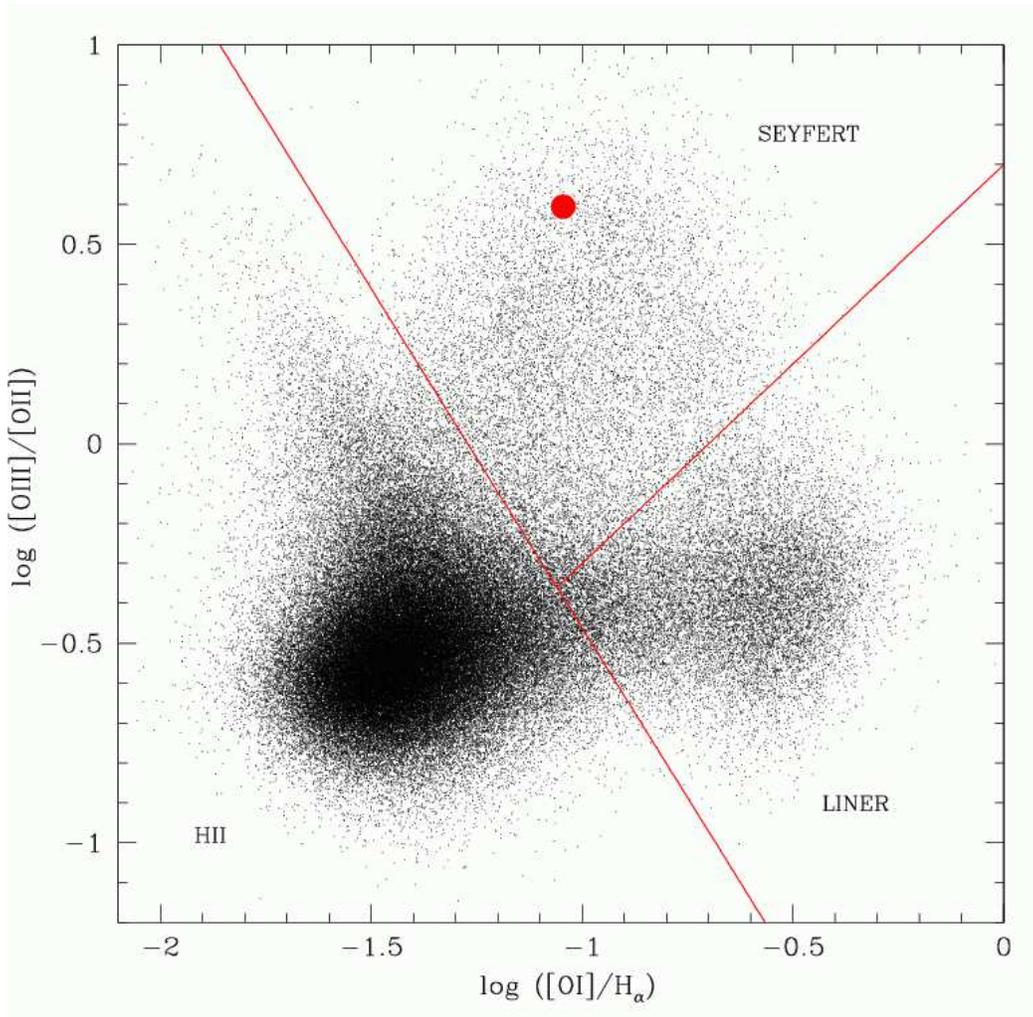}
\caption{\label{AGN2}
Comparison of the host galaxy abundances relative to SDSS emission-line galaxies.
The large solid dot shows the location for CSS100217 based on values from Table 4. 
The lines show the Kewley et al.~(2006) devision between Seyfert, Liner and 
Starburst galaxies.  
} 
}
\end{figure}

In Figure \ref{AGN2}, we present the location of the host in relation to the new emission-line diagnostic diagram of
Kewley et al.~(2006). Here the SDSS source lies within the Seyfert region.  However, of the 921 NLS1 with $\rm H_\alpha$
values we selected from Zhou et al.~(2006), only 32 lie in the starburst region of Kauffman et al.~(2003) in
[OIII]/$H_{\beta}$ vs.~[NII]/$H_{\alpha}$ and 16 lie in this region for the stricter starburst-AGN separation line of
Stasinska et al.~(2006). Clearly the host is not a typical NLS1. Indeed, the location of the object to the left of the
AGN-starburst composite objects suggests that the amount of star formation is significant.  Three similar objects were
investigated by Mao et al.~(2009). In that case one of the three objects was within the starburst region on the
diagrams, whereas two were within the composite region in the $[OIII]/H_\beta$ vs.~$[NII]/H_\alpha$ diagram. Mao et al.
explain the reason for the locations of these objects as AGN buried in HII regions where they are transitioning from a
starburst dominated phase to an AGN-dominated phase.

The presence of broad Balmer components within the SDSS spectrum is clear and strongly suggests the presence of an AGN.
Core-collapse type II supernovae and stellar winds from massive stars can also produce such broad lines.  Izotov et
al.~(2007) discussed the possibility of multiple SN events and massive star forming regions and noted that broad lines
can arise from these sources. The SDSS spectrum was taken on December 29th 2003, while our first CSS observations of the
object were in December 2004.  If a supernova had occurred in the galaxy in 2003 it would have faded by the time we first
observed it. However, the likelihood of such an event is not large. Mao et al.~(2009) also discussed this possibility for
their three objects.

The NLS1 nature of the host galaxy is further supported by the $\rm [Ne\,{V}] \lambda $3426 emission line.  NLS1 exhibit
strong FeII emission features as seen in the SDSS spectrum. However, type IIn supernovae also exhibit strong Fe-II lines
in late spectra. The width of the broad $\rm H_{\alpha}$ and $\rm H_{\beta}$ emission lines from model-II is 
$\rm \sim 2900 km s^{-1}$. This is significantly broader than the 2000 $\rm km s^{-1}$ limit expected for NLS1. 
The observed line width is consistent with that observed in the late time spectra of type IIn. Nevertheless, the
combination of the line diagnostics and spectral features with the low likelihood of observing two 
type-IIn supernovae in the same galaxies within a period of a few years suggests that the SDSS source 
is a NLS1 and also has significant ongoing star formation.

\subsubsection{Follow-up Spectroscopy}

\begin{figure}[ht]{
\epsscale{0.7}
\plotone{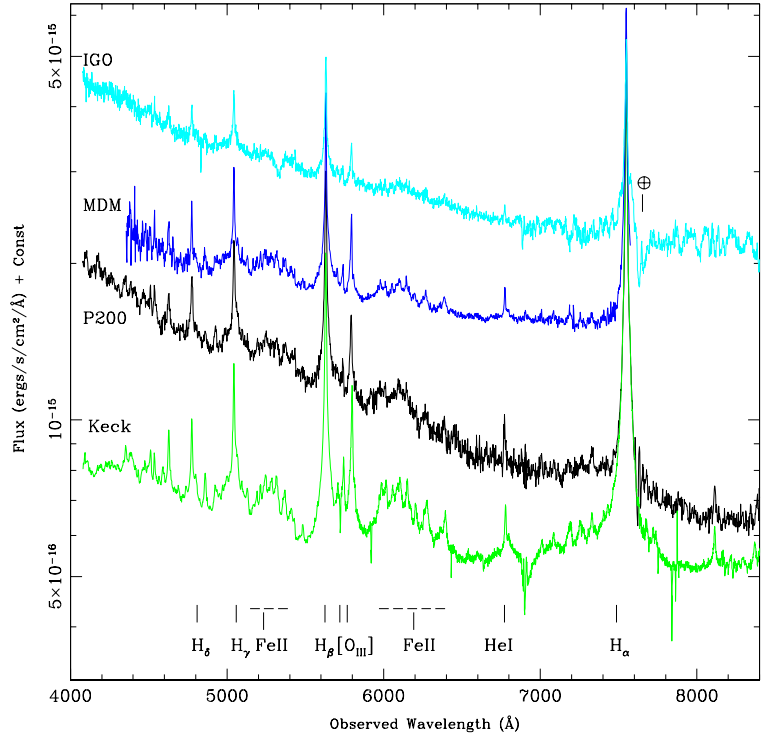}
\caption{\label{Spec}
Spectra of CSS100217 at $\tau = 1$ (IGO), $\tau=26$ (P200),$\tau=76$ (MDM) and $\tau=90$ days (Keck) after discovery.
Maximum light occurred at $\tau \sim 6$. Data shown were obtained with the IGO 2m, Palomar 5m + DBSP, MDM 2.4m and Keck
10m + LRIS.
}
}
\end{figure}

\begin{figure}{
\epsscale{0.7}
\plotone{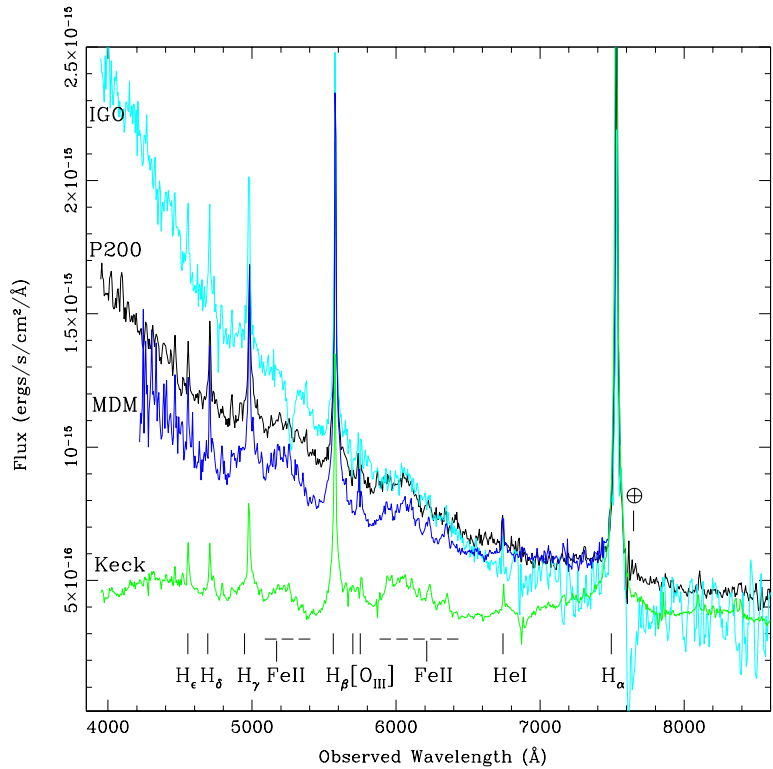}
\caption{\label{SP}
The host galaxy-subtracted follow-up spectra of CSS100217, taken between February 18th and May 18th 2010,
as per figure \ref{Spec}.
}
}
\end{figure}

Following the discovery of CSS100217 we immediately scheduled spectroscopic observations with the IGO 2m telescope.
These observations were taken on February 18th and showed an outburst spectrum similar to the archival SDSS
spectrum. The spectrum clearly exhibited a much bluer continuum and Balmer lines that were stronger by a factor of $\sim
5$, consistent with a type II supernova. Upon subtraction of the SDSS spectrum from IGO data there was no evidence for 
a change in [OII] and [OIII] and Fe emission lines.  However, the spectrum did not appear to exhibit a
significant new broad $H_{\alpha}$ component relative to the SDSS spectrum as expected for IIn supernovae.  Interpretation
of the event was complicated by its location near what was suspected to be an AGN. In order to confirm the initial
result we scheduled additional IGO spectra and took a Palomar 5m spectrum on March 15th. The Palomar spectrum was found
to be very similar to the initial IGO one but exhibited a slightly shallower continuum slope. Subsequently we obtained
spectroscopic follow-up with IGO, APO, MDM and Keck. These spectra do show signs of a new broad component not present
in the SDSS spectrum. In Figure \ref{Spec}, we present follow-up spectra of CSS100217 spanning the period from 
2010 February 18th to May 18th.  In Figure \ref{SP}, we present these same spectra after subtracting the archival 
SDSS spectrum of the host galaxy.

A test for whether CSS100217 may be due to AGN activity can be derived from changes in the emission line
characteristics. Narrow emission lines are potentially more useful for this test than broad emission lines because they
are powered by the average ionizing flux over decades, rather than over days for broad emission lines (Halpern et al.
2003).

To examine the spectroscopic changes in more detail, for each spectrum, we subtracted the fits to the SDSS emission
lines from the $H_\alpha$ and $H_\beta$ line profiles and fit the residual emission lines. In Table 5, we
present the fluxes and line widths for the residuals. The removal of the constant [OIII] lines from the host galaxy was
found to be relatively complete for IGO, Palomar, MDM and Keck spectra. This suggests that the flux calibration is good.
We believe that the errors in the remaining emission-line fluxes are $\sim$ 20\% due to calibration uncertainties.
However, the measurements may be higher for Balmer components since the CSS and SDSS photometry suggests 
that the AGN was in a lower state when it was spectroscopically observed by SDSS. The spectra with uncertain flux
calibration were not analyzed in this work.

\begin{figure}{
\epsscale{0.9}
\plotone{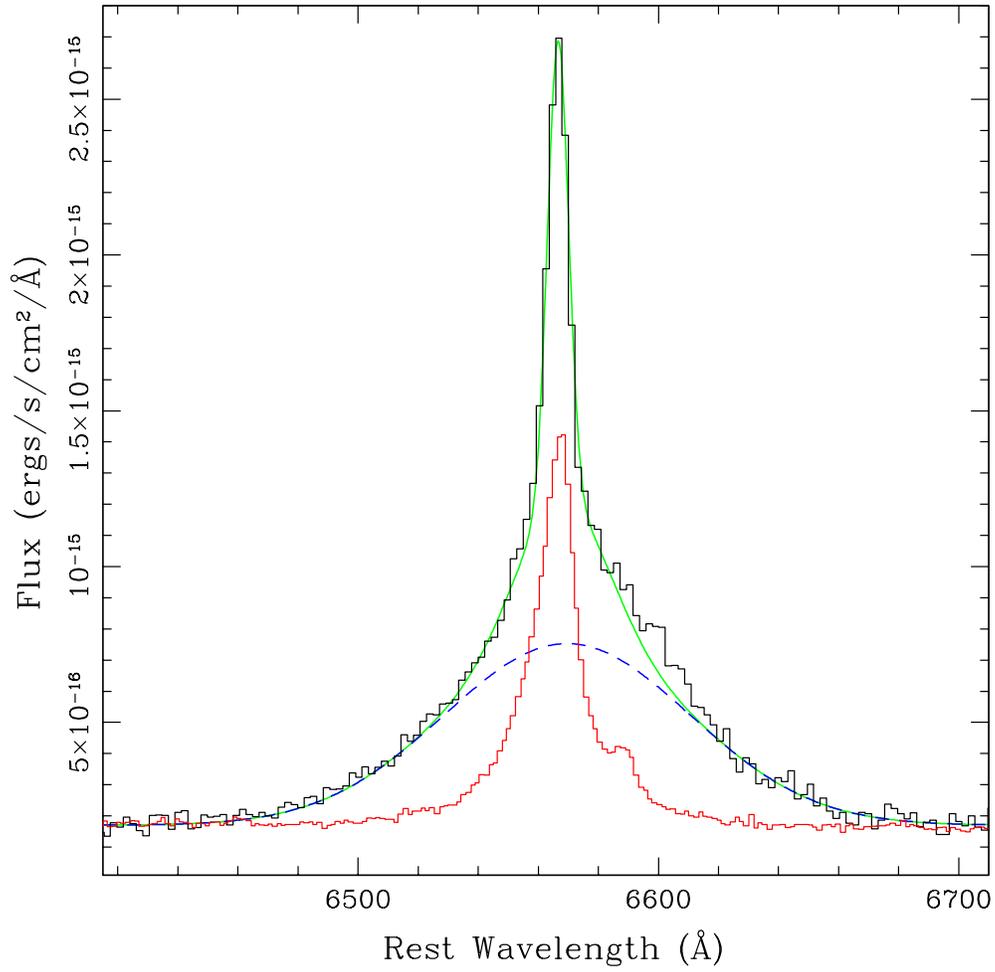}
\caption{\label{Emiss3}
Fit to outburst $H_\alpha$ emission observed with Palomar 5m on 2010 November 9th.
The red line shows the SDSS profile, the black line the Palomar data after
subtracting the SDSS fit. The green line shows the three component fit.
The dashed blue line the broad $H_{\alpha}$ component.
}
}
\end{figure}

In Figure \ref{Emiss3}, we display the fit to the residual $H_{\alpha}$ flux from our November 9th Palomar spectrum 
when the event had faded significantly. Clearly, unlike the SDSS spectrum, the emission exhibits a strong 
broad component.
However, the narrow component is of the same width as the SDSS spectrum and the broad component is significantly
narrower than that seen in the SDSS spectrum.  The flux ratio of $H_{\alpha\, b}$ to $H_{\alpha\, n}$ in the late
Palomar data is 4.6 after subtracting the SDSS contribution. This is very different from that found in the SDSS
spectrum (1.4). The total $H_{\alpha}$ flux is observed to increase with time, while the continuum and optical
luminosity decreases.  Both the $H_{\beta}$ and $H_{\alpha}$ exhibit a new broad component with width $\rm \sim 4000 km s^{-1}$
that increases in velocity with time. However, unlike $H_{\alpha}$, the $H_\beta$ shows little change in total flux 
with time.

As noted above, the lightcurve (Figure \ref{LC}), shows an increased brightness at the end of 2004. This variation is
interpreted as largely due AGN activity and suggests that the source was fainter at the time that the SDSS spectrum was
observed. Therefore, some of the flux observed in our SDSS-subtracted spectra is likely due to the host flux being
incompletely subtracted. Slight differences between the SDSS and follow-up spectra may also arise from the $3\arcsec$
diameter fibre used by SDSS.  This fibre is large enough to contain the bulk of the light from the host, whereas
long-slit spectra with widths $\sim 1\arcsec$ were used for transient follow-up observations.  However, our HST 
images suggest that almost all of the flux from the host and CSS100217 lies within the central $1\arcsec$.
No second contributing source is seen within $3\arcsec$.

In addition to optical spectra we obtained GALEX NUV grism observations on 2010 April 17 and 2010 April 29. We detected
emission near rest wavelength 1910 \AA which likely corresponds to [CIII] ($\lambda$1909). This emission is clearly
detected and has an intrinsic width of $3900 \pm 500$ km.s$^{-1}$. Such emission is seen in NLS1 galaxies (Leighly \&
Moore 2004) but is also associated with circumstellar emission and ejecta in type IIn supernovae (Fransson et al.~2005;
Cooke et al.~2010). We did not detect any variability over the 12 days between the two epochs.

\subsection{Energetics of CSS100217}

The extreme luminosity of CSS100217 is a clear sign of the energy powering this event.  In order to determine the amount
of optical energy expended we follow the calculations of the energetic type IIn supernova SN 2003ma given by Rest
et al.~(2009).  The bolometric luminosity in band $X$ can be defined as

\begin{equation}
L_{bol,X} = b_{X} L_{\sun,X} 10^{(M_{\sun,X} - (M_X)/2.5}
\end{equation}

where the solar constants for filters X $=$ (V, R), are $M_{\sun,X} = (4.83, 4.42)$ and $L_{\sun},X = (4.64, 6.94)
\times 10^{32}$ erg/s, respectively (Binney \& Merrifield 1998).  The absolute magnitude of the event in filter X,
$M_X$, is determined from the corrected photometry and b$_{X}$ is the bolometric correction.  As noted in the previous
section, the event exhibits significant spectroscopic evolution during the event. Thus the bolometric correction, and
the K-correction applied to determine $M_X$, vary with time. In order to account for this variability we use the IGO,
P200, APO, MDM and Keck calibrated spectra from Table 1.  The spectra are spread out over intervals of roughly 20 days,
accounting for 90 days when the event was emitting the most energy near peak luminosity. K-corrections are determined
for $V$ and $R$ filters by integrating the event spectra combined with the known filter transmissions. As the redshift
is relatively small, the spectra cover both the rest and observed $V$ passband wavelength range and we calculate
in-band ($V_{obs}$ to $V_{rest}$) K-corrections, rather than cross-band corrections (eg. $R_{obs}$ to $V_{rest}$).

The early spectra of CSS100217 show the clear presence of a hot continuum component that is a good fit to a Blackbody of
temperature $1.6 \times 10^4$ K in the rest frame. The Galex UV photometry from 29 January 2010 also supports this
result.  At this temperature much of the energy is emitted at rest frame wavelengths shorter than observed in optical
spectra ($\lambda < 3440 \AA$). At later times the Blackbody emission component has cooled to $\sim 8 \times 10^3$ and
far less flux is emitted at blue wavelengths. Therefore, in order to determine the bolometric correction, we must account
for this emission and its evolution. We fit the continuum component of the host-galaxy subtracted spectra taken
between $Tp-5$ to $Tp+84$ with a blackbody. A linear to the temperature of the blackbody component in these spectra 
gives a cooling rate of $73.0\pm0.5$ K day$^{-1}$. At the time the second Palomar spectrum was taken, $Tp+164$, the 
amount of additional continuum flux from CSS100217 is too small to provide an accurate blackbody fit, so we adopt the 
Keck values.

To approximate the full SED for the event, we combine our observed spectra with the blackbody for wavelengths $\lambda <
3440$ \AA.  We then integrate the complete model fluxes and those expected with response expected within the $V$ and $R$
passbands. In Table 6, we present the bolometric corrections and K-corrections determined from the SED models.  At early
times the bolometric correction is large because of the hot component and at later is much closer to values from
a solar SED.

In order to determine bolometric luminosity of the event we correct the filtered $V$ and $R$ photometry for extinction
and K-corrections from the spectra taken nearest the observing time. In all cases these corrections are small. To
determine the total bolometric luminosity we integrated the values of $L_{bol,V}$ and $L_{bol,R}$ during the period when the
filtered and unfiltered photometry completely overlap. The resulting values were: $E_{bol,Vp}=5.4\times 10^{51}$ and 
$E_{bol,Rp}=4.3 \times 10^{51}$ erg. The subscript $p$ denotes that this is only for part of the full lightcurve. 
The corresponding values without bolometric correction are $E_{Vp}=5.5 \times 10^{50}$ and $E_{Rp} =5.6 \times 10^{50}$ erg,
respectively.

In order to estimate values for the entire event we first assumed the V-band corrections for CSS photometry covering the
entire $V_{CSS}$ lightcurve. The total integrated bolometric luminosity is $E_{bol,VCSS}=1.4 \times 10^{52}$ erg, and
$E_{VCSS}=8.7 \times 10^{50}$ erg without bolometric correction.  The bolometric correction averaged over 
the light curve ($\bar b_{VCSS} \sim 16$) is large due to much of the energy being expended at short wavelengths 
near peak when the event was $\sim 1.6\times 10^4$.
Next, we determined the integrated luminosity for CSS data taken during period when both filtered and unfiltered 
observations were taken, $E_{bol,VCSSp}=5.5\times 10^{51}$ erg. This measurement is very close to the values 
from filtered photometry ($E_{bol,Vp}$ and $E_{bol,R}$). The ratio of $E_{bol,VCSS}$ to $E_{bol,VCSSp}$ ($\sim 2.5$) 
provides and estimate of the fraction of energy expended during the entire event, relative to the period of 
filtered photometry observations. We thus estimate the total bolometric luminosities, $E_{bol,V} \sim 1.3 \times 10^{52}$ 
and $E_{bol,R} \sim 1.1 \times 10^{52}$ erg by multiplying by this ratio.
The uncertainty in these values is expected to be of order 25\% due to photometric errors, sparse sampling of 
the CSS lightcurve near peak, variation of the spectra with time, uncertainty in spectroscopic flux 
calibration, and variations in the flux ratio for filtered and unfiltered response.
As these measurements are based on 287 rest-frame days it is likely that a slightly higher value would
be found for the full lightcurve.

\begin{figure}{
\epsscale{0.9}
\plotone{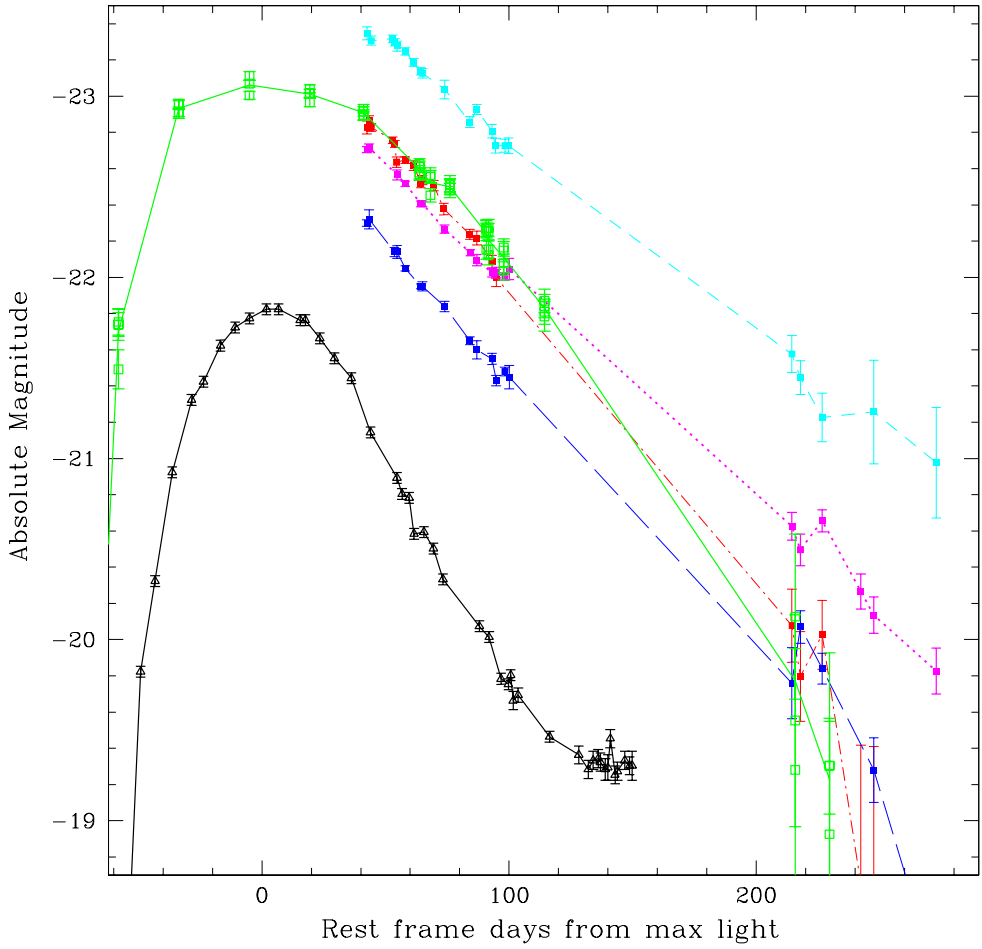}
\caption{\label{LC2}
The lightcurve of the CSS100217. Here we show the luminosity of transient event CSS100217 compared to the
the luminous and energetic type IIn SN 2006gy from Smith et al.~(2007) (open-triangles).
The CSS100217 lightcurves are given for $VCSS$, solid-line with open-boxes; $I$ short-dashed line; $R$ dot-dashed line;
$V$ dotted-line and $B$ long-dashed line.
The maximum brightness occurred at $\rm MJD = 55250$ which corresponds to $\sim 6$ days after discovery.
See text for further details.
}
}
\end{figure}

In comparison to CSS100217, a number of very energetic type IIn supernovae have recently be discovered. Rest et
al~(2009) found the past type IIn supernovae, SN 2003ma, expended $4\!\times\!10^{51}$ erg (with a bolometric correction
of $\sim3.4$) and Drake et al.~(2010a) obtained a value of $>1.4\!\times\!10^{51}$ erg (without bolometric correction) for
the type IIn supernova SN 2008fz.  Similarly, Kozlowski et al.~(2010) found that the type-IIn supernova, SDWFS-MT-1,
expended $> 10^{51}$ erg in the $4.5\mu m$ the Spitzer/IRAC band.  Smith et al.~(2010a) found the type IIn supernova SN
2006gy had an integrated bolometric luminosity of at least $5 \times 10^{51}$ erg.  In Figure \ref{LC2}, we plot the
filtered and unfiltered absolute magnitude lightcurves of CSS100217 along with that of the type IIn supernova 2006gy
(Smith et al.~2007). Considering the difference absolute luminosity between these two events ($\delta R >$ 1 mag), 
the approximate factor of two difference in energy appears as expected.

\subsection{High Resolution Imaging}

In order to discern whether CSS100217 was due to an event occurring in the nucleus of SDSS J102912.58+404219.7, as
required for a tidal disruption event or other AGN variability, we investigated the location of the event.  Firstly we
determined the location of the flux in the difference images relative to the centroid of SDSS J102912.58+404219.7.  Such
astrometry can be used to disambiguate variable sources that are not resolved because of blending or high background
light (Nelson et al.~2009). A significant offset between the two centroids would very strongly suggest the object was a
supernova. By stacking the sequence of difference images we found the additional flux was within $0.3\arcsec$ of the
galaxy's nucleus. In order to check this result we undertook follow-up observations with the Large Binocular Telescope
plus LUCIFER near-IR camera and spectrograph.  No second source was resolved in our Ks-band images with resolution of
$0.4\arcsec$.

To obtain a higher level of precision, and possibly separate the source for the AGN, we obtained HST Director's
Discretionary Time with the HST WFC3. We obtained WFC3 images in F390W, F555W, and F763M filters spanning one orbit. The
filters were chosen to separate the blue, continuum and $H_{\alpha}$ components of the spectra, since we expected a type
IIn supernova would be a strong source of $H_{\alpha}$ and this might enable a clearer separation from the hot blue
central AGN. A customized seven-pointing dither was chosen in each band to maximize the subsampling of the HST PSF
(Anderson \& King 2000).

\begin{figure}{
\epsscale{0.8}
\plottwo{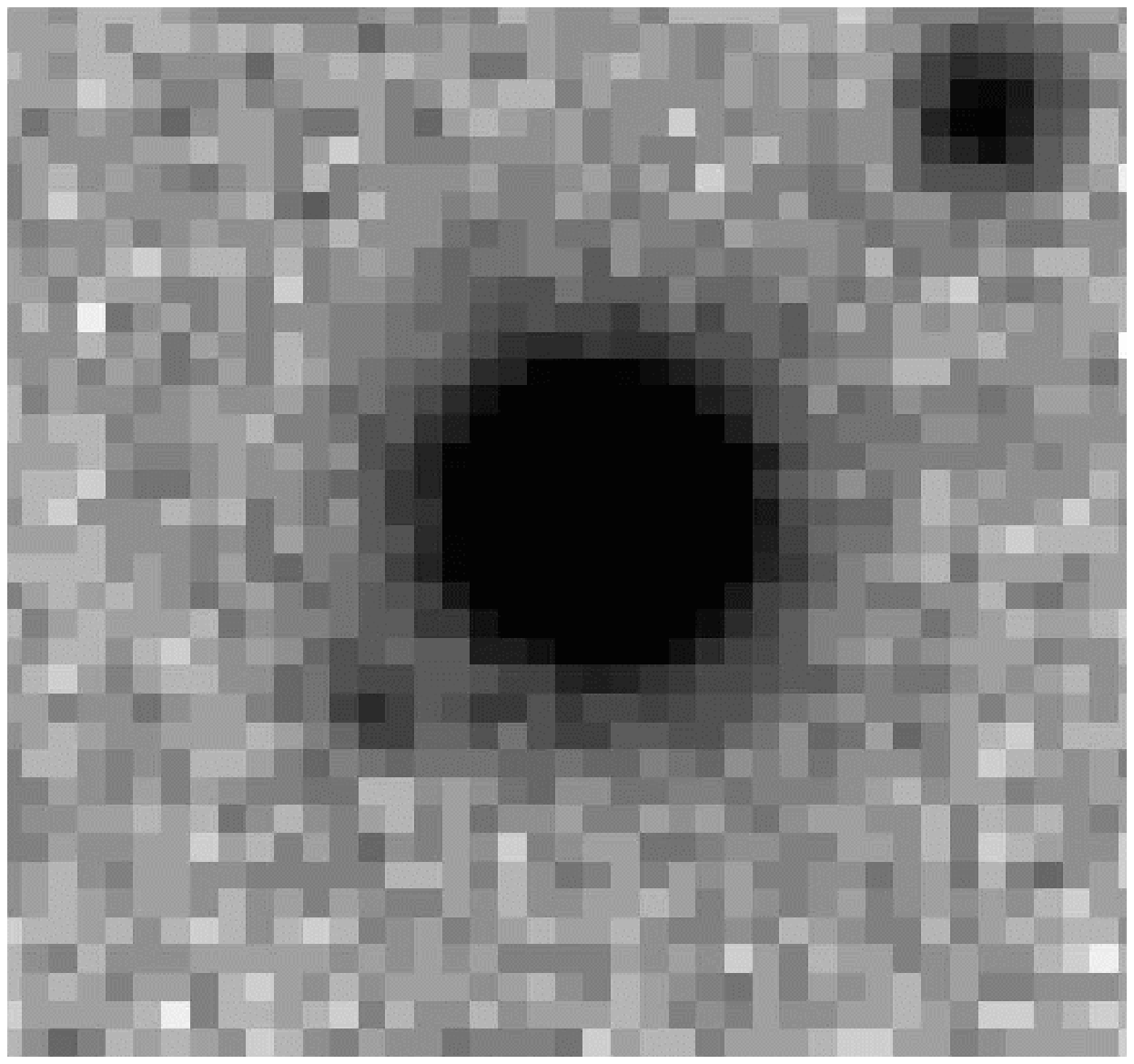}{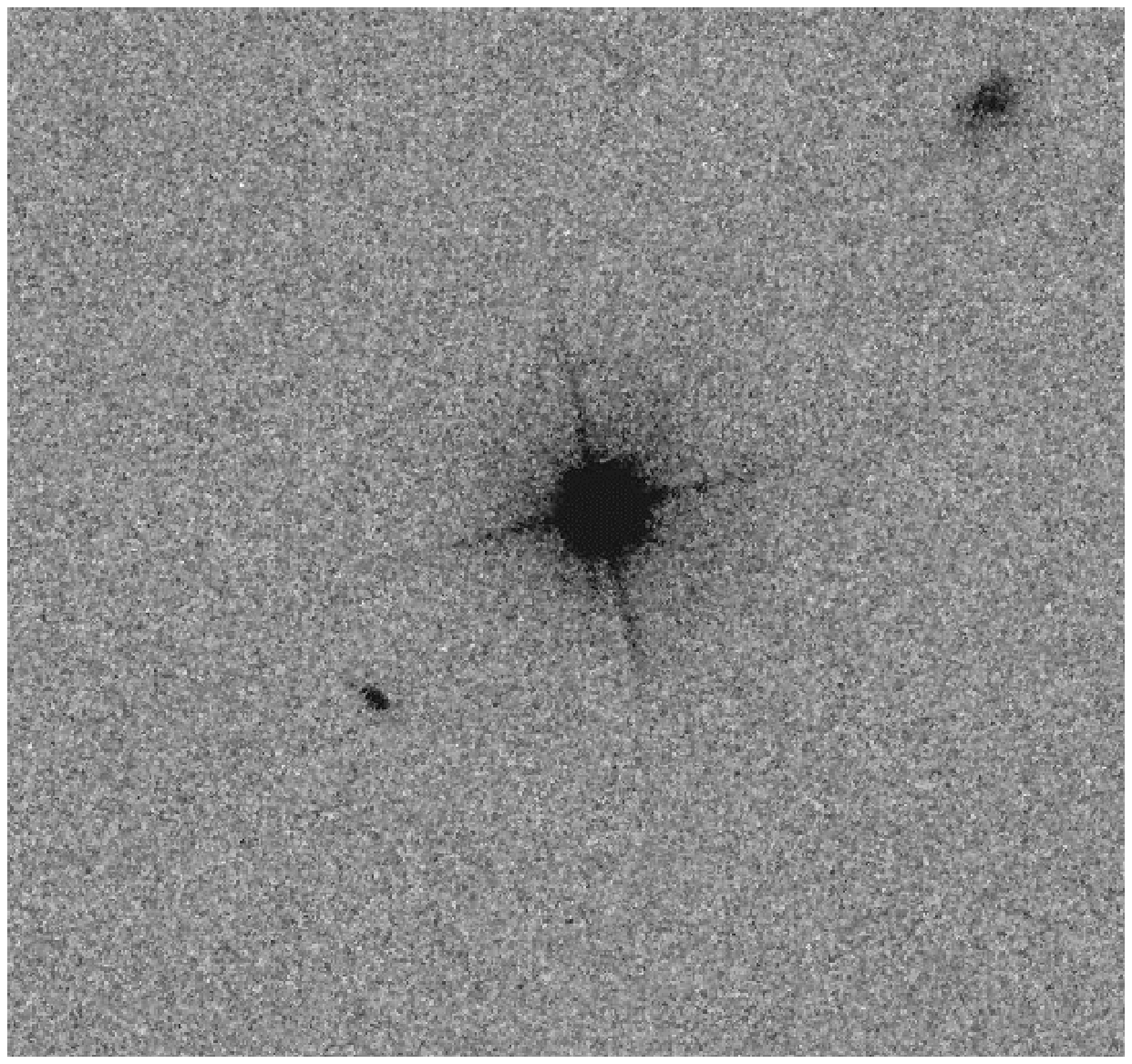}
\caption{\label{HST} 
The SDSS image of the host galaxy of CSS100217, SDSS J102912.58+404219.7 (left)
and HST F555W image of the event during outburst (right). The images are
$15.5\arcsec$ across with North up and East to the left.
}
}
\end{figure}

In Figure \ref{HST}, we show a coadded SDSS image of the host galaxy along with the dithered HST WFC3 F555W image of the
host including transient CSS100217.  The SDSS pre-event coadd consists of the median of 15 images (u,g,r,i,z) of the
galaxy observed on 3 nights (2002-12-31, twice, and 2003-03-26), whereas the HST image is derived from seven dithered
images. The SDSS and HST observations show that the galaxy is completely dominated by the central nucleus with no clear
sign of extension of the host galaxy.  Subtraction of HST PSF models did not reveal the galaxy or any sign of a second
separated source. In comparison, Deo et al.~(2006) examined HST data for a sample of 87 Seyfert galaxies and found
almost all occurred in clear spiral galaxies.  Only two objects appeared point-like (HEAO 2106-099 \& Mrk 335). However,
Bentz et al.~(2009) examined HST data for 35 AGN, including Mrk 335, and was able to fit the faint host galaxies for
all. It is therefore likely that the host galaxy simply has a very low surface brightness relative to the nucleus.  

\begin{figure}{
\epsscale{0.8}
\plottwo{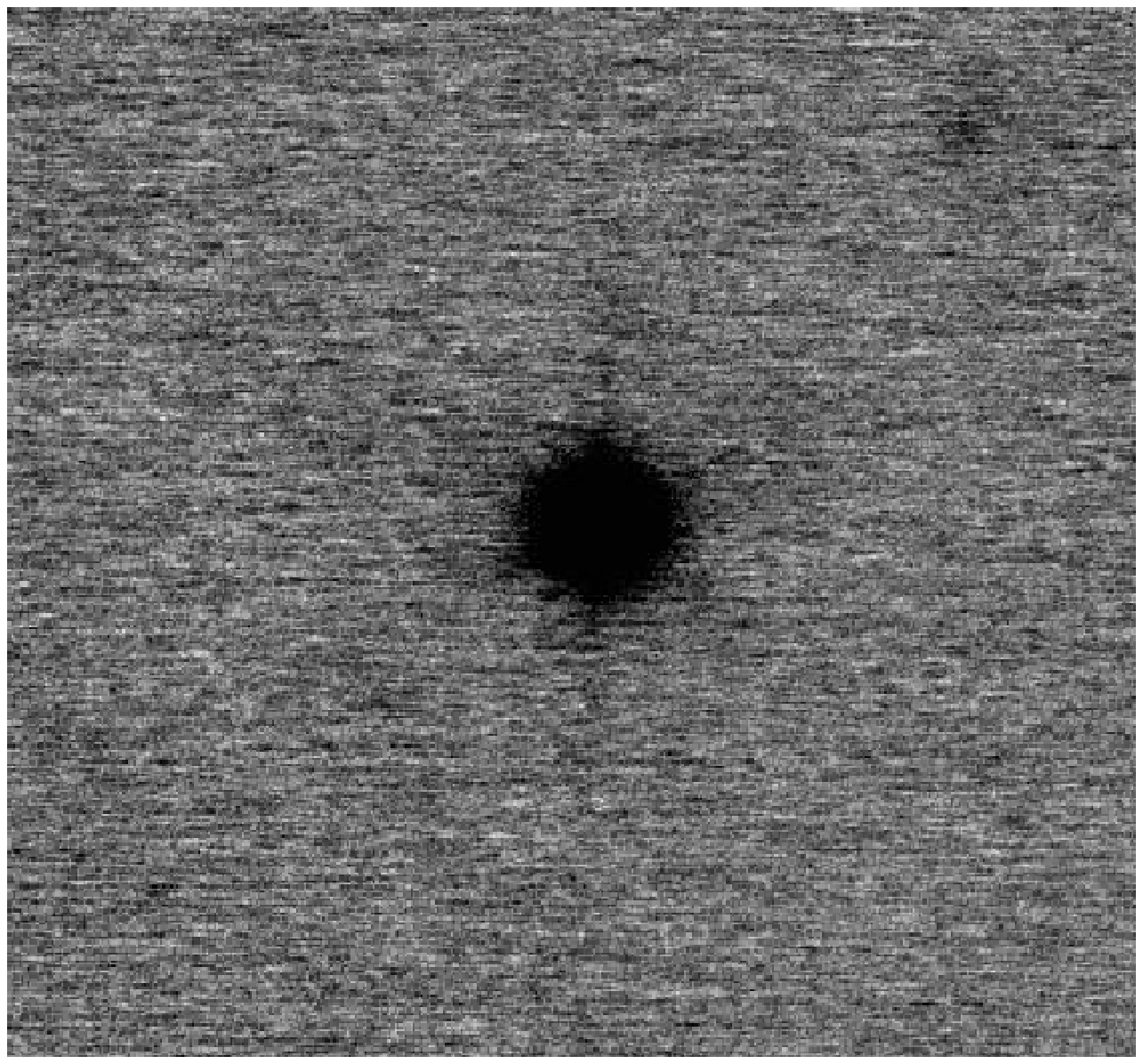}{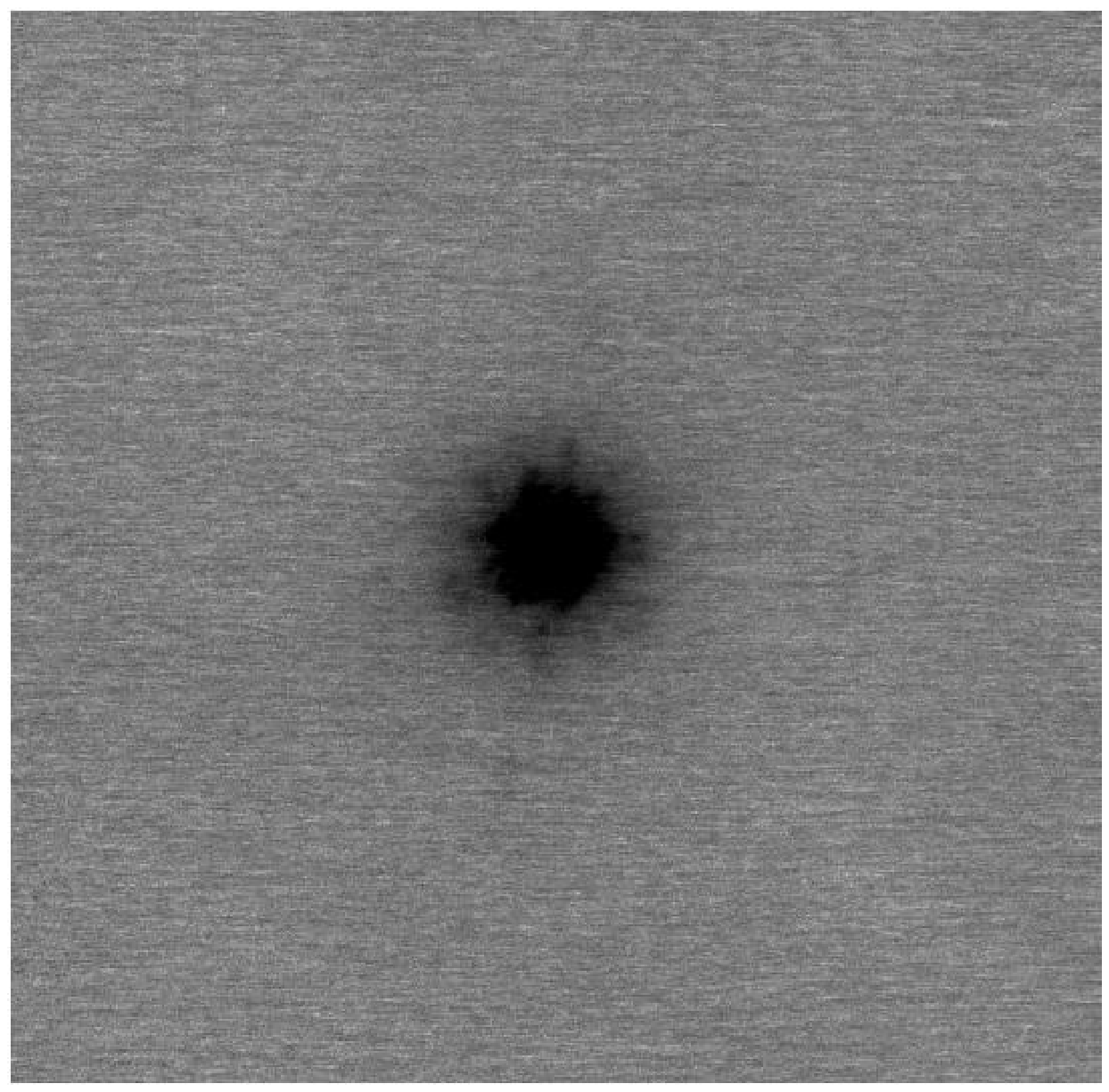}
\caption{\label{KeckAO} 
Keck NIRC2 images of CSS100217 with North up and East to the left. Left:  
The ``wide field`` image with scale $\sim 15\arcsec$ across. Right: 
The ``narrow field'' image with scale $\sim 10\arcsec$ across.
}
}
\end{figure}

We obtained images of the object using the NIRC2 imager and laser guide-star adaptive-optics (LGSAO; 
Wizinowich et al.~2006) at the Keck II 10-m telescope on Mauna Kea, Hawaii, on 2010 June 3 UT. We used the $K^\prime$ 
band and obtained multiple dithered exposures both in the NIRC2 ``wide field'' mode, with the sampling of 40 mas
pixel$^{-1}$, and the ``narrow field'' mode, with the sampling of 10 mas pixel$^{-1}$, followed by the observations of 
a nearby bright star to define the PSF. For our tip-tilt configurations and seeing conditions, the estimated
$K^\prime$-band Strehl ratio is $\sim 0.2$.  A coadd of a number of dithered images is shown in Figure \ref{KeckAO}.  At the 
redshift of the transient, the pixel sizes correspond to projected linear sizes of $\sim 103$ and $\sim 26$ pc.  Thus
the effective angular resolution of these observations is comparable to, or slightly better than that of the HST images, 
and the source appears unresolved even with the higher resolution images.

\subsection{Radio Observations}

\subsubsection{EVLA}

We observed CSS100217 with the Expanded Very Large Array (EVLA) for three separate epochs between April and May 2010.
Each observation was one hour in duration. We observed simultaneously at central frequencies of 4496 MHz and 7916 MHz
using the new wide C-band feeds for a total bandwidth of 256 MHz. The EVLA was in the compact D configuration, yielding
synthesized beams of 13$^{\prime\prime}$ and 7.5$^{\prime\prime}$ at 4496 MHz and 7916 MHz, respectively. Phase
calibration was carried out by making short observations of the nearby point source J1033+4166 every 10 minutes, while
amplitude and bandpass calibration was achieved using an observation of 3C\,147 at the end of each observing run. The
data were reduced following standard practice in the Astronomical Image Processing System (AIPS) software package.

A single, unresolved radio source was detected on all three epochs.  Table 7 summarizes the results of these
observations. The flux density varied between 0.3 and 0.5 mJy, but the spectrum was relatively flat between these two
radio frequencies. Inspection of the archival FIRST images made with data taken in the early 1990's (Becker et al.~1995)
shows a possible detection at this position with a 1.4 GHz flux density of 300$\pm$145 $\mu$Jy. At the redshift of
z=0.15 ($d_L$=705 Mpc) these flux densities correspond to a spectral luminosity L$_R\simeq 2.3 \times 10^{29}$ erg
s$^{-1}$ Hz$^{-1}$. The variation in spectral index of the 4.5 GHz and 7.9 GHz data suggests that the distribution is
getting flatter and eventually inverting. The origin of this effect is unknown.

The radio luminosity of CSS100217 plus the host galaxy exceeds that of the most luminous Type Ib/c and II supernovae by
factors of several (Chevalier, Fransson, \& Nymark 2006) and is approaching values more typical of gamma-ray burst
afterglows.  The absolute near-IR luminosity from 2MASS is $M_K=-24.65$. Mauch \& Sadler (1995) studied the properties
of star-forming galaxies and radio loud AGN. The combination of the 2MASS K-band luminosity and initial FIRST radio power 
place the object in the region of overlap between both star-forming galaxies and AGN.  Based on the local luminosity 
function star forming galaxies are slightly more common at this radio power (Mauch \& Sadler 1995). 
However, during outburst the near-IR luminosity is $M_k=-26$, making it brighter than most star-forming radio galaxies.
Additionally, when considered with the flat spectral index and the modest variability, the simplest explanation 
is that the radio emission originates from nuclear activity from the central black hole in the galaxy.

\subsubsection{GMRT}

The source $CSS100217$ was observed with the Giant Meterwave Radio Telescope (GMRT) on May 23, 2010 at a center
frequency of 608 MHz. Data were recorded using a new software correlator with a band-width of 33 MHz divided into 512
channels. The total observing time was 5 hrs (including the calibration overheads). A total of 2 hrs 45 mins was spent
on the target, interspersed with 4.5 min scans on the phase calibrator every 45 mins. The source 0834+555, which is an
unresolved point source at 610 MHz, was chosen as the phase calibrator, while 3C\,147 and 3C\,286 were used to calibrate
the flux as well as the bandpass. Their fluxes were determined using the Baars et al.~(1977) flux scale. 
Data were reduced using standard AIPS processing.  

The image was made using the central 30.5 MHz centered at 607.95 MHz. The synthesized beam achieved was
$5.78^{\prime\prime}\times4.21^{\prime\prime}$. An unresolved point source of flux 873$\pm$80 $\mu$Jy was detected. The
rms near to the source was about 50 $\mu$Jy, which implies a detection significance of about 17$\sigma$.  
The combination of EVLA data with GMRT suggests a spectral slope $\alpha$ between -0.4 and -0.5 depending on EVLA
observation date. This is similar to values commonly observed in low luminosity Seyfert galaxies (Ulvestad \& Ho 2001).

\subsection{X-ray Observations}

X-ray observations of the transient were taken on 6 April 2010 with Swift XRT (0.2-10 keV) to determine whether the
object was a strong soft X-ray source as expected for a TDE.  The object was clearly detected with flux $\rm \sim 10^{43}
ergs\, s^{-1}$ and is a soft source consistent with a TDE or NLS1 galaxy (Boller et al.~1996).

\subsection{Fermi-LAT follow-up}

We searched for emission in the gamma-ray band that was spatially and temporally coincident with the optical flare from
CSS100217 using data from the Large Area Telescope (LAT) aboard the {\it Fermi Gamma-ray Space Telescope} satellite.  The
{\it Fermi}-LAT instrument is sensitive to gamma-ray photons with energies in the range 20\,MeV to $>\,300$\,GeV.  With its
$\sim 2.4$\,sr field-of-view (FoV) and scanning mode of operation, {\it Fermi}-LAT provides all-sky monitoring coverage on 3
hour time scales (Atwood et al. 2009).

We extracted data within a $20^\circ$ acceptance cone centered on the CSS100217 position during the period 20 November
2009 to 04 July 2010, covering the nominal duration of the outburst.  We analyzed these data using the standard {\it Fermi}-LAT 
data analysis software, running the unbinned maximum likelihood analysis tasks; we fit the data over 1 week
intervals and over the entire extraction period. (See Abdo et al. 2010a for descriptions of the fitting methods.)  The
source model for these analyses comprised a power-law point source at the CSS100217 position (with unconstrained photon
index and normalization), all point sources from the 1FGL catalog (Abdo et al.~2010a) within the data extraction region,
and the models for the Galactic ({\tt gll\_iem\_v02.fit}) and isotropic diffuse emission ({\tt isotropic\_iem\_v02.fit})
that were recommended by the {\it Fermi}-LAT team.  The software and diffuse files are available from the Fermi Science
Support Center ({http://fermi.gsfc.nasa.gov/ssc/}).  The flux normalizations of the Galactic and isotropic components, the
point sources within $5^\circ$ of CSS100217, and the flaring blazar Mrk 421, which is $7.25^\circ$ from CSS100217, were
allowed to be fit freely. All other point source parameters were held fixed at their 1FGL catalog values.  No evidence
for a point source at the location of CSS100217 was found in the full dataset or in any of the 1-week intervals.  The
95\% confidence limit (C.L.) flux upper limit in the energy band 100\,MeV--100\,GeV for the full dataset is $3.1 \times
10^{-8}$\,photons\,cm$^{-2}$s$^{-1}$.  For the 1-week intervals, the 95\% C.L. upper limits ranged from $1.1 \times
10^{-7}$ to $3.4 \times 10^{-7}$\,photons\,cm$^{-2}$s$^{-1}$.  Near the time of the peak optical flux (MJD 55272), the 1
week 95\% C.L. upper-limit for CSS100217 was $2.1 \times 10^{-7}$\,photons\,cm$^{-2}$s$^{-1}$.  We note that the two
nearest point sources, 1FGL J1033.2+4116 ($0.95^\circ$ offset from CSS100217) and 1FGL J1023.6+3937 ($1.52^\circ$
offset), are both associated with blazars (Abdo et al.  2010b).  Neither source showed significant emission during any
of the weekly time intervals that we considered, and the fitted fluxes for each were 
$< 6 \times 10^{-8}$\,photons\,cm$^{-2}$s$^{-1}$ at the time of the peak optical flux of CSS100217.

Since luminous optical transients such as type Ib/c supernovae have been associated with gamma-ray bursts (GRBs;
Paczynski 1997; Stanek et al.~2003) we considered the possibility the CSS100217 was associated with such an event.  We
queried the GCN Notices archive (http://gcn.gsfc.nasa.gov/) for any bursts that occurred around the time of the onset of
the optical flare.  From 20 November to 29 December 2009, GCN Notices for 49 GRB triggers were issued. Of those 49
triggers, the closest GCN location to CSS100217 was $14.6^\circ$ away (RA, Dec = $155.667^\circ$, $55.233^\circ$, J2000)
for trigger number 281250934 issued by the Gamma-ray Burst Monitor (GBM) aboard {\it Fermi} with trigger time
05:15:32.98 UT, 30 November 2009.  The other 48 GRB candidate locations were more than $20^\circ$ away from CSS100217.
No obvious transient at the CSS10027 location is present in the LAT data stream in the 200 seconds bracketing the GBM
trigger time.  However, at the time of the trigger, the GBM location was at the edge of the LAT FoV (off-axis angle of
$70^\circ$). So to search for possible extended, longer timescale emission (e.g., Abdo et al. 2009), we extracted data
centered on the CSS100217 position for the $10^4$\,s following the GBM trigger.  An unbinned likelihood analysis of
those data finds no evidence for a point source and yields a 95\% C.L. flux upper limit of $3.7\times
10^{-6}$\,photons\,cm$^{-2}$s$^{-1}$ for energies 100\,MeV--100\,GeV.

\section{The Nature of Transient CSS100217}

Determination of the nature of transient CSS100217 bares distinct similarities to that of transient SDSS
J095209.56+214313.3 discovered by Komossa et al. (2008). Following the initial interpretation of SDSS J09522.56+2143 as
a tidal disruption event (TDE) by Komossa et al (2009) obtained additional data and reinterpreted their discovery as either
a TDE, AGN variability or a luminous type IIn supernova.  In this case we have the same three likely causes for the observed
outburst.  However, unlike SDSS J095209.56+214313.3, which was only discovered and followed two years after the outburst, we
obtain a spectrum before the event and were aware of the distinctive nature of the event from well before it reached
its optical peak. We were thus able obtained significant data covering the event. Nevertheless, the interpretation of
CSS100217 remains unclear. Below we will outline the cases for and against each of the possibilities.

\subsection{Type IIn Supernovae}

As noted above CSS100217 bares distinct similarities to type IIn supernova explosions.  Therefore it is useful to
compare the event to known sources of this type.  Type IIn supernovae are known to have a five-magnitude range in their
peak brightness and outburst timescales from months to years (Richardson et al. 2002).  In recent years luminous and
energetic type IIn supernova, such as SN 2006gy (Quimby et al.~2006), have been discovered and analyzed (Smith et
al.~2007). Comparison between the lightcurve of CSS100217, and the energetic type IIn supernova 2006gy (Smith et al.~2007)
is given in Figure \ref{LC2}. The similarity of these curves suggests a possible relation between the two events. 
Additional type IIn supernova with very long timescales have also been recently discovered. For example, SN 2008iy 
(Catelan et al.~2009) and 2003ma (Rest et al.~2009) have been observed over many years.  The decline rate measured from our
difference image photometry for CSS100217 at late times is 0.0175 mag day$^{-1}$.  This is much faster than observed for
the longest events, but slower than observed for most type IIn supernovae.

Additionally, the spectra of CSS100217 show evolution of the broad Balmer components.  In the initial IGO spectrum the
$\rm H_{\alpha}$ component had FWHM $\sim 2800$ km s$^{-1}$ and 3 months later during the Keck observations the width
was $\sim 4800$ km/s. This is close to that observed for $H_{\alpha}$ in SN 2006gy which initially had FWHM$\rm =2500 
km\, s^{-1}$ and later 4000 $\rm km\, s^{-1}$ (Smith et al.~2007).  The $H_{\beta}$ emission shows the similar evolution.
In the case of CSS100217, as in known type IIn supernovae, we see that the $H_{\alpha}$ emission component strengthens
with time, while the observed luminosity declines. This suggests that the emission will continue for an extended period
of time.  The spectrum of CSS100217 exhibits relatively strong high-energy Balmer lines ($H_{\gamma}$, $H_{\delta}$,
$H_{\epsilon}$) that are more commonly observed in LBV outbursts like SN 2009ip (Smith et al.~2010b).  This may be
attributed to the surrounding medium being hotter than normally observed for type IIn supernovae.  The emission features
observed in type IIn supernovae are due to the expansion of explosion ejecta into a dense circumstellar medium (CSM)
surrounding massive eta Carina-like stars (Smith et al.~2010a). In such cases, the outbursts of LBVs in years prior to
explosion are believed to cause massive shells through which the blast shock travels.  The location of the event near the
active galactic nucleus may be responsible for causing additional heating of the CSM surrounding a massive progenitor
star.

\subsubsection{Supernovae in Nuclear Regions}

Modern optical searches for supernovae have been tuned to avoid events occurring near galactic nuclei. The main reason for
this is the high likelihood that such variability is due to variation caused by an AGN.  AGN are thought to be present in
43\% of galaxies (Ho et al.~1997).  Therefore, the chance of finding a variation in any given galaxy due to an AGN
is much larger than the chance that this is due to a nuclear supernova.  In addition, supernovae lying far from the
crowded cores of galaxies can also be more readily discovered and spectroscopically confirmed.
Furthermore, supernovae in the dense cores of regular galaxies are likely to suffer from significant extinction ($A_V >
10$ in many cases), so that few are visible in optical wavelengths.  However, in the cores of luminous infrared galaxies
(LIRGs), such as the interacting system Arp299, at least one CCSN is expected every year (Mattila et al.~2004).  Such
supernovae can be discovered using near-IR imaging or radio observations (Perez-Torres et al.~2007).

Perez-Torres et al.~(2007) discovered 26 sources that they believed to be radio supernovae or supernova remnants within
the central 150pc of Arp299A. Additional follow-up by Perez-Torres et al.~(2010) showed that one of the central sources
was in fact a low-luminosity AGN. However, Perez-Torres et al.~(2010) identified one companion source within a projected
separation of 2pc as a radio supernova. In many cases the supernova observed near the core of a galaxy will lie in front
of the dust extinction layer making them visible in optical surveys.  Additionally, SN lying within the narrow line
region near an AGN (10pc - 1kpc) may not be obscured by the dense gas and dust disk.

Strubbe \& Quataert (2010) studied how one could discern nuclear supernovae from TDEs. They found that Hubble Space
Telescope or ground-based adaptive optics were needed to reduce contamination of supernovae near the galactic nucleus to
a level close to that expected for TDEs. However, even with high-resolution imaging the number of type Ia and II
supernovae will exceed the number of TDEs.

\subsection{Tidal Disruption Events}

The search for tidal disruption events has recently led to the discovery of a number of candidate TDEs ( Kompsa et al.
2002; Esquej et al. 2007; Gezari et al.~2009a; Cappelluti et al.~2009; Maksym et al.~2010).  Dynamical models of galaxy
nuclei predict that these events should occur at a rate of $\rm 10^{-4} - 10^{-5} galaxy^{-1} yr ^{-1}$ (Magorrian \&
Tremiane 1999; Wang \& Merritt 2004).  In comparison these events are 100 times less commonly observed than supernovae,
but still within range of transient surveys such as CRTS, PTF, PanSTARRS and LSST.

The signature of such events is bright UV and soft X-ray emission from the high-temperature outburst event that occurs
as the star is shredded and accreted by a massive black hole. Models for the optical lightcurves of these events show
significant variation with black hole mass.  TDEs are expected to have temperatures of $\sim 10^5$K and a luminosity
that declines with time as $t^{-5/3}$.  However, recent detailed models for the light curves of these events by Strubbe
\& Quataert (2009) and Ludato \& Rossi (2010) found the optical and UV lightcurve to be significantly shallower than 
$t^{-5/3}$. Ludato \& Rossi (2010) find a very slow $\rm t^{-5/12}$ decline at late times.

Although the clearest signature for TDEs is a flare near the center of an otherwise dormant galaxy, such events are very
likely to occur in galaxies that exhibit significant nuclear activity from black holes in the right mass range.  In
particular, SMBHs seen in blazars and QSOs are thought to be too massive for such disruption events to occur (Hills
1975).  Instead, in these cases the stars are swallowed whole. For black holes with masses $< 10^8 M_{\sun}$, as in
Seyfert galaxies, such events events should occur. Clearly the presence of X-ray, UV, and radio variability in 
AGN make the discovery of TDEs associated with AGN a much more complex task.

The slow rise and decline of the optical lightcurve of CSS100217 is inconsistent with the $t^{-5/3}$ decline 
expected for TDEs (Komossa et al.~1999) as well as the more recently theorized $\rm t^{-5/12}$ dependence.
A continuum fit to the IGO spectrum shifted to rest gives a temperature of $1.6\pm0.2 \times 10^4K$, much lower 
than the expected TDE value of $\sim 10^5$K. As noted earlier, the NUV photometry taken on 29 January 2010 by GALEX gives 
$NUV =17.08$. This value is 1.9 magnitudes brighter than when observed on 24 Jan 2004 (Gezari et al.~2010).
At this time the object was $\sim 1.5$ magnitudes brighter in CSS observations. The combination of extinction 
corrected GALEX NUV flux (centered at 2315\AA) along with the IGO data is consistent with a temperature of 
$\sim 10^4 K$. However, this early temperature is poorly constrained because of the three week period between 
these observations.

Recently, Strubbe \& Quataert (2010) studied the spectroscopic signatures of TDE events.  The results suggest that most
events will give rise to featureless blackbody spectra at wavelengths above 2000 \AA. In an otherwise inactive galaxy
the signature of a disruption event by a quiescent black hole would be similar to the early spectrum of a supernova.
However, the strong spectroscopic features would not evolve as time passed.  In presence of an AGN the combination of a
featureless spectrum with that of the AGN would be very difficult to discern from that of just an AGN.  Strubbe \&
Quataert (2010) note that TDEs, unlike type IIn SNe, become hotter with time and are not expected to produce strong
optical emission lines.  The theoretical predictions for spectra, temperature, and variability all appear inconsistent
with the observations of CSS100217.

\subsection{AGN Variability}

As the host galaxy appears to contain an AGN it is important to consider whether CSS100217 might be due to an accretion
event involving the massive black hole. In the previous section we determined that the event was very unlikely to be due
to the tidal disruption of a star by the black hole. However, as AGN are by nature variable it is necessary to
understand the limits of this variability in comparison to CSS100217. Recently, Ai et al.~(2010) investigated the
variability of a small sample of 58 NLS1s and 217 BLS1s. They found that the variability in three years of SDSS
stripe-82 data was less than 0.4 magnitudes.  As the host galaxy for CSS100217 is a NLS1, the sample of Ai et al.~(2010)
is too small to for a meaningful understanding of the true limits of variability in NLS1s.

\subsubsection{Comparison of CSS100217 with Known NLS1}

In order to assess how the variability we observed in CSS100217 matches with that naturally observed for NLS1-type AGN
we decided to investigate the light curves of a large, uniform sample of spectroscopically confirmed NLS1 galaxies.
We first selected the spectroscopically confirmed sample of NLS1 from Zhou et al.~(2006).  This set consists of $\sim
2000$ NLS1, chosen from emission line galaxies in the SDSS dr5 catalog. We found matches to most of the Zhou et
al.~NLS1s in the CSS data archive and for each object we extracted the photometry from the same 5 year period as
CSS100217. Many of the lightcurves show clear, significant variation over this period.

After removing the NLS1 lightcurves that were affected by blends or bad photometry (due to being on image edges, etc.),
we were left with 1541 objects. We matched the objects against the FIRST radio catalog using a $5\arcsec$ radius.  Of
the 143 matches, only three had an offset larger than $2\arcsec$. The NLS1 host of CSS100217 was not
detected by FIRST or NVSS radio surveys.

\begin{figure}{
\epsscale{1.1}
\plottwo{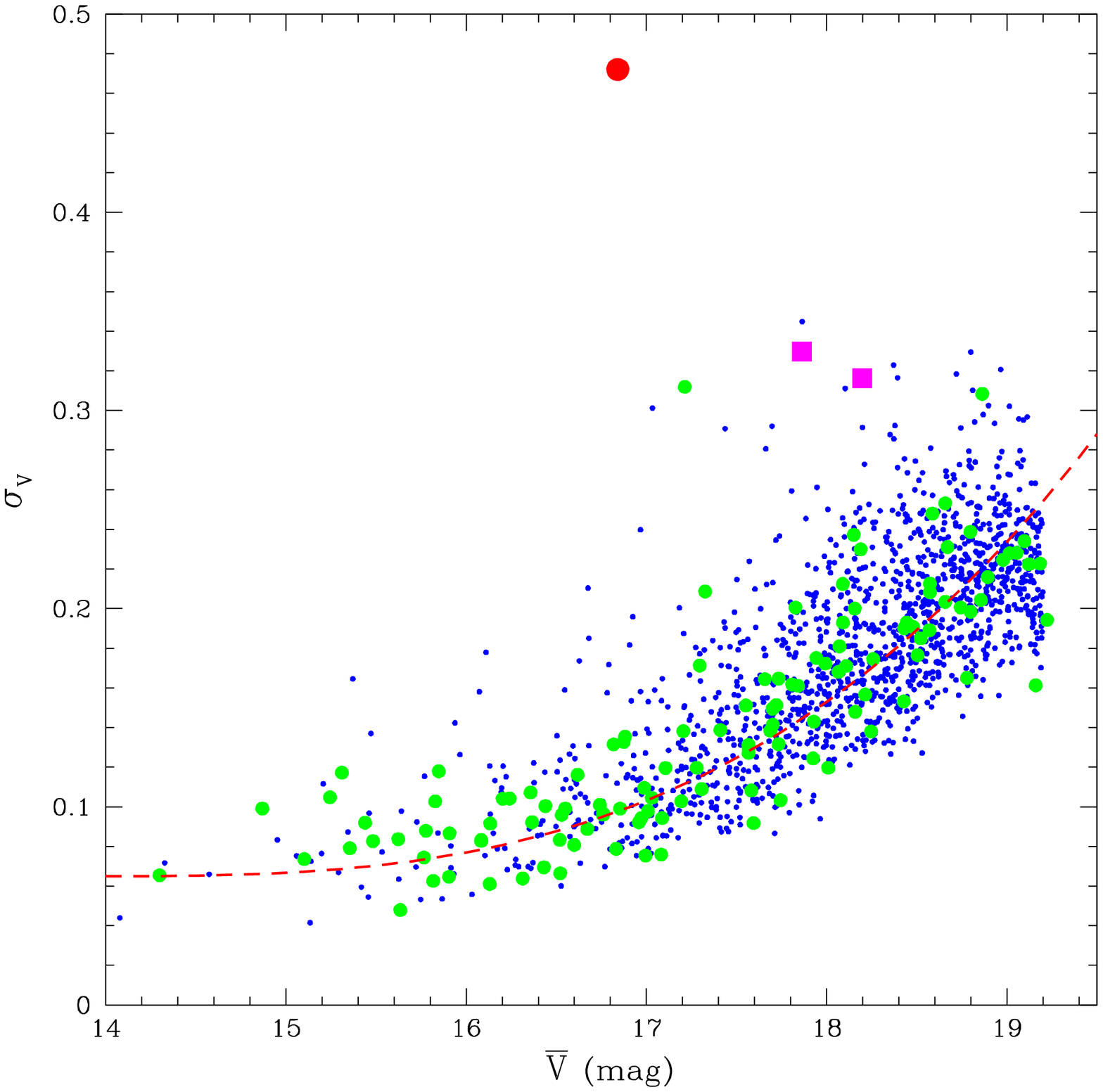}{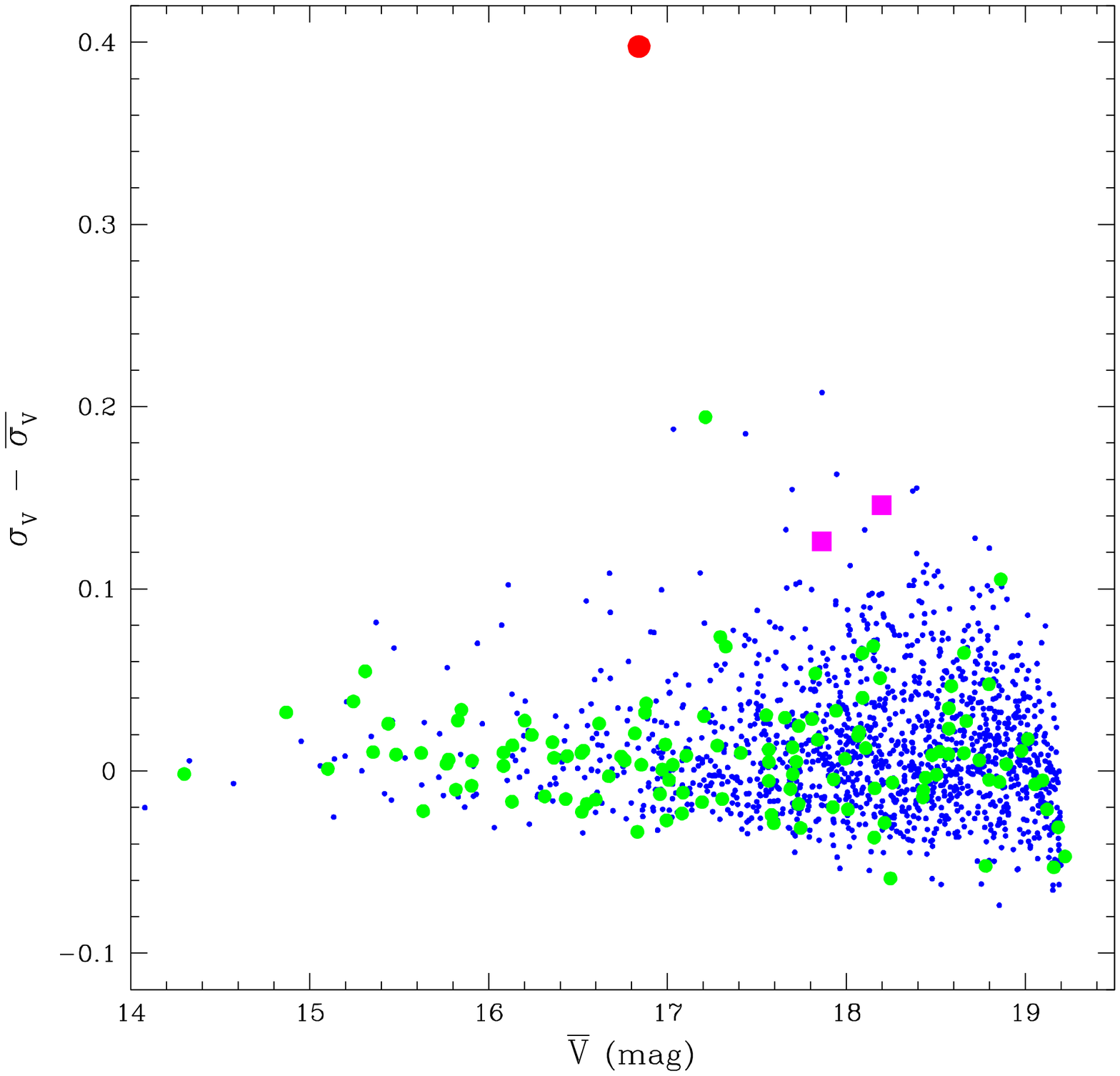}
\caption{\label{Var} 
  Comparison of the variability of CSS100217 (large dot) with known NLS1s selected from Zhou et al.~(2006).  Radio
  sources detected by FIRST are marked with medium-size dots.  The location of radio loud NLS1 SDSS J150506.47+032630.8
  and SDSS J094857.3+002225.5 are marked with boxes  Left: Standard deviation of the light curve based on central 90\% of the data.  The
  dotted-line shows the trend of increasing scatter that is mainly due to decreasing brightness. Right: The scatter of
  the data after removing the trend of increasing scatter with decreasing magnitude.  
} 
}
\end{figure}

For each light curve we iteratively determined the mean and range of values for the 90\% of data lying nearest the mean.
This process removes photometric outliers due to bad photometry caused by artifacts, satellite trails, bad seeing, etc.
We then calculate the standard deviation assuming that the remaining data follows the normal distribution.  In Figure
\ref{Var}, we plot the scatter of the data points for the NLS1 sample.  As the scatter is dominated by increasing
photometric uncertainty with decreasing brightness we also plot the data after removing this trend. Clearly CSS100217 is
an extreme outlier, whether we correct for the photometric uncertainty trend or not. 

\begin{figure}{
\epsscale{1.1}
\plottwo{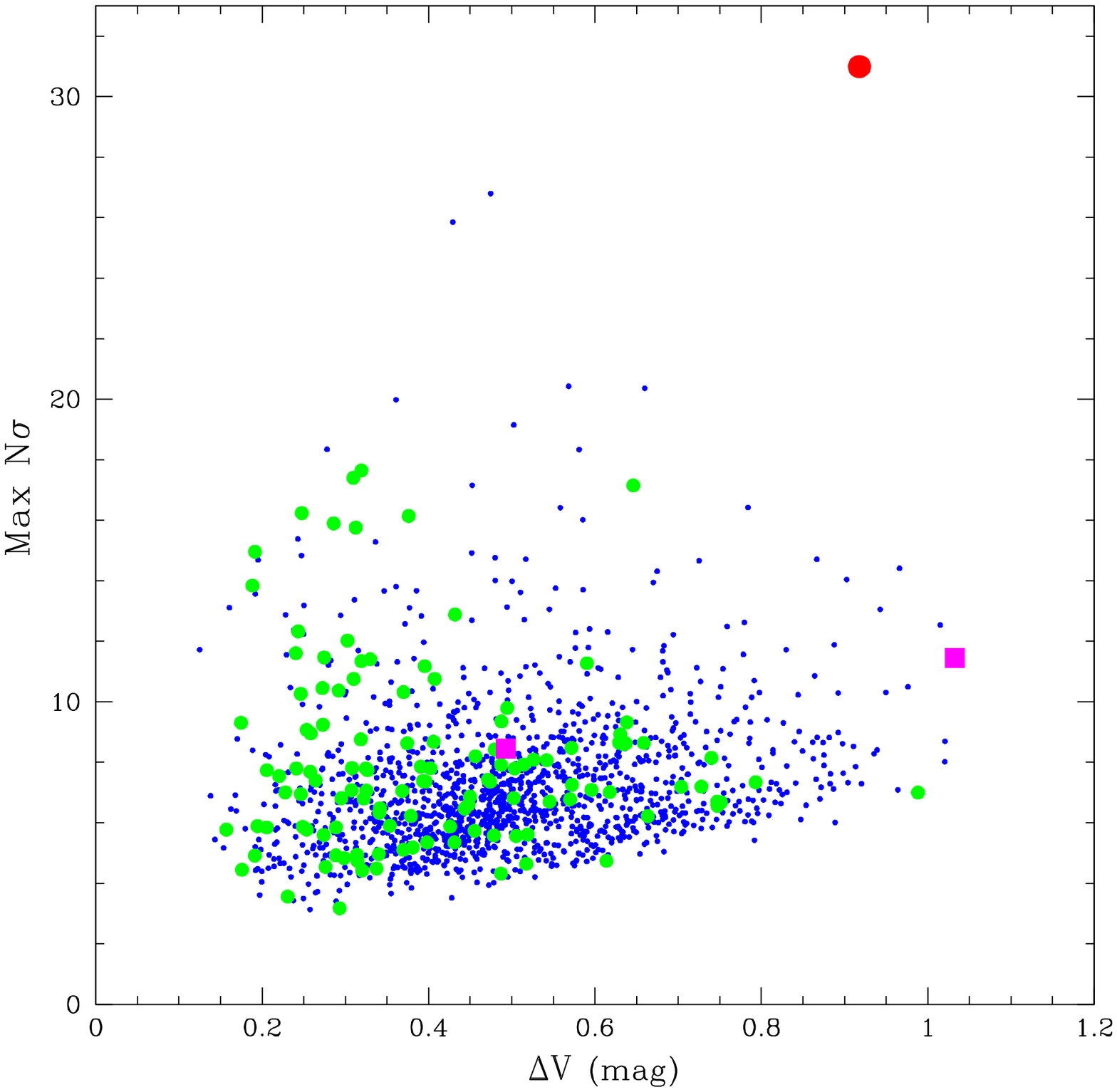}{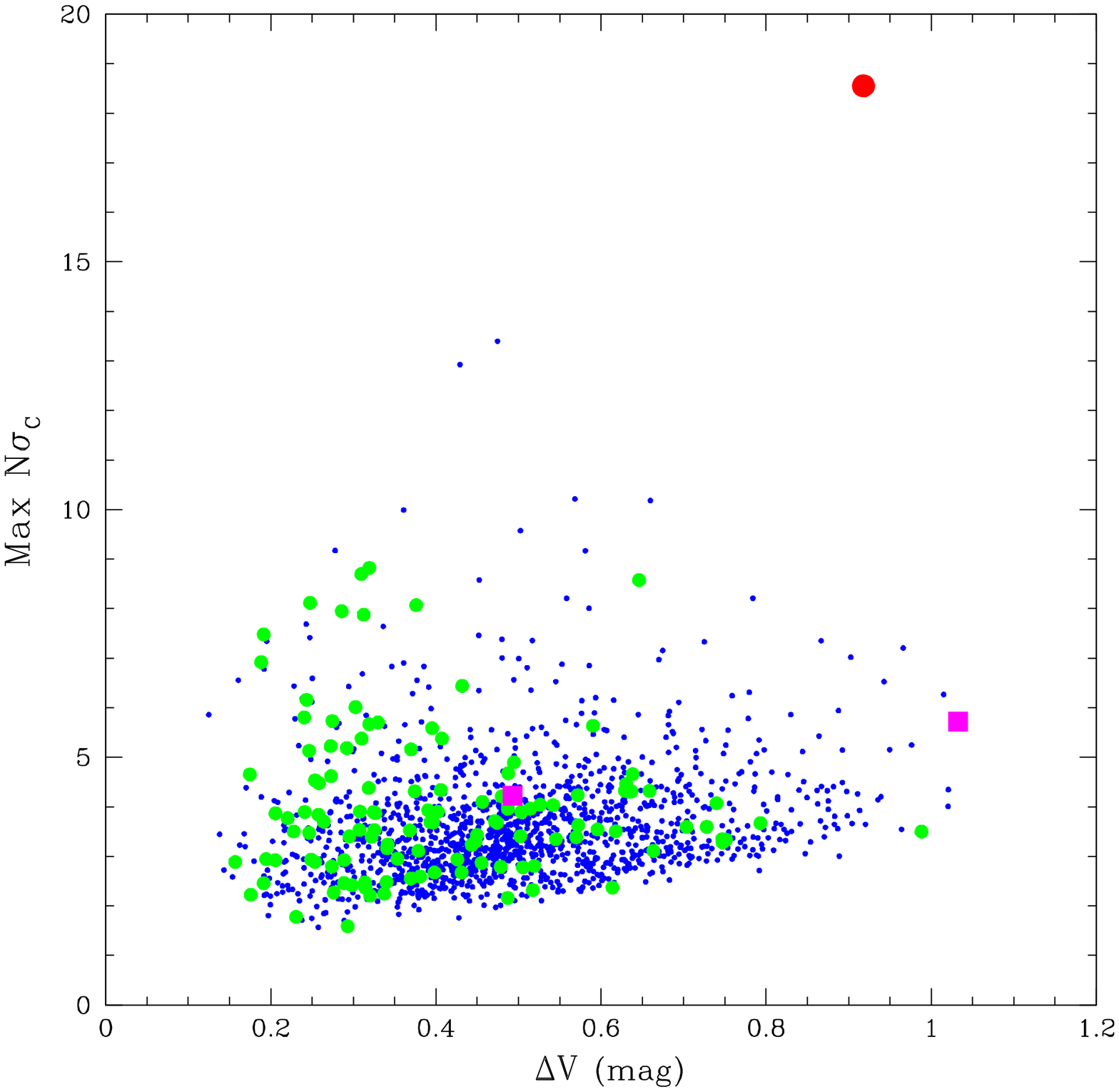}
\caption{\label{Var2} 
Variability of CSS100217 (large dot) relative to known NLS1s selected from Zhou et al.~(2006).
Radio sources detected by FIRST are marked with medium-size dots.
The location of radio loud NLS1 SDSS J150506.47+032630.8 and SDSS J094857.3+002225.5 
are marked with boxes. Left: Maximum deviation from median magnitude in terms of sigma based on NLS1 data.
Right: Maximum deviation from median magnitude in sigma based on NLS1 data that
has been corrected for increasing scatter with decreasing brightness.
}
}
\end{figure}

To further investigate the variability we determined the median magnitude for each light curve and the value that varies
most from the median in terms of photometric uncertainties.  Here we included only those data points lying within $\pm 3
\sigma$ as determined from the scatter in each lightcurve. This sigma cut was necessary because of outliers in other
NLS1 photometry, due to artifacts, etc. In Figure \ref{Var2}, we plot the peak variation versus the number of sigma that
the point represents.  The size of the deviations are mainly less than $5\sigma$, suggesting that the photometric
uncertainties are approximately correct. We also show the result when we have removed the effect of increasing
photometric scatter with increasing magnitude.  This figure shows that CSS100217 is once again an extreme outlier in
both plots.  As the host of CSS100217 is brighter than most of the NLS1 sample, the offset is most clearly seen in the
corrected plots.  We note that the largest deviation ($\Delta V$) from the median is not shown in the figures for
CSS100217 as the true maximum lies $> 3\sigma$ from the median magnitude. The level of observed NLS1 variation is
consistent with that found by Ai et al.~(2010).

\begin{figure}{
\epsscale{1.1}
\plottwo{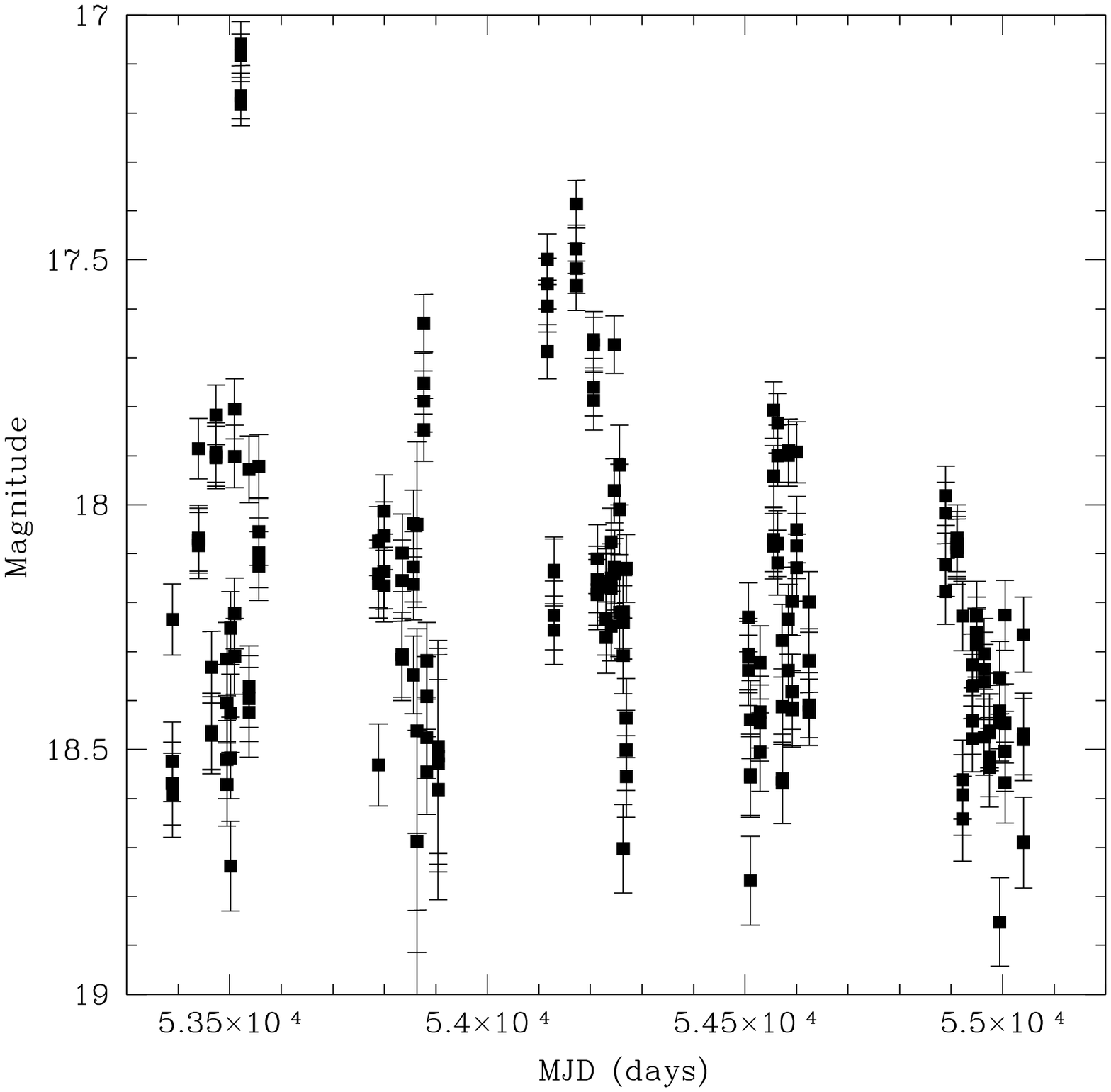}{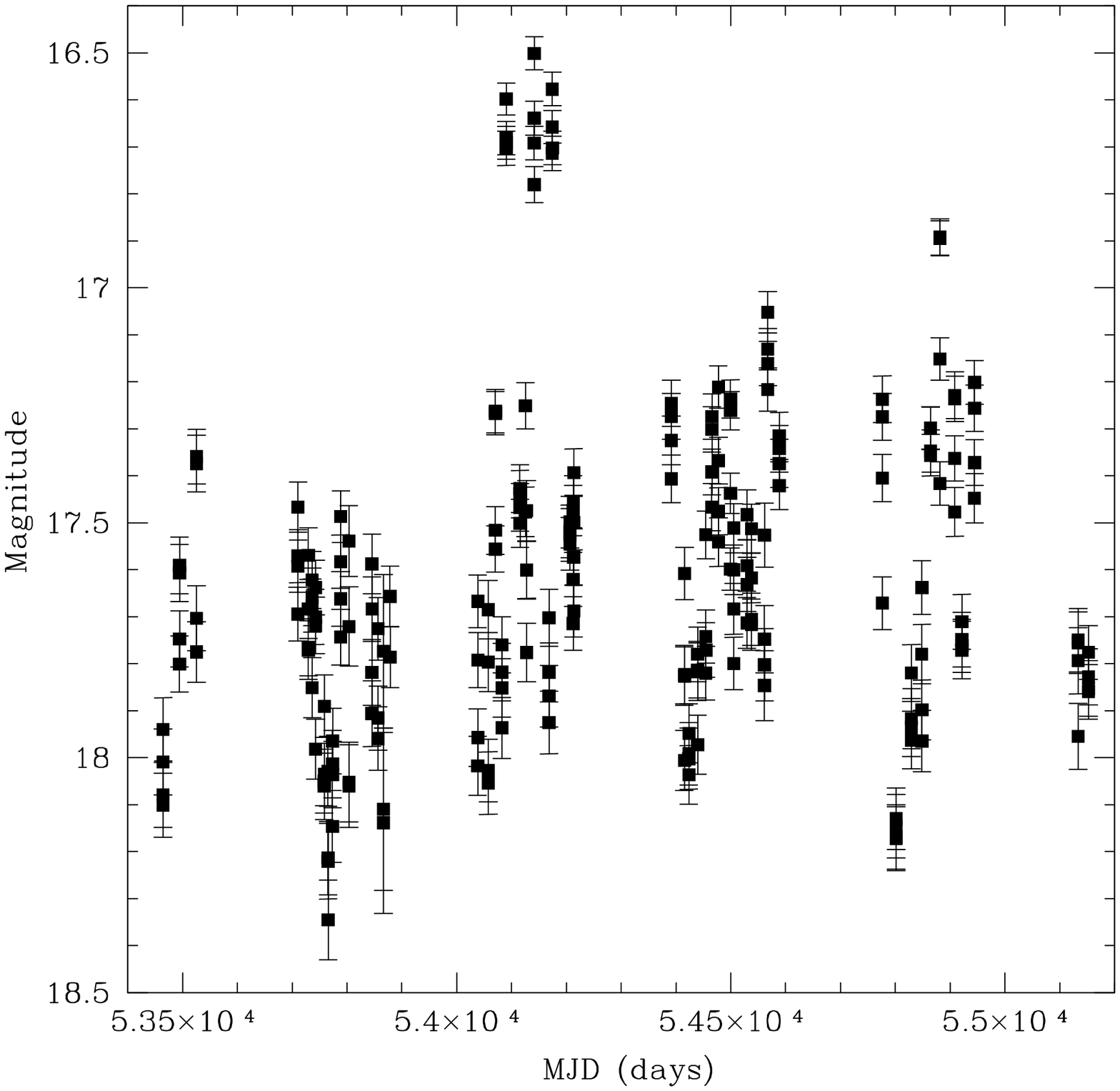}
\caption{\label{NLS1Rad} 
The CSS V-band light curves of two highly variable radio-loud NLS1 galaxies.  
Left: the lightcurve of SDSS J150506.47+032630.8. Right: the lightcurve of SDSS J094857.3+002225.5.  
} 
}
\end{figure}

Among the Zhou et al.~(2006) spectroscopically selected NLS1 two are known to exhibit significant variability on short
timescales. However, both of these NLS1, SDSS J150506.47+032630.8 (QSO B1502+036, Yuan et al.~2008) and SDSS
J094857.3+002225.5 (Zhou et al.~2003), are radio-loud NLS1 (380.49 mJy \& 111.46 mJy, respectively in FIRST data).
In Figure \ref{NLS1Rad}, we present the light curves of these two NLS1. Both objects exhibit erratic 
variability which is completely unlike CSS100217 or the other radio quiet NLS1 examined.
In these cases, like BL Lacs, the variability has been attributed to a jet lying along our line-of-sight and emanating
from the central black hole. The optical variability is attributed largely to this source.  The maximum observed
variability for these objects in CSS data of SDSS J150506.47+032630.8 and SDSS J094857.3+002225 is $\sim 1.5$ and $\sim
1.2$ magnitudes, respectively.  The CSS lightcurves of these NLS1 resemble those of blazars monitored by CRTS.
SDSS J094857.3+002225 exhibits intra-day variability (Liu et al.~2010).  These two sources are $\sim 500$ times
brighter than CSS100217 at radio wavelengths and neither exhibits a prolonged outburst like that observed for CSS100217.

From our analysis we find that a small fraction of the NLS1 exhibit significant variability.  In most cases the
variability is a slow change in brightness over a timescale of years, as previously observed for QSOs.  A couple of NLS1
in our large sample exhibit a high level of variability but none like that observed in CSS100217. In the cases where a
rapid event lasting months is seen the source is consistent with a luminous supernova.  This strongly suggests that
CSS100217 is not due to regular AGN variability.  However, it is not possible to completely exclude very rare sources of
AGN variability.  Recent reports of AGN outbursts have been made in sparsely sampled data by Kankare et al.~(2010) and
Valenti et al.~(2010).  However, these have been found to be due to gradual changes rather than outbursts when examined
in better sampled data (Drake et al.~2010b, 2010c).  TDEs themselves are examples of rare AGN outbursts.  Although the
characteristics of CSS100217 do not follow current theoretical predictions for TDEs, recent revisions of both the
timescale (Lodato \& Rossi 2010) and spectroscopic signature of such events (Strubbe \& Quataert 2010) suggests a need
for constraints based on empirical data. Future surveys such as LSST should provide these constraints.

\section{Discussion and Conclusions}

We have analysed the multi-wavelength data covering the discovery of the unusual transient CSS100217. The coincidence of
the event's location with an NLS1 galaxy makes the event most consistent with a TDE, AGN variability, or a supernova.

\subsection{CSS100217 as a TDE}

The large outburst of CSS100217 within 150pc of of the nucleus of an AGN make it a good candidate for the tidal
disruption of a massive star by the central black hole. The object is also detected as an X-ray source in Swift
telescope follow-up. However, the event exhibits a slow rise over a period of months and a similarly slow decline. The
lightcurve is completely inconsistent with theoretical TDE models that predict an immediate rise and $\rm
t^{-5/3}$ decline.  However, the more recent lightcurve models for TDEs (Lodato \& Rossi 2010) show optical light curves
with rises times of a month followed by a flat peak and much slower decline following $\rm t^{-5/12}$. Such light curves
are very similar to those observed for supernovae, although with much longer tails.  The peak brightness $M_{V\,CSS}
\sim -23$ of the event is far greater than theorized for TDEs, which are predicted to reach the brightness of regular
SNe $M_{V} \sim -18$. Furthermore, the galaxy-subtracted spectra of the event exhibit strong Balmer emission and a
continuum consistent with $T=1.5 \times 10^4$K rather than the theorized $10^5$K value.  Based on theoretical
predictions CSS100217 is a poor TDE candidate.

\subsection{CSS100217 as AGN variability}

The host galaxy spectrum of SDSS J102912.58+404219.7 exhibits both broad and narrow Balmer lines consistent as well as
Fe-II with this being a NLS1 galaxy.  The HST follow-up of the event shows that the location of CSS100217 is consistent
with the location of the bright nucleus. The object is also found to be a an X-ray source if Swift data, a radio
source in EVLA and GMRT follow-up observations, and is brighter than known supernovae at these wavelengths.

It is well known that AGN variability can amount to variations of a magnitude or more over the period of a number of
years.  As AGN are more common than supernovae, surveys for supernovae specifically avoid follow-up of detections
occurring near the cores of galaxies. Indeed, the IAU recommend all supernova candidates are checked against the
V\'eron-Cetty and V\'eron (2010) AGN catalog before submission. Events near the core of galaxies are not announced or
given an official ID by the IAU's Central Bureau for Astronomical Telegrams unless they are also spectroscopically
confirmed to reduce the possibility that any given discovery is due to AGN variability. The peak outburst luminosity of
CSS100217 is well within the range observed for NLS1.
One model for AGN variability suggests that the cause is the superposition of supernova explosions in giant stellar
clusters (Trelevich et al.~1992). In this model the explosions interact with the high density circumnuclear
environment. If this model is correct, CSS100217 could be an example of such an event.

Analysis of host galaxy SDSS J102912.58+404219.7 on the BPT diagram suggests that it is not a typical NLS1.  The object
lies on the locus of starburst galaxies suggesting that the galaxy is also undergoing rapid star formation.  Comparison
of the optical variability of CSS100217 with 1500 other NLS1 galaxies selected from SDSS data strongly suggests that the
observed variability is inconsistent with normal AGN variability. The follow-up spectra of CSS100217 exhibits a Balmer
component that is significantly broader than the archival spectrum. This component is twice as broad as the $\rm \sim 2000
km\, s^{-1}$ defining limit for NLS1 galaxies (Osterbrock \& Pogge 1985). Furthermore, although the narrow and medium
$H_{\alpha}$ velocity components vary little in width with time the strength increased as the event faded rather than
decreasing. Variation of the narrow component on short timescales is not expected because of the size of narrow-line
region. The event also shows the presence of a hot continuum component not observed in the prior SDSS spectrum.  This
component cools as the outburst fades. Furthermore, although there are few examples of supernova occurring near the cores
of AGN, this is due to active selection against such events because of possible AGN variability.  Similar
events may have been detected in the past but dismissed because of their location near the core of a galaxy and the
presence of a NLS1-like spectrum.  Based on the observations of NLS1, CSS100217 is unlikely to be due to AGN variability.

\subsection{CSS100217 as a Supernova}

The distinctive smooth rise and fall of the lightcurve of CSS100217 closely matches the general shapes observed for
supernovae and type IIn, such as SN 2008iy and SN 2006gy. Spectroscopic follow-up of CSS100217 reveals strong narrow Balmer
features superimposed on broader features as is required for classification of type IIn supernovae.  The Balmer features
are observed to vary significantly from that observed in SDSS spectrum.  Some variation in the broad-line strength is
expected for NLS1 galaxies but such rapid variation in the narrow lines strength has not been observed.  Additionally
the follow-up spectra exhibit a broader, 4000 $\rm km\, s^{-1}$ component that is not seen in the SDSS spectra.
The velocity of this component increases with time in both $H_\alpha$ and $H_\beta$ until the most recent observations
while the narrow and medium components shows little change.  Also, the strength of the broad component is seen to
increase even as the event fades.  The host-subtracted spectrum exhibits strong Fe-II lines. Such lines are pronounced
in both NLS1 spectra and SN type-IIn like SN 2008iy, SN 2007rt (Trundle et al.~2009) and SN 1997ab (Hagen et
al.~1997). Such lines are not predicted by current models of TDEs.

\begin{figure}{
\epsscale{0.9}
\plotone{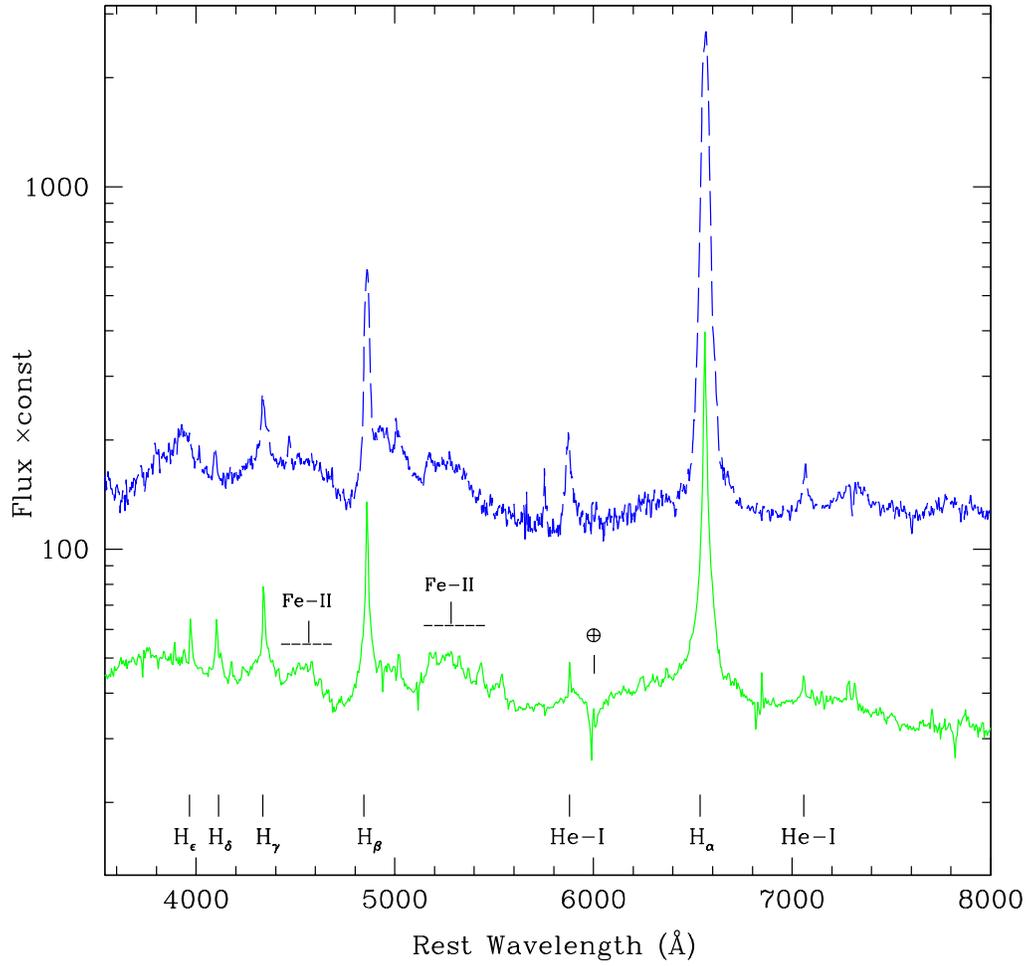}
\caption{\label{Emiss4} 
  A comparison between the host-subtracted Keck spectrum of CSS100217 and that of type-IIn supernova SN 2008iy.
  Solid-line the spectrum of CSS100217. Long dashed-line the spectrum of SN2008iy. In both case the events are observed
  well after maximum light.  
} 
}
\end{figure}

As is characteristic of type-IIn (Trundle et al.~2009) the spectra of CSS100217 do not exhibit the broad
P-Cygni features commonly observed in other type-II supernova. For example, the type-IIn prototype, SN 1998Z was determined 
not to exhibit a P-Cygni absorption component (Turatto et al.~1993).  Narrow P-Cygni features are commonly 
observed in high-resolution data of type-IIn supernova, but are not seen in low-resolution data (Trundle et al.~2009).
In Figure \ref{Emiss4}, we contrast the spectrum of CSS100217 with that of SN 2008iy.

The temperature derived from the continuum in the initial follow-up spectrum ($T=1.5\times 10^4$K) is consistent with
supernovae in general and very similar to that observed for the luminous type-IIn SN 2006gy (Smith et al.~2007).  The
brightness of this event, $\rm M_V= -22.7$, is the greater as observed for past IIn supernovae SN 2008fz (Drake et
al.~2010a) and SN 2008es (Gezari et al.~2009b). As type IIn supernova are known to exhibit a variation in peak
brightness of at least 5 magnitudes, this discovery is not very surprising, particularly since the most luminous among
these supernovae have only been discovered in the past decade as surveys have begun which search for transients in
intrinsically faint galaxies (Drake et al.~2009).

The presence of a luminous supernova near the core of an AGN is expected to be a rare occurrence.  However, the host
spectrum suggests a significant star formation rate, which would enhance the rate of type II supernova -- as seen in Arp299A by
Perez-Torres et al.~(2007). Additionally, based on Strubbe \& Quataert (2010), the rate of type II nuclear supernovae
will exceed the rate of TDEs by a factor of $\sim3$ at $0.05\arcsec$ resolution for a $\rm 10^7M_{\sun}$ black hole.
The radio detections by GMRT and EVLA are brighter than expected for a supernova, but are consistent with the presence
of the NLS1. These detections are below the threshold of past radio surveys like FIRST and are therefore consistent with
no change occurring due to CSS100217.  The lack of detection in {\it Fermi} data does not place a strong constraint on the
nature of this event as the follow-up spectra are not consistent with broad-line type Ib/c supernovae that have 
been linked to GRBs. Type-IIn supernova such as 2006gy have been detected in X-rays (Smith et al.~2007) but
once again the detection of CSS100217 by Swift is consistent with emission from the NLS1.

Massive $\eta$ Carinae-like LBVs undergo large outbursts and deposit material into the ISM that will eventually be
illuminated when the stars explode. In these cases much of the kinetic energy of these explosions is converted to
luminosity.  The brightness limit of such events is mainly constrained by the amount and distribution of circumstellar
material interacting with the supernova shock.  A series of dense circumstellar shells or a surrounding dense medium can
give rise to extreme luminosity so that the explosion appears like a super-massive star. Although the optical energy
expended by the event is higher than any past type-IIn supernova, it is within a factor of three of the bolometric values
for SN 2006gy ($5 \times 10^{51}$ ergs; Smith et al.~2010a) and SN 2003ma ($4 \times 10^{51}$ ergs; Rest et al.~2009).
The presence of strong continuum evolution and ongoing strong Balmer emission, along with a bright and distinctly
supernova-like lightcurve, suggest that CSS100217 is most likely an extremely luminous type IIn supernova near the
nucleus of an AGN/starburst galaxy. Further photometric and spectroscopic monitoring of CSS100217 should secure the
nature of this event and also give further insight into the nature of the host galaxy.

\subsection{A Supernova associated with the AGN?}

A key piece of evidence is the observed proximity of the transient to the AGN of the host, since both the HST and the
Keck AO observations indicate a single, unresolved point source, consistent with a single PSF.  It is worth noting that
they span a range of $\sim$ 5 - 6 in wavelength, so that the absence of a second point source in either one cannot be
attributed to a hypothetical extreme difference in colors.  For the HST, Sparrow's resolution limit is $0.043\arcsec$.
The resolution of the Keck AO images is comparable, with 0.04 arcsec pixels. Using the HST resolution limit and the
distance to CSS100217 based on the redshift we find that CSS100217 occurred within $\sim 150$ pc of the galactic nucleus.

The probability of a chance alignment seems small, although it cannot be rigorously excluded.  Since NLS1 are expected
to largely occur in spiral galaxies (Crenshaw et al.~2003), with typical half-light radii of a few kpc, the possibility
of a chance alignment along the line of sight but far from the AGN is a priori very small, depending on the unknown
central density profile of the host. We also note that there is no evidence for a redshift offset between the AGN
emission and the event emission.  Since the narrow line region in Seyfert-1 galaxies have been measured to extend to
sizes from 700 pc to 1.5kpc (Bennert et al.~2006), the event likely occurred well within the limits of the narrow-line
region.

Significant star formation has been indicated within the nuclear regions of number of Seyfert 1 galaxies and nuclear
starbursts have been predicted in the dusty tori of Seyfert galaxies (Imanishi \& Wada 2004; Riffel et al.~2007). The
presence of rapid star formation would naturally led to type-II supernovae in such regions. However, there is no
morphological evidence of star forming regions or superposed dust lanes outside of the nucleus, and again, the broad
range of wavelengths (U to K bands), as well as the lack of any reddening signature in our spectra argue against it
being hidden by dust.  This suggests that any star forming activity is likely within the region dominated by the AGN,
rather than in some unrelated region in its vicinity.

On the other hand, a strong UVX radiation field in the vicinity of the AGN would preclude star formation, unless it is
well shielded.  One interesting possibility is that the event is a SN associated with the outer edges of the accretion
disk, where it would be shielded from the AGN radiation.  These outer regions of the accretion disks are expected to be
violently unstable, naturally leading to fragmentation and star formation, as predicted originally by Shlosman \&
Begelman (1987, 1989), and further fortified by the modern numerical and semi-analytical models (Goodman 2003, Goodman
\& Tan 2004, Jiang \& Goodman 2010).  Moreover, these models favor formation of very massive stars, which would fit
naturally with the event as a hyperluminous supernova.  If so, this would be the first detection of such a SN from an
AGN accretion disk.

A question naturally arises, why such events have not been reported previously, either by us or by other groups?  A
likely cause is the deliberate bias of SN searches against AGN, where any observed variability could be naturally
assumed to be associated with the AGN itself.  With sufficient sampling transient surveys can now build empirical models
of nuclear optical variability.  This allows us to separate the slowly changing variability occurring over long
time-scales, common in AGN, from the rapid changes seen due to tidal disruption events and supernovae.  As we have shown
above, CSS100217 is highly unlikely to be an instance of a normal AGN variability.  The use of aperture photometry in
the CRTS survey produces bias against detecting optical transients such as regular supernovae in the brightest regions
of intrinsically luminous galaxies (Drake et al.~2009). For this reason only events brighter than $\rm M_{V\,CSS} \sim
-21$ would have been detected in this particular host galaxy. However, an event as luminous as CSS100217 would be
detected at any location within almost any galaxy to the distance of CSS100217.  We will report on a systematic search
for more such events in the CRTS data in a forthcoming paper.

\acknowledgements
We would like thank to Minjin Kim for help analyzing the SDSS spectrum.  The CRTS survey is supported by the
U.S.~National Science Foundation under grants AST-0909182 and CNS-0540369.  Support for program number GO proposal 12117
was provided by NASA through a grant from the Space Telescope Science Institute, which is operated by the Association of
Universities for Research in Astronomy, Inc., under NASA contract NAS 5-26555.  The work at Caltech was supported in
part by the NASA Fermi grant 08-FERMI08-0025, and by the Ajax Foundation. The CSS survey is funded by the National
Aeronautics and Space Administration under Grant No. NNG05GF22G issued through the Science Mission Directorate
Near-Earth Objects Observations Program. J.L.P. is supported by NSF grant AST-0707982.  The PQ survey is supported by
the U.S.~National Science Foundation under Grants AST-0407448 and AST-0407297.  Support for M.C. is provided by Proyecto
Basal PFB-06/2007, by FONDAP Centro de Astrof\'{i}sica 15010003 and by MIDEPLAN~Rs Programa Iniciativa Cient\'{i}fica
Milenio through grant P07-021-F, awarded to The Milky Way Millennium Nucleus.  GALEX (Galaxy Evolution Explorer) is a
NASA Small Explorer, launched in 2003 April. We gratefully acknowledge NASA's support for construction, operation, and
science analysis for the GALEX mission, developed in cooperation with the Centre National d'Etudes Spatiales of France
and the Korean Ministry of Science and Technology.  The Expanded Very Large Array is operated by the National Radio
Astronomy Observatory, a facility of the National Science Foundation operated under cooperative agreement by Associated
Universities, Inc.  We thank the staff of the GMRT that made these observations possible. GMRT is run by the National
Centre for Radio Astrophysics of the Tata Institute of Fundamental Research.  The Fermi LAT Collaboration acknowledges
support from a number of agencies and institutes for both the development and the operation of the LAT as well as
scientific data analysis. These include NASA and DOE in the United States, CEA/Irfu and IN2P3/CNRS in France, ASI and
INFN in Italy, MEXT, KEK, and JAXA in Japan, and the K. A. Wallenberg Foundation, the Swedish Research Council and the
National Space Board in Sweden. We thank all the observers at ARIES who provided their valuable time and support for the
observations of this event. The UBVRI observations presented here are included here by R. Roy in partial fulfillment of
the requirements for a Ph.D degree.

\begin{deluxetable}{lll}
\tablecaption{Observation Sequence}
\small
\tablewidth{0pt}
\tablehead{\colhead{Telescope + Instrument} & \colhead{} & \colhead{Observation Date}}
\small\startdata

{\bf Photometry}   & {\it Filters} \\
SDSS               & u,g,r,i,z                  &  2002-12-31$\times 2$, 2003-03-26\\
CSS + 4K CDD       & V (unfiltered)             &  2003 - 2010  \\
ARIES 1m           & U,B,V,R,I                  &  2010-04+\\
Palomar 1.5m       & g,r,i,z                    &  2010-03-02, 2010-03-11, 2010-06-22, 2010-06-23 \\
SWIFT + UVOT       & UVW1, UVM2, UVW2, U, B, V  &  2010-04-06, 2010-04-25, 2010-05-09, 2010-05-23\\
GALEX              & NUV, FUV                   &  2004-01-24, 2010-01-29, 2010-04-17, 2010-04-29 \\
HST + WFC3         & F390W, F555W, F763M        &  2010-05-31\\
Keck + AO          &  NIR                       & 2010-06-02\\
LBT + LUCIFER      &  $\rm K_{s}$               & 2010-05-02\\
GHO 2.1m +CANICA   &  J, H, $\rm K_{s}$         & 2010-04+ \\
------------------- \\
{\bf Spectra}      & {\it Wavelength range} ($\rm \AA$) \\
SDSS + MOS         & $3800-9100$    &  2002-12-29 \\
IGO + IFOSC        & $4000-8600$    &  2010-02-18, 2010-04-04, 2010-04-23\\
P200 + DBSP        & $3500-9100$   &                   2010-03-15, 2010-11-09\\
APO + DIS          & $3500-9600$   &  2010-04-09 \\  
GALEX + NUV grism            & $1900-2800$    &  2010-04-17, 2010-04-29 \\
MDM 2.4m + Modspec & $4200-7560$    &  2010-05-04\\
Keck-I + LRIS      & $3100-10100$   &                    2010-05-18 \\
-------------------\\
{\bf Radio}      & {\it Central Frequency}  \\
EVLA              &  4.5 GHz + 7.9 GHz  & 2010-04-29, 2010-05-14, 2010-06-01\\
GMRT              &  608 MHz  & 2010-05-23\\
----------------- \\
{\bf X-ray \& $\gamma$-ray} & {\it Energy}  \\ \\
Swift +XRT   &   $0.2$ - $10$\, $\rm keV$    & 2010-04-06\\
Fermi +LAT   &    20\,MeV - $300$\,GeV    &          2009-11-20 - 2010-07-04\\
\enddata
\end{deluxetable}

\begin{deluxetable}{lll}
\tablecaption{CSS photometry of CSS100217}
\footnotesize
\tablewidth{0pt}
\tablehead{\colhead{MJD} & \colhead{Phase} & \colhead{$V_{CSS}$}}
\startdata
  54884.246 & -365.75 &  16.82$\pm$ 0.08\nl 
  54884.251 & -365.75 &  16.87$\pm$ 0.06\nl 
  54884.263 & -365.74 &  16.76$\pm$ 0.18\nl 
  54892.171 & -357.83 &  16.92$\pm$ 0.07\nl 
  54892.205 & -357.80 &  16.95$\pm$ 0.04\nl 
  54911.194 & -338.81 &  16.87$\pm$ 0.05\nl 
  54911.211 & -338.79 &  16.88$\pm$ 0.05\nl 
  54922.299 & -327.70 &  16.91$\pm$ 0.05\nl 
  54922.323 & -327.68 &  16.82$\pm$ 0.07\nl 
  54945.252 & -304.75 &  16.97$\pm$ 0.04\nl 
  54945.272 & -304.73 &  16.95$\pm$ 0.05\nl 
  55160.445 &  -89.55 &  16.78$\pm$ 0.16\nl 
  55160.453 &  -89.55 &  16.82$\pm$ 0.08\nl 
  55160.461 &  -89.54 &  16.82$\pm$ 0.09\nl 
  55160.469 &  -89.53 &  16.75$\pm$ 0.17\nl 
  55183.410 &  -66.59 &  16.32$\pm$ 0.09\nl 
  55183.414 &  -66.59 &  16.32$\pm$ 0.07\nl 
  55183.419 &  -66.58 &  16.39$\pm$ 0.11\nl 
  55183.424 &  -66.58 &  16.32$\pm$ 0.07\nl 
  55211.379 &  -38.62 &  15.80$\pm$ 0.03\nl 
  55211.386 &  -38.61 &  15.83$\pm$ 0.03\nl 
  55211.394 &  -38.61 &  15.80$\pm$ 0.02\nl 
  55211.401 &  -38.60 &  15.83$\pm$ 0.03\nl 
  55244.284 &   -5.72 &  15.74$\pm$ 0.02\nl 
  55244.291 &   -5.71 &  15.77$\pm$ 0.02\nl 
  55244.303 &   -5.70 &  15.71$\pm$ 0.02\nl 
  55272.198 &   22.20 &  15.77$\pm$ 0.02\nl 
  55272.203 &   22.20 &  15.77$\pm$ 0.02\nl 
  55272.208 &   22.21 &  15.75$\pm$ 0.02\nl 
  55272.213 &   22.21 &  15.80$\pm$ 0.02\nl 
  55297.123 &   47.12 &  15.82$\pm$ 0.02\nl 
  55297.131 &   47.13 &  15.82$\pm$ 0.03\nl 
  55297.138 &   47.14 &  15.84$\pm$ 0.02\nl 
  55297.146 &   47.15 &  15.83$\pm$ 0.03\nl 
  55323.210 &   73.21 &  15.97$\pm$ 0.03\nl 
  55323.216 &   73.22 &  16.00$\pm$ 0.03\nl 
  55323.221 &   73.22 &  15.98$\pm$ 0.03\nl 
  55323.227 &   73.23 &  16.00$\pm$ 0.04\nl 
  55328.254 &   78.25 &  16.01$\pm$ 0.04\nl 
  55328.261 &   78.26 &  16.06$\pm$ 0.04\nl 
  55328.267 &   78.27 &  16.00$\pm$ 0.04\nl 
  55337.188 &   87.19 &  16.03$\pm$ 0.05\nl 
  55337.196 &   87.20 &  16.04$\pm$ 0.04\nl 
  55337.203 &   87.20 &  16.04$\pm$ 0.04\nl 
  55337.211 &   87.21 &  16.03$\pm$ 0.04\nl 
  55354.159 &  104.16 &  16.13$\pm$ 0.04\nl 
  55354.163 &  104.16 &  16.14$\pm$ 0.04\nl 
  55354.168 &  104.17 &  16.14$\pm$ 0.04\nl 
  55354.173 &  104.17 &  16.18$\pm$ 0.05\nl 
  55355.154 &  105.15 &  16.13$\pm$ 0.05\nl 
  55355.159 &  105.16 &  16.18$\pm$ 0.05\nl 
  55355.164 &  105.16 &  16.19$\pm$ 0.05\nl 
  55355.169 &  105.17 &  16.14$\pm$ 0.05\nl 
  55362.163 &  112.16 &  16.22$\pm$ 0.05\nl 
  55362.168 &  112.17 &  16.18$\pm$ 0.05\nl 
  55362.172 &  112.17 &  16.21$\pm$ 0.05\nl 
  55362.177 &  112.18 &  16.18$\pm$ 0.05\nl 
  55381.168 &  131.17 &  16.31$\pm$ 0.08\nl 
  55381.169 &  131.17 &  16.29$\pm$ 0.07\nl 
  55381.170 &  131.17 &  16.30$\pm$ 0.07\nl 
  55381.171 &  131.17 &  16.28$\pm$ 0.07\nl 
  55497.470 &  247.47 &  16.70$\pm$ 0.15\nl 
  55497.476 &  247.48 &  16.64$\pm$ 0.10\nl 
  55497.482 &  247.48 &  16.66$\pm$ 0.10\nl 
  55497.487 &  247.49 &  16.61$\pm$ 0.10\nl 
  55513.417 &  263.42 &  16.67$\pm$ 0.10\nl 
  55513.425 &  263.43 &  16.66$\pm$ 0.10\nl 
  55513.433 &  263.43 &  16.65$\pm$ 0.10\nl 
  55513.441 &  263.44 &  16.66$\pm$ 0.10\nl 
\enddata
\tablecomments{
The phase of the event is taken relative to adopted maximum bright at MJD 55250.
The photometry includes the flux measured for the host galaxy in the difference image
template.
}
\end{deluxetable}

\begin{deluxetable}{llllllll}
\tablecaption{Filtered photometry of CSS100217 and host galaxy}
\footnotesize
\tablewidth{0pt}
\tablehead{\colhead{MJD} & \colhead{Phase} & \colhead{Band1} & \colhead{Band2} & \colhead{Band3} & \colhead{Band4} & \colhead{Band5} & \colhead{Band6}}
\startdata

{\bf SDSS} &  & {\bf u} & {\bf g} & {\bf r} & {\bf i} & {\bf z} & \nl

 52639.380 & -2610.62 &  18.20$\pm$ 0.01 &  17.84$\pm$ 0.01 &  17.63$\pm$ 0.01 &  17.29$\pm$ 0.01  &  17.34$\pm$ 0.01 & \nl
 52639.470 & -2610.53 &  18.17$\pm$ 0.01 &  17.84$\pm$ 0.01 &  17.64$\pm$ 0.01 &  17.35$\pm$ 0.01  &  17.36$\pm$ 0.01 & \nl
 52724.240 & -2525.76 &  18.24$\pm$ 0.01 &  17.90$\pm$ 0.01 &  17.70$\pm$ 0.01 &  17.38$\pm$ 0.01  &  17.45$\pm$ 0.01 & \nl
--- \nl
{\bf Pal. 1.5m} &  & {\bf g} & {\bf r} & {\bf i} & {\bf z}\nl


55257.248 & 7.25   & 16.43$\pm$0.22  & 16.27$\pm$0.17 & 16.04$\pm$0.08 & 16.18$\pm$0.12 & \nl 
55257.381 & 7.38   & \nodata         & 16.48$\pm$0.08 & 16.10$\pm$0.12 & 16.07$\pm$0.19 & \nl 
55266.508 & 16.51  & 16.44$\pm$0.21  & 16.26$\pm$0.16 & 16.05$\pm$0.19 & 16.30$\pm$0.21 & \nl
55369.171 & 119.17 & 17.58$\pm$0.19  & 17.31$\pm$0.19 & 16.80$\pm$0.23 & 16.68$\pm$0.21 & \nl 
55370.174 & 120.17 & 17.73$\pm$0.15  & 17.31$\pm$0.18 & 16.75$\pm$0.21 & 16.68$\pm$0.30 & \nl

--- \nl
{\bf AIRES} &  & {\bf U} & {\bf B} & {\bf V} & {\bf R} & {\bf I} & \nl

 55297.771 &   47.77 &  16.45$\pm$ 0.05   &  16.94$\pm$ 0.02        &  16.56$\pm$ 0.01        &  16.38$\pm$ 0.02 &  15.92$\pm$ 0.02 &\nl
 55298.799 &   48.80 &  16.52$\pm$ 0.05   &  16.95$\pm$ 0.02        &  16.53$\pm$ 0.02        &  16.40$\pm$ 0.04 &  15.87$\pm$ 0.03  &\nl
 55299.637 &   49.64 &  16.41$\pm$ 0.06   &  16.93$\pm$ 0.05        &  16.52$\pm$ 0.02        &  16.36$\pm$ 0.02        & \nodata &\nl
 55300.643 &   50.64 &  16.59$\pm$ 0.06   & \nodata       & \nodata       &  16.40$\pm$ 0.02        &  15.91$\pm$ 0.03 &\nl
 55310.656 &   60.66 &  16.55$\pm$ 0.03   & \nodata       & \nodata       &  16.47$\pm$ 0.02        &  15.90$\pm$ 0.02 &\nl
 55311.613 &   61.61 &  16.61$\pm$ 0.05   &  17.11$\pm$ 0.02        & \nodata       &  16.49$\pm$ 0.02        & $\pm$ 0.02 &\nl
 55312.632 &   62.63 &  16.72$\pm$ 0.05   &  17.11$\pm$ 0.04        &  16.67$\pm$ 0.03        &  16.59$\pm$ 0.03 &  15.94$\pm$ 0.03 &\nl
 55316.604 &   66.60 &  16.73$\pm$ 0.05   &  17.20$\pm$ 0.01        &  16.72$\pm$ 0.01        &  16.58$\pm$ 0.02 &  15.97$\pm$ 0.02 &\nl
 55320.610 &   70.61 &  16.68$\pm$ 0.14   & \nodata       & \nodata       &  16.61$\pm$ 0.03        &  16.03$\pm$ 0.02 &\nl
 55323.646 &   73.65 &  16.85$\pm$ 0.04   &  17.30$\pm$ 0.01        &  16.83$\pm$ 0.02        &  16.71$\pm$ 0.02 &  16.08$\pm$ 0.02 &\nl
 55324.601 &   74.60 &  16.89$\pm$ 0.04   &  17.30$\pm$ 0.03        &  16.83$\pm$ 0.01        &  16.69$\pm$ 0.03 &  16.09$\pm$ 0.03 & \nl
 55329.625 &   79.63 & \nodata  & \nodata       & \nodata       &  16.72$\pm$ 0.03        & \nodata & \nl
 55334.624 &   84.62 &  16.92$\pm$ 0.03   &  17.41$\pm$ 0.03        &  16.97$\pm$ 0.02        &  16.85$\pm$ 0.03 &  16.18$\pm$ 0.05 &\nl
 55346.611 &   96.61 & \nodata  &  17.60$\pm$ 0.02        &  17.10$\pm$ 0.02        &  16.99$\pm$ 0.03        & $\pm$ 0.03 &\nl
 55349.629 &   99.63 &  17.28$\pm$ 0.07   &  17.65$\pm$ 0.05        &  17.14$\pm$ 0.03        &  17.01$\pm$ 0.04 &  16.29$\pm$ 0.03 &\nl
 55356.654 &  106.65 &  17.19$\pm$ 0.07   &  17.70$\pm$ 0.03        &  17.21$\pm$ 0.02        &  17.14$\pm$ 0.03 &  16.41$\pm$ 0.04 &\nl
 55358.620 &  108.62 &  17.25$\pm$ 0.05   &  17.82$\pm$ 0.03        &  17.21$\pm$ 0.03        &  17.23$\pm$ 0.05 &  16.49$\pm$ 0.04 &\nl
 55362.622 &  112.62 &  17.14$\pm$ 0.04   &  17.77$\pm$ 0.02        &  17.23$\pm$ 0.02        & \nodata       &  16.49$\pm$ 0.04 &\nl
 55364.619 &  114.62 &  \nodata       &  17.80$\pm$ 0.06        &  17.19$\pm$ 0.06        & \nodata       &  16.49$\pm$ 0.04 &\nl
 55495.983 &  245.98 &  \nodata       &  19.49$\pm$ 0.20    &  18.61$\pm$ 0.08  &  19.15$\pm$ 0.20 &  17.64$\pm$ 0.10 &\nl
 55499.932 &  249.93 &  18.97$\pm$ 0.20 &  19.18$\pm$ 0.09  &  18.74$\pm$ 0.09  &  19.43$\pm$ 0.25 &  17.77$\pm$ 0.09 &\nl
 55509.979 &  259.98 &  19.28$\pm$ 0.24 &  19.41$\pm$ 0.09  &  18.58$\pm$ 0.06  &  19.20$\pm$ 0.19 &  17.99$\pm$ 0.13 &\nl
 55527.971 &  277.97 &  18.84$\pm$ 0.19 &  \nodata          &  18.97$\pm$ 0.10  &  20.58$\pm$ 0.77 & \nodata &\nl
 55533.979 &  283.98 &  19.10$\pm$ 0.28 &  19.97$\pm$ 0.18  &  19.10$\pm$ 0.10  &  20.68$\pm$ 0.86 &  17.96$\pm$ 0.29 &\nl
 55562.971 &  312.97 &  19.41$\pm$ 0.30 &  21.11$\pm$ 0.49  &  19.41$\pm$ 0.13  &  \nodata         &  18.24$\pm$ 0.30 &\nl

--- \nl
{\bf Swift} & & {\bf UVW1} & {\bf UVM2} & {\bf UVW2} & {\bf U}  & {\bf B} & {\bf V} \nl

 55292.580 &   42.58 &  16.10$\pm$ 0.03 &  16.09$\pm$ 0.02 &  16.27$\pm$ 0.02 &  15.76$\pm$ 0.03  &  16.55$\pm$ 0.03 &  16.18$\pm$ 0.04\nl
 55311.620 &   61.62 &  16.19$\pm$ 0.10 &  16.44$\pm$ 0.07 &  16.43$\pm$ 0.03 &  15.90$\pm$ 0.03  &  16.80$\pm$ 0.03 &  16.32$\pm$ 0.04\nl
 55325.940 &   75.94 &  16.39$\pm$ 0.03 &  16.46$\pm$ 0.03 &  16.64$\pm$ 0.03 &  16.07$\pm$ 0.03  &  16.84$\pm$ 0.03 &  16.35$\pm$ 0.04\nl
 55339.120 &   89.12 &  16.78$\pm$ 0.06 &  16.66$\pm$ 0.08 &  16.71$\pm$ 0.02 &  16.17$\pm$ 0.06  &  16.98$\pm$ 0.03 &  16.56$\pm$ 0.04\nl

--- \nl
{\bf Near-IR} & & {\bf J} & {\bf H} & {\bf K$_{s}$} & \nl

 55312.176 &   62.18 &  15.06$\pm$ 0.11 &  14.19$\pm$ 0.13 &  13.54$\pm$ 0.09 & & & \nl
 55313.311 &   63.31 &  15.23$\pm$ 0.12 &  14.38$\pm$ 0.14 &  13.44$\pm$ 0.12 & & & \nl
 55333.230 &   83.23 &  15.03$\pm$ 0.10 &  14.28$\pm$ 0.12 &  13.17$\pm$ 0.11 & & & \nl
 55336.209 &   86.21 &  15.22$\pm$ 0.11 &  14.18$\pm$ 0.13 &  13.03$\pm$ 0.10 & & & \nl
 55338.175 &   88.18 &  14.99$\pm$ 0.11 &  14.38$\pm$ 0.14 &  13.69$\pm$ 0.09 & & & \nl
 55340.162 &   90.16 &  15.59$\pm$ 0.13 &  14.46$\pm$ 0.12 &  13.74$\pm$ 0.10 & & & \nl
 55342.155 &   92.16 &  15.41$\pm$ 0.11 &  14.32$\pm$ 0.12 &  13.51$\pm$ 0.10 & & & \nl
 55361.191 &  111.19 &  15.15$\pm$ 0.11 &  14.09$\pm$ 0.12 &  13.17$\pm$ 0.09 & & & \nl
 55369.147 &  119.15 &  15.52$\pm$ 0.11 &  14.33$\pm$ 0.12 &  13.72$\pm$ 0.09 & & & \nl

\enddata
\tablecomments{
The phase of the event is taken relative to adopted maximum bright at MJD 55250.
SDS photometry is that of the host galaxy. Swift photometry measurements include 
the flux from the host galaxy. All other photometry have host galaxy flux subtracted.
}
\end{deluxetable}





\begin{deluxetable}{lcc}
\tablecaption{SDSS Spectrum Emission Lines}
\small
\tablehead{\colhead{Feature} & \colhead{Flux (1E-17 erg cm$^{-2}$ $\rm s^{-1}$)} & \colhead{$\rm \lambda (\AA)$}}
\startdata
$\rm H_{\beta b}$     & 750    & 4863.0\nl 
$\rm H_{\beta m}$     & 300    & 4861.5\nl 
$\rm H_{\beta n}$     & 297    & 4864.6\nl 
$\rm [OIII]_{4959 m}$ & 160    & 4956.8\nl 
$\rm [OIII]_{4959 n}$ & 134    & 4963.2\nl 
$\rm [OIII]_{5007 m}$ & 470    & 5003.9\nl 
$\rm [OIII]_{5007 n}$ & 400    & 5010.8\nl 
$\rm [OI]$            & 81     & 6306.6 \nl
$\rm [SII]$           & 75     & 6722.1 \nl
$\rm [SII]$           & \nodata & 6736.2 \nl
$\rm H_{\alpha b}$    & 1400   & 6564.4\nl 
$\rm H_{\alpha m}$    & 1490   & 6563.7\nl 
$\rm H_{\alpha n}$    & 900    & 6567.5\nl 
$\rm [NII]_m$         & 250    & 6548.5\nl 
$\rm [NII]_n$         & 163    & 6552.8\nl 
\enddata
\tablecomments{
Subscripts {\it b}, {\it m} and {\it n} denote the broad, medium and narrow components, respectively.
The FWHM fit values for these componets are 2899, 911, and 376 $\rm km s^{-1}$, respectively.
The velocity measurements are corrected for instrumental dispersion.
}
\end{deluxetable}

\begin{deluxetable}{ccccccccc}
\tablecaption{CSS100217 Emission Line Properties}
\label{Tab4}
\small
\tablewidth{0pt}
\tablehead{\colhead{Observation} & \colhead{MJD} & \colhead{Phase} & \colhead{$\rm H_{\beta b}$} & \colhead{$\rm H_{\beta m}$ } & \colhead{$\rm H_{\beta n}$} &
\colhead{$\rm H_{\alpha b}$} & \colhead{$\rm H_{\alpha m}$} & \colhead{$\rm H_{\alpha n}$}}
\startdata
{\bf FWHM} (km $\rm s^{-1}$) \nl
IGO$\rm _{Feb}$  & 55245 & -5 & 2106    &  \nodata & 465 & 2814    & 1126 & 323 \nl
P200$\rm _{Mar}$ & 55270 & 20 & 2614    &  \nodata & 610 & 3190    & 1126 & 323 \nl
MDM              & 55320 & 70 &3630     & 2106     & 435 & \nodata & 1219 & 320 \nl
Keck             & 55334 & 84 & 4357    & 1423     & 407 & 4879    & 1219 & 347 \nl
P200$\rm _{Nov}$ & 55509 & 164 & 4304   & 1772  & 405 & 3843    & 1406 & 328 \nl
{\bf Flux} ($\rm 10^{-17} erg\, cm^{-2} s^{-1}$) \nl
IGO$\rm _{Feb}$  & 55245 & -5 & 2105  & \nodata & 1195 & 4136    & 1126 & 1259 \nl
P200$\rm _{Mar}$ & 55270 & 20 & 2210  & \nodata & 1334 & 4136    & 1126 & 1888 \nl
MDM              & 55320 & 75 & 2280  & 1228    & 1501 & \nodata & 1219 & 2727 \nl
Keck             & 55334 & 84 & 2315  & 842     & 438  & 7031    & 1219 & 2937 \nl
P200$\rm _{Nov}$ & 55509 & 164 & 2877    & 2105    & 702  & 8686    & 1406 & 1888 \nl
\enddata
\tablecomments{
Phase is taken relative to maximum light at MJD 55250.
Subscripts {it b}, {\it m} and {\it n} denote the broad, medium and narrow components, respectively. 
Velocity measurements are not corrected for instrumental dispersion.
Palomar spectra taken in March 2010 and November 2010 are noted
as P200$\rm _{Mar}$ and P200$\rm _{Nov}$, respectively
and the IGO spectrum from February 2010 is noted as IGO$\rm _{Feb}$.
}
\end{deluxetable}

\begin{deluxetable}{lllllll}
\tablecaption{Correction parameters}
\footnotesize
\tablewidth{0pt}
\tablehead{\colhead{Spectrum} & \colhead{Phase} & \colhead{BB$\rm _T$} & \colhead{K$_V$} & \colhead{K$_R$} & \colhead{$b_V$} & \colhead{$b_R$}}
\startdata
IGO  & -5 & 16.2$\pm$1.8 & -0.13  & -0.11 & 23.6 & 25.1 \nl
P200 & 20 & 13.5$\pm$0.8 & -0.01  &  0.04 & 15.2 & 14.2 \nl
APO  & 45 & 12.5$\pm$0.8 & -0.08  &  0.12 & 14.8 & 12.1 \nl
MDM  & 70 & 11.5$\pm$1.2 &  0.02  &  0.08 & 12.2 & 10.0 \nl
Keck & 84 &  7.8$\pm$2.0 &  0.21  &  0.36 &  7.4 &  4.1 \nl
\enddata
\tablecomments{
Col. (1), gives the source of the spectrum used to determine corrections.
Col. (2), gives the time of the spectrum relative adopted maximum light at MJD 55250.
Col. (3), gives the Blackbody fit temperature in kilo kelvin.
Cols. (4 \& 5), give the K-correction in $V$ and $R$ bands respectively.
Cols. (6 \& 7), give the bolometric correction in $V$ and $R$ bands respectively.
}
\end{deluxetable}

\begin{deluxetable}{llll}
\tablecaption{EVLA Radio Measurements}
\label{radio}
\small
\tablewidth{0pt}
\tablehead{\colhead{2010 UT Date } & \colhead{Flux 4.5 GHz ($\mu$Jy)} & \colhead{Flux 7.9 GHz ($\mu$Jy)} & \colhead{$\alpha$}}
\startdata
Apr. 29.22 & 447 $\pm$ 26 & 399$\pm$24 &  -0.20 $\pm$ 0.02\\
May 14.07 & 312 $\pm$ 24 & 349 $\pm$ 24 &  +0.19 $\pm$ 0.02\\
Jun 01.05 & 408 $\pm$ 16 & 506  $\pm$ 28 &  +0.38 $\pm$ 0.03\\
\enddata
\end{deluxetable}

\newpage
\begin{thebibliography}{}
\bibitem[\protect\citeauthoryear{{Abadie} et~al.}{{Abadie} et~al.}{2010}]{Aba10}
{Abadie}, J., et al. 2010, astro-ph 1001.0165, ApJ, in press.
\bibitem[\protect\citeauthoryear{{Abazajian} et al.}{{Abazajian} et al.}{2009}]{Aba09}
{Abazajian}, K.~N., et al. 2009, ApJS, 182, 543
\bibitem[\protect\citeauthoryear{{Abdo} et al.}{{Abdo} et al.}{2009}]{Adb09}
{Abdo}, A. A., {Ackermann}, M., {Ajello}, M., et al. 2009, ApJ, 706, L138
\bibitem[\protect\citeauthoryear{{Abdo} et al.}{{Abdo} et al.}{2010b}]{Adb10a}
{Abdo}, A. A., {Ackermann}, M., {Ajello}, M., et al. 2010a, ApJS, 188, 405
\bibitem[\protect\citeauthoryear{{Abdo} et al.}{{Abdo} et al.}{2010a}]{Adb10b}
{Abdo}, A. A., {Ackermann}, M., {Ajello}, M., et al. 2010b, ApJ, 715, 429
\bibitem[\protect\citeauthoryear{{Adelman-McCarthy} et al.}{{Adelman-McCarthy} et al.}{2006}]{Ald06}
{Adelman-McCarthy}, J. K. et al. 2006, ApJS, 162, 38
\bibitem[\protect\citeauthoryear{{Ai} et~al.}{{Ai} et~al.}{2010}]{Ai10}
{Ai}, Y.L., {Yuan}, W., {Zhou}, H.Y., {Wang}, T.G., {Dong}, X.-B. et al. 2010, ApJ, 716, 31
\bibitem[\protect\citeauthoryear{{Alcock} et~al.}{{Alcock} et~al.}{2003}]{Alc03}
{Alcock}, C., et al.~2003, Nature, 365, 621
\bibitem[\protect\citeauthoryear{{Anderson}, \& {King}}{{Anderson}, \& {King}}{2000}]{And00}
{Anderson}, J., \& {King}, I.R., 2000, PASP, 112, 1360
\bibitem[\protect\citeauthoryear{{Atwood} et al.}{{Atwood} et al.}{2009}]{Atw10}
{Atwood}, W. B., {Abdo}, A. A., {Ackermann}, M., et al. 2009, ApJ, 697, 1071
\bibitem[\protect\citeauthoryear{{Aubourg} et~al.}{{Aubourg} et~al.}{1993}]{Aub93}
{Aubourg}, E., et al. 1993, Nature, 365, 623
\bibitem[\protect\citeauthoryear{{Baars} et~al.}{{Baars} et~al.}{1977}]{Bar77}
{Baars}, J.W.M., {Genzel}, R., {Pauliny-Toth}, I.I.K., {Witzel}, A. 1977, A\&A, 61, 99
\bibitem[\protect\citeauthoryear{{Baldwin}, {Phillips} \&  {Terlevich}}{{Baldwin} et~al.}{1981}]{Bal81}
{Baldwin}, J. A., {Phillips}, M. M., \& {Terlevich}, R. 1981, PASP, 93, 5
\bibitem[\protect\citeauthoryear{{Barthelmy} et~al.}{{Barthelmy} et~al.}{2005}]{Bar05}
{Barthelmy}, S.D., et al. 2005, SSRv, 120, 143
\bibitem[\protect\citeauthoryear{{Bauer} et~al.}{{Bauer} et~al.}{2009}]{Bau09}
{Bauer}, A., et al. 2009, ApJ, 699, 1732
\bibitem[\protect\citeauthoryear{{Becker}, {White} \& {Helfand}}{{Becker} et~al.}{1995}]{Bec95}
{Becker}, R. H., {White}, R. L., {Helfand}, D. J. 1995, ApJ, 450, 559
\bibitem[\protect\citeauthoryear{{Bennert} et al.}{{Bennert} et al.}{2006}]{Ben06}
{Bennert}, N. et al. 2006, A\&A, 459, 55
\bibitem[\protect\citeauthoryear{{Bessell}, \& {Brett}}{{Bessell}, \& {Brett}}{1988}]{Bes88}
{Bessell} M.S., \&  {Brett}, J.M., 1988, PASP, 1134
\bibitem[\protect\citeauthoryear{{Bentz}, et~al.}{{Bentz}, et~al.}{2009}]{Ben09}
{Bentz}, M.C., {Peterson}, B.M., {Netzer}, H., {Pogge}, R.W. \& {Vestergaard}, M. 2009, ApJ, 697, 160
\bibitem[\protect\citeauthoryear{{Binney} \& {Merrifield}}{{Binney} \& {Merrifield}}{1998}]{Bin98}
{Binney}, J., \& {Merrifield}, M. 1998, Galactic Astronomy, Ed. J.P. Ostriker \& D.N. Spergel.
\bibitem[\protect\citeauthoryear{{Boller}, {Brandt} \& {Fink}}{{Boller} et~al.}{1996}]{Bol96}
{Boller},R., {Brandt}, W.N., \& {Fink}, H. 1996, A\&A, 305, 53
\bibitem[\protect\citeauthoryear{{Cappelluti} et al.}{{Cappelluti} et al.}{2009}]{Cap09}
{Cappelluti}, N., et al. 2009, A\&A, 495, L9
\bibitem[\protect\citeauthoryear{{Catalan} et~al.}{{Catalan} et~al.}{2009}]{Cat09}
{Catelan}, M. et~al. 2009, CBET, 1780, 1
\bibitem[\protect\citeauthoryear{{Chevalier}, {Fransson}, \& {Nymark}}{{Chevalier} et al.}{2006}]{Cge06}
{Chevalier}, R.~A. and {Fransson}, C. and {Nymark}, T.~K., 2006, ApJ, 641, 1029
\bibitem[\protect\citeauthoryear{{Figer}, {McLean}  \& {Morris}}{{Figer} et~al.}{1995}]{Fig95}
{Figer}, D. F., {McLean}, I. S., \& {Morris}, M. 1995, ApJL, 447, 29
\bibitem[\protect\citeauthoryear{{Condon} et~al.}{{Condon} et~al.}{1998}]{Con98}
{Condon}, J.J., et al. 1998, AJ, 115, 1963
\bibitem[\protect\citeauthoryear{{Cooke} et~al.}{{Cooke} et~al.}{2010}]{Coo10}
{Cooke}, J., et al.~2010, ATel, 2491
\bibitem[\protect\citeauthoryear{{Crenshaw}, {Kraemer} \&  {Gabel}}{{Crenshaw} et~al.}{2003}]{Cre03}
{Crenshaw}, D.N., {Kraemer}, S.B., \& {Gabel} J.R. 2003, AJ, 126, 1690 
\bibitem[\protect\citeauthoryear{{Croft} et al.}{{Croft} et al.}{2009}]{Cro09}
{Croft}, S., et al. 2009, BAAS, 41, 402 
\bibitem[\protect\citeauthoryear{{Deo}, {Crenshaw} \& {Kraemer}}{{Deo} et~al.}{2006}]{Deo06}
{Deo}, R.P., {Crenshaw}, D. M., \&  {Kraemer}, S.B. 2006, AJ, 132, 321
\bibitem[\protect\citeauthoryear{{Djorgovski} et~al.}{{Djorgovski} et~al.}{2008}]{Djo01a} {Djorgovski}, S.G., et al.
  2001a, Exploration of Large Digital Sky Surveys, in: Mining the Sky, eds. A.J. Banday et al., ESO Astrophysics
  Symposia, (Berlin: Springer Verlag), 305
\bibitem[\protect\citeauthoryear{{Djorgovski} et~al.}{{Djorgovski} et~al.}{2008}]{Djo01b}
{Djorgovski}, S.G., et al. 2001b, Exploration of Parameter Spaces in a Virtual Observatory, in: Astronomical Data
  Analysis, eds. J.-L. Starck \& F. Murtagh, Proc. SPIE, 4477, 43.
\bibitem[\protect\citeauthoryear{{Djorgovski} et~al.}{{Djorgovski} et~al.}{2008}]{Djo08}
{Djorgovski}, S.G., et al. 2008, AN, 329, 263
\bibitem[\protect\citeauthoryear{{Drake}, et~al.}{{Drake}, et~al.}{2009}]{Dra09}
{Drake}, A.J., et al. 2009, ApJ, 696, 870
\bibitem[\protect\citeauthoryear{{Drake}, et~al.}{{Drake}, et~al.}{2010}]{Dra10a}
{Drake}, A.J., et al. 2010a, ApJL, 718, 127
\bibitem[\protect\citeauthoryear{{Drake}, et~al.}{{Drake}, et~al.}{2010b}]{Dra10b}
{Drake}, A.J., et al. 2010b, ATel, 2725
\bibitem[\protect\citeauthoryear{{Drake}, et~al.}{{Drake}, et~al.}{2010c}]{Dra10c}
{Drake}, A.J., et al. 2010c, ATel, 2784
\bibitem[\protect\citeauthoryear{{Esquej} et~al.}{{Esquej} et~al.}{2007}]{Esq07}
{Esquej}, P. et al. 2007, A\&A, 462, L49
\bibitem[\protect\citeauthoryear{{Fransson} et~al.}{{Fransson} et~al.}{2005}]{Fra05}
{Fransson}, C., et al. 2005, ApJ, 622, 991
\bibitem[\protect\citeauthoryear{{Gal-Yam} et~al.}{{Gal-Yam} et~al.}{2009}]{Gal09b}
{Gal-Yam}, A., et al. 2010, Nature, 465, 322
\bibitem[\protect\citeauthoryear{{Gerardy} et al.}{{Gerardy} et al.}{2002}]{Ger02}
{Gerardy}, C.L., et al. 2002, ApJ, 575, 1007
\bibitem[\protect\citeauthoryear{{Gezari} et al.}{{Gezari} et al.}{2009}]{Gez09a}
{Gezari}, S., et al. 2009a, ApJ, 698, 1367
\bibitem[\protect\citeauthoryear{{Gezari} et al.}{{Gezari} et al.}{2009}]{Gez09b}
{Gezari}, S., et al. 2009b, ApJ, 690, 1313
\bibitem[\protect\citeauthoryear{{Gezari} et~al.}{{Gezari} et~al.}{2010}]{Gez10}
{Gezari}, S., {Forster}, K., {Neill}, J.D., {Martin}, D.C., Astron. Tel. 2010, 2554, 1
\bibitem[\protect\citeauthoryear{{Goodman}}{{Goodman}}{2003}]{Gol03}
{Goodman}, J. 2003, MNRAS, 339, 937
\bibitem[\protect\citeauthoryear{{Goodman} \& {Tan}}{{Goodman} \& {Tan}}{2003}]{Gol04}
{Goodman}, J. \& {Tan}, J. 2004, ApJ, 608, 108.
\bibitem[\protect\citeauthoryear{{Hagen}, {Engels} \& {Reimers}}{{Hagen} et~al.}{1997}]{Hag97}
{Hagen}, H.J., {Engels}, D. \& {Reimers}, D., 1997, A\&A, 324, L29
\bibitem[\protect\citeauthoryear{{Halpern}, {Leighly}, \& {Marshall}}{{Halpern} et al.}{2003}]{Hal03}
{Halpern}, J. P., {Leighly}, K. M., \& {Marshall}, H. L., 2003, ApJ, 585, 665
\bibitem[\protect\citeauthoryear{{Ho} et~al.}{{Ho} et~al.}{1997}]{Ho97}
{Ho}, L.C., et al.~1997, ApJ, 487, 568
\bibitem[\protect\citeauthoryear{{Hodapp}, et~al.}{{Hodapp}, et~al.}{2004}]{Hod04}
{Hodapp}, K.W., et al.~2004, AN, 325, 636 
\bibitem[\protect\citeauthoryear{{Hills}}{{Hills}}{1975}]{Hil75}
{Hills}, J.G. 1975, Nature,  254, 295
\bibitem[\protect\citeauthoryear{{Imanishi}, \& {Wada}}{{Imanishi}, \& {Wada}}{2004}]{Ima04}
{Imanishi}, M., {Wada}, K., 2004, ApJ, 617, 214
\bibitem[\protect\citeauthoryear{{Ivezic}, et~at.}{{Ivezic}, et~at.}{2008}]{Ive08}
{Ivezic}, Z., et~al. 2008, astro-ph 0805.2366
\bibitem[\protect\citeauthoryear{{Izotov} et al.}{{Izotov} et al.}{2007}]{Izo07}
{Izotov}, Y. I., {Thuan}, T. X., \& {Guseva}, N. G. 2007, ApJ, 671, 1297
\bibitem[\protect\citeauthoryear{{Jahoda} et al.}{{Jahoda} et al.}{1996}]{Jah96}
{Jahoda}, K. et al. 1996, SPIE, 2808, 59 
\bibitem[\protect\citeauthoryear{{Jester}, {Schnider}, {Richards}}{{Jester} et al.}{2005}]{Jah05}
{Jester}, S., Schnieder, D. P., \& Richards, G.T. 2005, AJ, 130, 873 
\bibitem[\protect\citeauthoryear{{Jiang} \& {Goodman}}{{Jiang} \& {Goodman}}{2010}]{Jia10}
{Jiang}, Y., \& {Goodman}, J. 2010, ApJ, submitted (arXiv/1011.3541)
\bibitem[\protect\citeauthoryear{{Johnston}}{{Johnston}}{2007}]{Joh07}
{Johnston}, S. et al.~2007, PASP, 24, 174
\bibitem[\protect\citeauthoryear{{Kankare} et~al.}{{Kankare} et~al.}{2010}]{Kan10}
{Kankare}, E., et al. 2010, ATel, 2716
\bibitem[\protect\citeauthoryear{{Kauffmann} et~al.}{{Kauffmann} et al.}{2003}]{Kau03}
{Kauffmann}, G. et al. 2003, MNRAS, 346, 1055
\bibitem[\protect\citeauthoryear{{Keller}, et~al.}{{Keller}, et~al.}{2007}]{Kel07}
{Keller}, S.C., et~al. 2007, PASA, 24, 1
\bibitem[\protect\citeauthoryear{{Kewley} et~al.}{{Kewley} et~al.}{2001}]{Kew01}
{Kewley}, L. J. et al. 2001, ApJ, 556, 121
\bibitem[\protect\citeauthoryear{{Kewley} et~al.}{{Kewley} et~al.}{2006}]{Kew06}
{Kewley}, L.J., {Groves}, B., {Kauffmann}, G., \& {Heckman}, T. 2006, MNRAS, 372, 961
\bibitem[\protect\citeauthoryear{{Klebesadel}, R. W., {Strong}, I. B., \& {Olson}}{{Klebesadel} et~al.}{1973}]{Kle73}
{Klebesadel}, R. W., {Strong}, I. B., \& {Olson}, R. A. 1973, ApJ, 182, L85
\bibitem[\protect\citeauthoryear{{Kozlowski} et~al.}{{Kozlowski} et~al.}{2010}]{Koz10}
{Kozlowski}, S., et al. 2010, ApJ, 722, 1624
\bibitem[\protect\citeauthoryear{{Komossa} et~al.}{{Komossa} et~al.}{1999}]{Kom99}
{Komossa}, S., {Bohringer}, H., {Huchra}, J.P. 1999, A\&A, 349, 45
\bibitem[\protect\citeauthoryear{{Komossa} et~al.}{{Komossa} et~al.}{2002}]{Kom02}
{Komossa}, S., et al. 2002,  Lighthouses of the Universe: Proc.~of the MPA/ESO/MPE/USM 
Joint Astronomy Conf., Ed. M. Gilfanov, R. Sunyaev, and E. Churazov, 436
\bibitem[\protect\citeauthoryear{{Komossa} et~al.}{{Komossa} et~al.}{2008}]{Kom08}
{Komossa}, S., et al. 2008, ApJL, 678, 13
\bibitem[\protect\citeauthoryear{{Komossa} et~al.}{{Komossa} et~al.}{2009}]{Kom09}
{Komossa}, S., et al. 2009, ApJ, 701, 105
\bibitem[\protect\citeauthoryear{{Landolt}}{{Landolt}}{1992}]{Lan92}
{Landolt}, A.U., 1992, AJ, 194, 3401 
\bibitem[\protect\citeauthoryear{{Larson}}{{Larson}}{2003}]{Lar03}
{Larson}, S., et~al. 2003, BAAS, 35, 982 
\bibitem[\protect\citeauthoryear{{Leighly}, \& {Moore}}{{Leighly}, \& {Moore}}{2004}]{Lei04}
{Leighly}, K. \& {Moore}, J. R. 2004, ApJ, 611, 107
\bibitem[\protect\citeauthoryear{{Li} et~al.}{{Li} et~al.}{2006}]{Li06}
{Li}, W., et al. 2006, PASP, 118, 37.
\bibitem[\protect\citeauthoryear{{Liu} et~al.}{{Liu} et~al.}{2010}]{Liu10}
{Liu}, H., {Wang}, J., {Mao}, Y., \& {Wei}, J., 2010, astro-ph 1005.0916
\bibitem[\protect\citeauthoryear{{Ludato} \& {Rossi}}{{Ludato} \& {Rossi}}{2010}]{Lud}
{Lodato}, G., \& {Rossi}, E.M., 2010, MNRAS, in press. 
\bibitem[\protect\citeauthoryear{{Maksym}, {Ulmer}, {Eracleous}}{{Maksym} et~al.}{2010}]{Mak10}
{Maksym}, W.P., {Ulmer}, M.P., {Eracleous}, M., 2010, ApJ, 722, 1035
\bibitem[\protect\citeauthoryear{{Mao}, {Wang}, \& {Wei}}{{Mao} et~al.}{2009}]{Mao09}
{Mao}, Y., Wang, J., Wei, J. 2009, ApJ, 698, 859
\bibitem[\protect\citeauthoryear{{Martin}, et~al.}{{Martin}, et~al.}{2005}]{Mar05}
{Martin}, D.C., et~al. 2005, ApJ, 619, L1
\bibitem[\protect\citeauthoryear{{Mattila}, {Meikle} \& {Greimel}}{{Mattila} et al.}{2004}]{Mat04}
{Mattila}, S., {Meikle}, W. P. S. \& {Greimel}, 2004, R. Astronomy Reviews, 48, 595
\bibitem[\protect\citeauthoryear{{Magorrian} \& {Tremaine}}{{Magorrian} \& {Tremaine}}{1999}]{Mag99}
{Magorrian}, J., {Tremaine}, S. 1999, MNRAS, 309, 447 
\bibitem[\protect\citeauthoryear{{Mauch}, \& {Sadler}}{{Mauch}, \& {Sadler}}{2007}]{Mau07}
{Mauch}, T. \& {Sadler}, E.~M., 2007, 375, 931
\bibitem[\protect\citeauthoryear{{Miller}, et~al.}{{Miller}, et~al.}{2009}]{Mil09}
{Miller}, A.A., et al. 2009, ApJ, 690, 1303
\bibitem[\protect\citeauthoryear{{Monet}, et~al.}{{Monet}, et~al.}{2003}]{Mon03}
{Monet}, D., et al. 2003, AJ, 125, 984
\bibitem[\protect\citeauthoryear{{Nakar}, {Piran} \& {Granot}}{{Nakar} et~al.}{2002}]{Nak02}
{Nakar}, E., {Piran}, T., \& {Granot}, J.~2002, ApJ, 579, 699
\bibitem[\protect\citeauthoryear{{Nelson} et al.}{{Nelson} et al.}{2009}]{Nel09}
{Nelson}, C. A., et al. 2009, arXiv:0902.2213
\bibitem[\protect\citeauthoryear{{Osterbrock} \& {Pogge}}{{Osterbrock} \& {Pogge}}{1985}]{Ost85}
{Osterbrock}, D.E., \& {Pogge} R.W., 1985, ApJ, 297, 166
\bibitem[\protect\citeauthoryear{{Paczynski}}{{Paczynski}}{1986}]{Pac86}
{Paczynski}, B., 1986, ApJ, 304, 1
\bibitem[\protect\citeauthoryear{{Paczynski}}{{Paczynski}}{1997}]{Pac97}
{Paczynski}, B., 1997, atro-ph 9712123
\bibitem[\protect\citeauthoryear{{Paczynski}}{{Paczynski}}{2000}]{Pac00}
{Paczynski}, B. 2000, PASP, 112, 1281.
\bibitem[\protect\citeauthoryear{{Perez-Torres} et~al.}{{Perez-Torres} et~al.}{2007}]{Per07}
{Perez-Torres}, M.A., et al., 2007, ApJ, 671, L21 
\bibitem[\protect\citeauthoryear{{Perez-Torres} et~al.}{{Perez-Torres} et~al.}{2010}]{Per10}
{Perez-Torres}, M.A., et al., 2010, astro-ph 1008.4466
\bibitem[\protect\citeauthoryear{{Quimby}}{{Quimby}}{2006}]{Qui06}
{Quimby}, R.M. et~al. 2006, CBET, 644, 1
\bibitem[\protect\citeauthoryear{{Rau} et~al.}{{Rau} et~al.}{2009}]{Rau09}
{Rau}, A., et al. 2009, PASP, 121, 1334
\bibitem[\protect\citeauthoryear{{Rest} et~al.}{{Rest} et~al.}{2009}]{Res09}
{Rest}, A., et al. 2009, astro-ph 0911.2002
\bibitem[\protect\citeauthoryear{{Richardson}, et~al.}{{Richardson}, et~al.}{2002}]{Ric02}
{Richardson}, D., et al. 2002, ApJ, 123, 745
\bibitem[\protect\citeauthoryear{{Riffel}, {Pastoriza}, {Rodriguez-Ardila}, {Maraston}}{{Riffel}, et~al.}{2007}]{Rif07}
{Riffel}, R., {Pastoriza}, M. G., {Rodriguez-Ardila}, A., {Maraston}, C. 2007, ApJ, 659, 103
\bibitem[\protect\citeauthoryear{{Rengelink} et~al.}{{Rengelink} et~al.}{1997}]{Ren97}
{Rengelink}, R. B., et al 1997, A\&AS, 124, 259
\bibitem[\protect\citeauthoryear{{Rottgering}}{{Rottgering}}{2003}]{Rot03}
{Rottgering}, H.J.A., 2003, NewAR, 47, 405
\bibitem[\protect\citeauthoryear{{Sesar}, et~al.}{{Sesar}, et~al.}{2007}]{Ses07}
{Sesar}, B., et al.~2007 (the SDSS team), AJ, 134, 2236
\bibitem[\protect\citeauthoryear{{Schawinski} et~al.}{{Schawinski} et~al.}{2008}]{Sch08}
{Schawinski}, K. et al. 2008, Science, 321, 223
\bibitem[\protect\citeauthoryear{{Schlegel}, et~al.}{{Schlegel}, et~al.}{1998}]{Smi98}
{Schlegel}, D., et al., 1998, ApJ, 500, 525
\bibitem[\protect\citeauthoryear{{Shlosman} \& {Begelman}}{{Shlosman} \& {Begelman}}{1987}]{Shl87}
{Shlosman}, I., \& {Begelman}, M. 1987, Nat, 329, 810.
\bibitem[\protect\citeauthoryear{{Shlosman} \& {Begelman}}{{Shlosman} \& {Begelman}}{1989}]{Shl89}
{Shlosman}, I., \& {Begelman}, M. 1989,  ApJ, 341, 685.
\bibitem[\protect\citeauthoryear{{Smith}, et~al.}{{Smith}, et~al.}{2007}]{Smi07}
{Smith}, N., et al., 2007, ApJ, 666, 1116
\bibitem[\protect\citeauthoryear{{Smith}, et~al.}{{Smith}, et~al.}{2010}]{Smi10a}
{Smith}, N., et al., 2010a, ApJ, 709, 856
\bibitem[\protect\citeauthoryear{{Smith}, et~al.}{{Smith}, et~al.}{2010}]{Smi10b}
{Smith}, N., et al., 2010b, AJ, 139, 1451
\bibitem[\protect\citeauthoryear{{Stanek}, et~al.}{{Stanek}, et~al.}{2003}]{sta03}
{Stanek}, K.Z., et al., 2003, ApJL, 591, 17
\bibitem[\protect\citeauthoryear{{Soderberg} et al.}{{Soderberg} et al.}{2008}]{Sod08}
{Soderberg}, A.M., et al. 2008, Nature, 453, 469
\bibitem[\protect\citeauthoryear{{Stasinska} et~al.}{{Stasinska} et~al.}{2006}]{Sta06}
{Stasinska}, G., et al. 2006, MNRAS, 371, 972
\bibitem[\protect\citeauthoryear{{Stetson}}{{Stetson}}{1987}]{Ste87}
{Stetson}, P.B., 1987, PASP, 99, 191
\bibitem[\protect\citeauthoryear{{Strubbe} \& {Quataert}}{{Strubbe} \& {Quataert}}{2009}]{Str09}
{Strubbe}, L.E., \& {Quataert}, E., 2009, MNRAS, 400, 2070
\bibitem[\protect\citeauthoryear{{Strubbe} \& {Quataert}}{{Strubbe} \& {Quataert}}{2010}]{Str10}
{Strubbe}, L.E., \& {Quataert}, E., 2010, astro-ph 1008.4131
\bibitem[\protect\citeauthoryear{{Skrutskie}, et~al.}{{Skrutskie}, et~al.}{2006}]{Skr06}
{Skrutskie}, M. F., et al.~2006, AJ, 131, 1163
\bibitem[\protect\citeauthoryear{{Tomaney} \& {Crotts}}{{Tomaney} \& {Crotts}}{1996}]{Tom96}
{Tomaney}, A. B., \& {Crotts}, A.P.S., 1996, AJ, 112, 2872
\bibitem[\protect\citeauthoryear{{Trelevich} et~al.}{{Trelevich} et~al.}{1992}]{Tre92}
{Trelevich}, R., et al. 1992, MNRAS, 255, 713
\bibitem[\protect\citeauthoryear{{Trundle} et~al.}{{Trundle} et~al.}{2009}]{Tru09}
{Trundle}, C., et al. 2009, A\&A, 504, 945
\bibitem[\protect\citeauthoryear{{Turatto} et~al.}{{Turatto} et~al.}{1993}]{Tur93}
{Turatto}, M., {Cappellaro}, E., {Danziger}, I. J., {Benetti}, S., {Gouiffes}, C., 
{della Valle}, M. 1993, MNRAS, 262, 128
\bibitem[\protect\citeauthoryear{{Ueno} et~al.}{{Ueno} et~al.}{2008}]{Uen08}
{Ueno}, S. et al. 2008, SPIE, 7011, 75 
\bibitem[\protect\citeauthoryear{{Ulrich}, {Maraschi} \& {Urry}}{{Ulrich} et al.}{1997}]{Ulr97}
{Ulrich}, M.-H., {Maraschi}, L., \& {Urry}, C. M., 1997, ARAA, 35, 445
\bibitem[\protect\citeauthoryear{{Ulvestad} \& {Ho}}{{Ulvestad} \& {Ho}}{2001}]{Ulv01}
{Ulvestad}, J.S., \& {Ho}, L.C. 2001  ApJ, 558, 561
\bibitem[\protect\citeauthoryear{{Valenti} et~al.}{{Valenti} et~al.}{2010}]{Val10}
{Valenti}, S., et al. 2010, ATel, 2773
\bibitem[\protect\citeauthoryear{{van Velzen} et~al.}{{van Velzen} et~al.}{2010}]{Vel10}
{van Valzen}, S., et al. 2010, astro-ph 1009.1627
\bibitem[\protect\citeauthoryear{{V\'eron-Cetty} \& {V\'eron}}{{V\'eron-Cetty} \& {V\'eron}}{2010}]{Ver10}
{V\'eron-Cetty}, M.-P. \& {V\'eron}, P., 2010, A\&A, 518, 10 
\bibitem[\protect\citeauthoryear{{Wang}, \& {Merritt}}{{Wang}, \& {Merritt}}{2004}]{Wan04}
{Wang}, J. \& {Merritt}, D. 2004, ApJ, 600, 149
\bibitem[\protect\citeauthoryear{{Webb} \& {Malkan}}{{Webb} \& {Malkan}}{2000}]{Web00}
{Webb}, W., \& {Malkan}, M. 2000, ApJ, 540, 652
\bibitem[\protect\citeauthoryear{{Wizinovich} et~al.}{{Wizinovich} et~al.}{2006}]{Wiz06}
{Wizinovich}, P., et al. 2006, PASP, 118, 297
\bibitem[\protect\citeauthoryear{{Yuan}, et~al.}{{Yuan}, et~al.}{2008}]{Yuan09}
{Yuan}, W., {Zhou}, H. Y., {Komossa}, S., {Dong}, X. B.; {Wang}, T. G., {Lu}, H. L., {Bai}, J. M., 2008, ApJ, 685, 801
\bibitem[\protect\citeauthoryear{{Zhou} et~al.}{{Zhou} et~al.}{2006}]{Zha06}
{Zhao} F.Y., {Strom}, R.G., {Jiang}, S.Y., 2006, Chinese J Astron Astrophys, 6, 635
\bibitem[\protect\citeauthoryear{{Zhou} et~al.}{{Zhou} et~al.}{2003}]{Zho03}
{Zhou}, H.-Y., {Wang}, T.-G., {Dong}, X.-B., {Zhou}, Y.-Y., \& {Li} C. 2003, ApJ, 584, 147.
\bibitem[\protect\citeauthoryear{{Zhou} et~al.}{{Zhou} et~al.}{2006}]{Zho06}
{Zhou}, H. Y., et al. 2006, ApJS, 166, 128 
\end{thebibliography}
\end{document}